\documentclass[preprint]{aastex62}
\usepackage{epsfig, natbib, graphicx, color, amsmath, cancel, icomma, bm}
\usepackage{booktabs}
\usepackage{hyperref}

\usepackage{bookmark}
\bookmarksetup{}

\graphicspath{{cluster-images/}{figures/}{flags/}{flowchart/}}

\newcommand*{\matcha}{MATCha}
\newcommand*{\chandra}{\emph{Chandra}}
\newcommand*{\xmm}{\emph{XMM}}
\newcommand*{\redmapper}{redMaPPer}
\newcommand*{\sdss}{SDSS}
\newcommand*{\memmatch}{MEM\_MATCH\_ID}
\newcommand*{\rtfh}{$r_{2500}$}
\newcommand*{\rfh}{$r_{500}$}
\newcommand*{\lx}{$L_X$}
\newcommand*{\tx}{$T_X$}
\newcommand*{\sigmaintr}{\sigma_\mathrm{intr}}

\newcommand*{\colorinfosingle}{As with every \chandra{} image in this paper, this image obeys the coloring conventions described in the caption for \autoref{fig:matcha-typical}.}
\newcommand*{\colorinfo}{As with every \chandra{} image in this paper, these images obey the coloring conventions described in the caption for \autoref{fig:matcha-typical}.}
\newcommand*{\txrelation}[3][]{\ln{\left(\frac{k_{B}T_{X \ifx\\#1\\\else,\fi #1}}{1.0 \mathrm{keV}}\right)} = \left(#3\right)\ln{\left(\frac{\lambda{}_{#1}}{70}\right)} + \left(#2\right)}
\newcommand*{\lxrelation}[3][]{\ln{\left(\frac{L_{X \ifx\\#1\\\else,\fi #1}}{E\left(z\right)\cdot{}{10^{44}}\mathrm{ergs / s}}\right)} = \left(#3\right)\ln{\left(\frac{\lambda{}_{#1}}{70}\right)} + \left(#2\right)}
\newcommand*{\txlxrelation}[3][]{\ln{\left(\frac{k_{B}T_{X \ifx\\#1\\\else,\fi #1}}{1.0 \mathrm{keV}}\right)} = \left(#3\right)\ln{\left(\frac{L_{X \ifx\\#1\\\else,\fi #1}}{E\left(z\right)\cdot{}10^{44}\mathrm{ergs / s}}\right)} + \left(#2\right)}
\newcommand*{\lxtxrelation}[3][]{\ln{\left(\frac{L_{X \ifx\\#1\\\else,\fi #1}}{E\left(z\right)\cdot{}{10^{44}}\mathrm{ergs / s}}\right)} = \left(#3\right)\ln{\left(\frac{k_{B}T_{X \ifx\\#1\\\else,\fi #1}}{2.0 \mathrm{keV}}\right)} + \left(#2\right)}

\begin{document}

\title{\chandra{} Follow-Up of the \sdss{} DR8 \redmapper{} Catalog Using the \matcha{} Pipeline}

\author{Devon~L.~Hollowood}
\affiliation{Santa Cruz Institute for Particle Physics, Santa Cruz, CA 95064, USA}
\affiliation{University of California Santa Cruz, Santa Cruz, CA 95060, USA}
\author{Tesla~Jeltema}
\affiliation{Santa Cruz Institute for Particle Physics, Santa Cruz, CA 95064, USA}
\affiliation{University of California Santa Cruz, Santa Cruz, CA 95060, USA}
\author{Xinyi~Chen}
\affiliation{Department of Physics, University of Michigan, Ann Arbor, MI 48109, USA}
\affiliation{Department of Physics, Yale University, New Haven, CT 06511, USA}
\author{Arya~Farahi}
\affiliation{Department of Physics, University of Michigan, Ann Arbor, MI 48109, USA}
\author{August~Evrard}
\affiliation{Department of Physics, University of Michigan, Ann Arbor, MI 48109, USA}
\author{Spencer~Everett}
\affiliation{Santa Cruz Institute for Particle Physics, Santa Cruz, CA 95064, USA}
\affiliation{University of California Santa Cruz, Santa Cruz, CA 95060, USA}
\author{Eduardo~Rozo}
\affiliation{Department of Physics, University of Arizona, Tucson, AZ 85721, USA}
\affiliation{SLAC National Accelerator Laboratory, Menlo Park, CA 94025, USA}
\author{Eli~Rykoff}
\affiliation{SLAC National Accelerator Laboratory, Menlo Park, CA 94025, USA}
\author{Rebecca~Bernstein}
\affiliation{Observatories of the Carnegie Institution of Washington, 813 Santa Barbara St., Pasadena, CA 91101, USA}
\author{Alberto~Bermeo-Hernandez}
\affiliation{Department of Physics and Astronomy, Pevensey Building, University of Sussex, Brighton, BN1 9QH, UK}
\author{Lena~Eiger}
\affiliation{Santa Cruz Institute for Particle Physics, Santa Cruz, CA 95064, USA}
\affiliation{University of California Santa Cruz, Santa Cruz, CA 95060, USA}
\author{Paul~Giles}
\affiliation{Department of Physics and Astronomy, Pevensey Building, University of Sussex, Brighton, BN1 9QH, UK}
\author{Holger~Israel}
\affiliation{Fakult\"at f\"ur Physik, LMU M\"unchen, Scheinerstr. 1, 81679 M\"unchen, Germany}
\author{Renee~Michel}
\affiliation{Santa Cruz Institute for Particle Physics, Santa Cruz, CA 95064, USA}
\affiliation{University of California Santa Cruz, Santa Cruz, CA 95060, USA}
\author{Raziq~Noorali}
\affiliation{Santa Cruz Institute for Particle Physics, Santa Cruz, CA 95064, USA}
\affiliation{University of California Santa Cruz, Santa Cruz, CA 95060, USA}
\author{A. Kathy~Romer}
\affiliation{Department of Physics and Astronomy, Pevensey Building, University of Sussex, Brighton, BN1 9QH, UK}
\author{Philip~Rooney}
\affiliation{Department of Physics and Astronomy, Pevensey Building, University of Sussex, Brighton, BN1 9QH, UK}
\author{Megan~Splettstoesser}
\affiliation{Santa Cruz Institute for Particle Physics, Santa Cruz, CA 95064, USA}
\affiliation{University of California Santa Cruz, Santa Cruz, CA 95060, USA}

\correspondingauthor{Devon Hollowood}
\email{devonhollowood@gmail.com}

\shorttitle{\matcha{} \redmapper{} Follow-Up}
\shortauthors{Hollowood,~Jeltema,~Chen,~et.~al.}

\begin{abstract}
  In order to place constraints on cosmology through optical surveys of galaxy clusters, one must first understand the properties of those clusters.
  To this end, we introduce the \textbf{M}ass \textbf{A}nalysis \textbf{T}ool for \textbf{Cha}ndra (\matcha{}), a pipeline which uses a parallellized algorithm to analyze archival Chandra data.
  \matcha{} simultaneously calculates X-ray temperatures and luminosities and performs centering measurements for hundreds of potential galaxy clusters using archival X-ray exposures.
  We run \matcha{} on the \redmapper{} \sdss{} DR8 cluster catalog and use \matcha{}'s output X-ray temperatures and luminosities to analyze the galaxy cluster temperature-richness, luminosity-richness, luminosity-temperature, and temperature-luminosity scaling relations.
  We detect 447 clusters and determine 246 \rtfh{} temperatures across all redshifts.
  Within $0.1 < z < 0.35$ we find that \rtfh{} \tx{} scales with optical richness ($\lambda$) as $\txrelation{1.85 \pm{} 0.03}{0.52 \pm{} 0.05}$ with intrinsic scatter of $0.27 \pm{} 0.02$ ($1 \sigma$).
  We investigate the distribution of offsets between the X-ray center and \redmapper{} center within $0.1 < z < 0.35$, finding that $68.3 \pm 6.5$\% of clusters are well-centered.
  However, we find a broad tail of large offsets in this distribution, and we explore some of the causes of \redmapper{} miscentering.
\end{abstract}

\section{Introduction}%
\label{sec:introduction}
The formation history of galaxy clusters is a powerful probe of cosmology~\citep[e.g.][]{Voit, Frieman, MantzI, Allen11, Weinberg, McClintock18}.
In particular, one may place strong constraints on the dark energy equation of state by examining the evolution across redshift of the number density of galaxy clusters as a function of mass~\citep{Mohr, ChandraCosmoIII}.
Upcoming and in-progress optical imaging surveys, such as the Dark Energy Survey (DES)~\citep{DES}, the Hyper Suprime Cam (HSC)~\citep{HSC}, Euclid~\citep{Euclid}, and the Large Synoptic Survey Telescope (LSST)~\citep{LSST}, are expected to observe of tens of thousands of galaxy clusters, thus dramatically expanding our ability to use clusters to place these constraints~\citep{Cunha09, Sanchez, Oguri11, Weinberg, EuclidCosmology}.

The galaxy cluster mass function is the key observable predicted by theory for galaxy-cluster-based studies of dark energy.
Ideally galaxy cluster masses would be measured directly via lensing.
However, because large surveys rarely produce the depth of data required to directly measure the mass of an individual galaxy cluster via lensing, one must instead use some other observable as a mass proxy, and then use an observable-mass relation in order to relate that observable to a distribution of potential masses.
Any given observable-mass relation for massive halos will have some intrinsic scatter distribution driven by recent dynamical activity as well as the full assembly history of each specific halo.
Thus, in order to turn a measured distribution of observables into a distribution of masses, one must understand both the mean observable-mass relation and the intrinsic scatter distribution of this relation.
Stacked weak lensing, which allows one to look at the average mass of many ``similar'' galaxy clusters, is a powerful method by which to determine a mean observable-mass relation~\citep[e.g.][]{Leauthaud, WeighingTheGiantsI, Simet, Melchior, McClintock18}.
The remaining task for the cosmologist is then to understand the intrinsic scatter distribution of the given observable-mass relation.

For the purposes of this paper, we will examine the \emph{richness} optical mass proxy~\citep{Bahcall83, Andreon, redmapperI} and the intrinsic scatter distribution of its relation with other cluster mass proxies.
The precise definition of richness differs from cluster finder to cluster finder, but in essence it is some measure of the number of galaxies in a cluster.
The intrinsic scatter distribution of the richness-mass relation is currently one of the largest sources of uncertainty in using cluster richness to place cosmological constraints~\citep{Wu}.
One may constrain this scatter distribution and improve these constraints by following up a subset of these optically-selected clusters to obtain mass proxies in other wavelengths.
To this end, we have developed a pipeline to perform automated, massively parallelized X-ray follow-up on galaxy clusters which fall within archival \chandra{} data.
This pipeline is called \textbf{\matcha{}}: the \textbf{M}ass \textbf{A}nalysis \textbf{T}ool for \textbf{Cha}ndra.
\matcha{} attempts to measure gas temperatures and X-ray luminosities for these clusters, which can then be compared with their richnesses to help better understand the intrinsic scatter distribution of the richness-mass relation.

Additionally, \matcha{} produces two measures of the ``center'' of a galaxy cluster: the X-ray centroid (i.e.\ center-of-flux) and the X-ray peak.
Miscentering by galaxy cluster finders is a major source of systematic uncertainty in stacked weak-lensing analyses~\citep{Johnston, Melchior}, and without accurate centering information it is difficult for stacked weak-lensing pipelines to produce masses accurate to the level required to realize the full potential of cluster cosmology~\citep{Weinberg}.
By comparing our X-ray centering information with that produced by a given cluster finder, it is possible to understand the centering characteristics of said cluster finder and calibrate for their effects on cosmological analyses.

In this paper, we present the \matcha{} algorithm and describe its application to galaxy clusters identified in the \sdss{} DR8~\citep{DR8} \redmapper{} optical cluster catalog~\citep{redmapperSDSS}.
We use the resulting X-ray temperatures, luminosities, and centering information to explore scatter distributions of richness--mass-proxy relations as well as \redmapper{}'s ability to correctly assign galaxy cluster centers.
In \autoref{sec:cluster-selection}, we give a brief overview of the \redmapper{} galaxy cluster finder.
In \autoref{sec:matcha}, we outline \matcha{}, a pipeline which uses archival \chandra{} data to study the X-ray properties of clusters.
In \autoref{sec:results}, we present temperature-richness and luminosity-richness scaling relations derived from the data produced by \matcha{}, compare \redmapper{} centering with the centering information produced by \matcha{}, and discuss ramifications for stacked weak lensing analyses that use \redmapper{} galaxy cluster locations.
Finally, in \autoref{sec:summary}, we summarize the paper and discuss future work to be done.
In \autoref{sec:example-images}, we present sample images of galaxy clusters produced by \matcha{}.
In \autoref{sec:flag-effects}, we visually highlight various subsamples of our data and their effects on our scaling relations.
In \autoref{sec:matcha-data}, we outline the structure of three machine-readable tables, available online, which contain data used in this paper.

Throughout this paper, we assume a flat $\Lambda$CDM cosmology with $\Omega_m = 0.3$, $H_0 = 70 \mathrm{km \cdot{} s^{-1} \cdot{} Mpc^{-1}}$.
Luminosities are scaled by $E\left(z\right) \equiv H\left(z\right) / H\left(0\right) = \sqrt{\Omega_R {\left(1 + z\right)}^4 + \Omega_M {\left(1 + z\right)}^3 + \Omega_k {\left(1 + z\right)}^2 + \Omega_\Lambda}$, where H is the (redshift-dependent) Hubble parameter; and $\Omega_R$, $\Omega_M$, $\Omega_k$, and $\Omega_\Lambda$ are the densities due to radiation, matter, curvature, and a cosmological constant, respectively, all normalized by the critical density.

\section{Cluster Selection}%
\label{sec:cluster-selection}
In our analysis, we use cluster richnesses and positions from the \textbf{red}-sequence \textbf{Ma}tched-filter \textbf{P}robabilistic \textbf{Per}colation (\textbf{\redmapper{}}) cluster finding algorithm (version 6.3.1, richness $> 20$), found in the Sloan Digital Sky Survey (\sdss{}) Data Release 8 (DR8) catalog.
\redmapper{} is an optical cluster finder designed for use in cluster cosmology by surveys such as DES or LSST\@.
A brief summary of the \redmapper{} algorithm is given here; full details can be found in \citet{redmapperI}.
For a full description of the \redmapper{} v.\ 6.3.1 \sdss{} DR8 catalog, see \citet{redmapperSV}.

The \redmapper{} cluster finder is a two-stage iterative process.
In the first stage, \redmapper{} takes a series of galaxies with known spectroscopic redshifts and uses them as a seeds to find overdensities of galaxies of similar colors.
These overdensities are then used to create a model for the colors of red-sequence galaxies as a function of redshift.
The second stage applies this empirical red-sequence model to group galaxies into clusters, and assign a photometric redshift to the clusters.
The clusters with spectroscopic central galaxies are selected, and the training of the red-sequence is iterated until convergence.

Once the red-sequence model is converged, \redmapper{} uses this model to calculate the number of nearby red-sequence galaxies centered on every galaxy in the photometric catalog.
Galaxies that show an excess of nearby galaxies are ranked according according to the likelihood of the potential cluster centered on that galaxy.
The richness of the top-ranked cluster is measured, and the members probabilistically removed from the other candidate clusters.
The algorithm then moves on to the next highest ranked candidate central galaxy, and the procedure is iterated.
This process is called \emph{percolation}.
The \redmapper{}-assigned richness ($\lambda$) is the sum of the membership probabilities of galaxies within a richness-scaling radius $R_{\lambda} = \left(1.0 h^{-1} \mathrm{Mpc}\right) {\left( \frac{\lambda}{100.0} \right)}^{0.2}$.
This radius scaling is empirically determined to minimize scatter in the mass-richness relation~\citep{Rykoff12}.
Richnesses are corrected for missing galaxy data via Monte Carlo sampling; this primarily effects high-redshift clusters ($z > 0.35$).

In the first generation of the catalog, central galaxies are selected as the brightest members.
The statistical properties of these candidate centrals are then used to define a set of filters that can be used to recenter clusters onto their most-likely-to-be-central galaxy.
This procedure is iterated until convergence is achieved.
The end result is a cluster catalog with central galaxies that are not simply the brightest cluster members, but also take into consideration the local galaxy density in the immediate neighborhood of the galaxy.
The final catalog contains a list of galaxy clusters with their associated positions, redshifts, richnesses, membership probabilities, and top-five most-likely centers (and their centering probabilities).

The \redmapper{} v.\ 6.3.1 \sdss{} DR8 $\lambda{} > 20$ catalog contains 26,308 potential galaxy clusters, 863 of which fell within a public archival \chandra{} observation as of the time at which we ran the \matcha{} pipeline (see \autoref{subsec:data-prep}).

\section{Overview of \chandra{} Pipeline}%
\begin{figure*}[h!]
  \centering
  \includegraphics[width=0.5\textwidth{}]{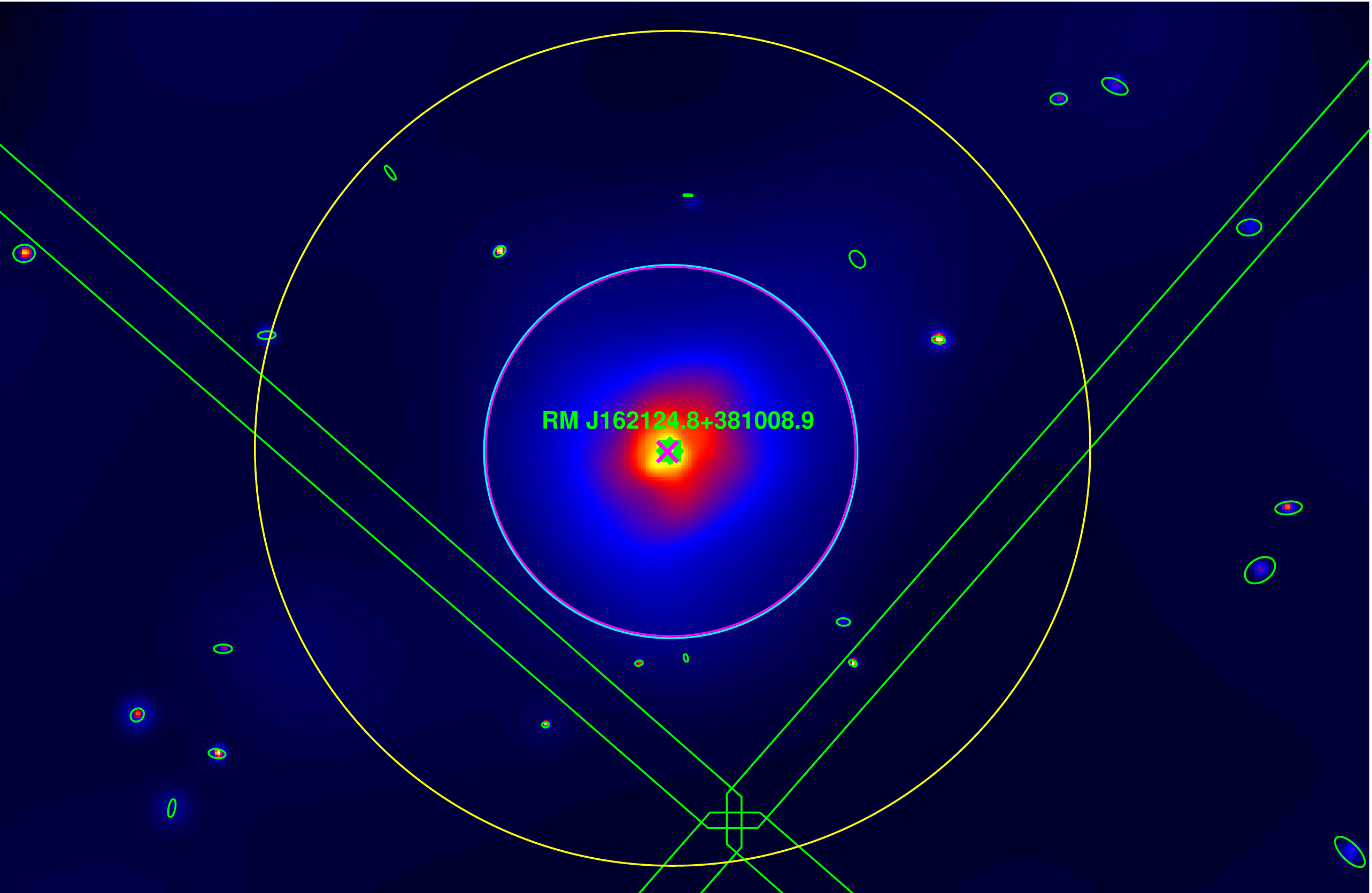}%
  \caption{%
    (a) RM J162124.8+381008.9 (\memmatch{} 2573, $z = 0.48$), ObsID 10785, ACIS-I detector.
    This is a typical example of the output of \matcha{} for a relaxed cluster.
    In each galaxy cluster image in this paper, small green circles represent \redmapper{} clusters, with the green text above this circle giving the name of the cluster.
    The magenta X marks the X-ray peak (see \autoref{subsec:peak-finding}).
    The cyan circle marks a 500 kpc aperture centered on the final location of the iterated 500 kpc centroid.
    The magenta circle marks a \rtfh{} aperture centered at the cluster's \rtfh{} centroid.
    The yellow circle marks a \rfh{} aperture centered at the cluster's \rfh{} centroid.
    Green ellipses mark X-ray point sources.
    Finally, the large green polygons in each image mark the boundaries of the \chandra{} CCDs.
    Each image has been smoothed, and has the point sources left in.
    Any given image may contain multiple \redmapper{} clusters, but these clusters are analyzed separately and we only present the information for one cluster at a time.\label{fig:matcha-typical}
  }
\end{figure*}

\label{sec:matcha}
The \chandra{} analysis is performed using \matcha{}, a custom pipeline which is described in this section.
\matcha{} takes a series of (right ascension, declination, redshift) coordinates (hereafter RA, Dec, and $z$ respectively) from a galaxy cluster catalog and returns a list of cluster centroids, peaks, temperatures and luminosities (hereby referred to as \tx{} and \lx{} respectively) by running a series of CIAO version 4.7 (CALDB version 4.6.7)~\citep{CIAO} and HEASOFT version 6.17 tools.
All spectral fitting is performed using \emph{XSPEC} version 12.9.0~\citep{XSPEC}.
A visual representation of the output for a typical, relaxed cluster may be seen in \autoref{fig:matcha-typical}.
For visual representations of more complex cases, see \autoref{sec:example-images}.

  In the interests of performance, \matcha\ uses a parallel algorithm and features minimal data duplication.
  \matcha\ uses a worker-pool model, in which ``tasks'' encompassing the analysis of each cluster are inserted into a single-producer-multiple-consumer queue.
  Worker threads then perform tasks from this queue until all analysis is complete.

  In order to minimize data duplication, a single ``DataManager'' object is shared between all worker threads.
  This object manages the automatic downloading and analysis of observations.
  Because of the shared nature of the DataManager each observation is downloaded exactly once, even in the case that multiple clusters appear in that observation.
  This drastically reduces the amount of data which needs to be downloaded and stored during \matcha{}'s analysis.

  Each observation is cleaned (as described in \autoref{subsec:data-prep}) in its own sub-thread.
  Once every observation of a given cluster is cleaned, the worker thread in charge of that cluster's analysis may proceed.
  This holds even if a subset of those observations are shared with another cluster which is still waiting for further observations to be cleaned.
  Thus clusters and observations are cleaned as if each cluster and each observation were fully independent, yet \matcha\ as a whole enjoys the full performance and memory benefits of sharing of observations between clusters.

\subsection{Data Preparation}%
\label{subsec:data-prep}
In the data preparation phase, \matcha{} starts with a list of sky coordinates and redshifts for \redmapper{} clusters.
It then uses the \emph{find\_chandra\_obsid} CIAO tool to query the \chandra{} archive for the relevant sky coordinates, determining which of these clusters lie within one or more \chandra{} observations.
\matcha{} then downloads these relevant observations and re-reduces them using the \emph{chandra\_repro} CIAO tool.

\matcha{} then cleans the observations, as follows.
First, \matcha{} cuts the energy range to 0.3--7.9 keV and removes flares from each observation with the \emph{deflare} CIAO tool.
\emph{deflare} is set to use the \emph{lc\_clean} algorithm and a lightcurve time interval of 259.28 seconds.
This time interval is chosen to match the best practices espoused in the CIAO cookbook\footnote{See http://cxc.harvard.edu/contrib/maxim/acisbg/COOKBOOK}.

Next, \matcha{} produces images and exposure maps for the observation.
\matcha{} then identifies point sources using the \emph{wavdetect} CIAO tool and removes these from the observation.
In this process, the ACIS-I chips are cleaned together, separate from the ACIS-S chips.
The ACIS-S chips are cleaned individually, separate from the ACIS-I chips and from the other ACIS-S chips.
We choose to clean ACIS-S chips individually because of their significantly differing instrumental responses, e.g.\ the ACIS S1 and S3 CCDs are backside-illuminated whereas the other CCDs are frontside-illuminated.

At this point \matcha{} is ready to start the analysis of individual clusters, determining whether they are are detected in X-ray, attempting to fit a \tx{} and \lx{} for detected clusters, and attempting to fit an upper-limit \lx{} for undetected clusters.
A few visual examples of the output of \matcha{} are given in \autoref{sec:example-images}, \autoref{fig:matcha-visual}.

\subsection{Finding \texorpdfstring{\tx{} and \lx{}}{Tx and Lx}}%
\label{subsec:finding-tx}
After the observations are downloaded and cleaned, \matcha{}'s next step is to find X-ray centroids, temperatures, and luminosities within \rtfh{}, and \rfh{} regions.
A key strength of \matcha{} is the parallel nature of this computation, allowing for fully concurrent analysis of galaxy clusters.
Care is taken to ensure that cluster may be analyzed soon as all of its observations are cleaned, and that each observation is downloaded and cleaned only once even when multiple clusters lie within it.

In this section we enumerate the steps involved in the analysis of a single cluster; this algorithm is additionally presented as a flowchart in \autoref{fig:matcha}.
Note that \rtfh{} is defined as the radius around a halo at which the average density is 2500 times the cosmological critical density; \rfh{} is the radius at which the average density is 500 times the critical density.
\matcha{} uses the temperature-radius relation from \citet{ArnaudMT} to calculate these radii when they are needed\footnote{%
  These calculations are only meant as an approximation to the $r_{\delta{}}$ radius.
  \citet{ArnaudMT} uses core-cropped temperatures, and uses \xmm{} instead of \chandra{}, introducing a systematic bias in the input of this $T_{X}-r_{\delta{}}$ relation of a few keV~\citep{Nevalainen, Schellenberger}.
  However, in the \citet{ArnaudMT} relation, moderate changes in input temperature have little effect on the resulting radii.
  Conversely, when we fit our temperatures we find that moderate changes in source radius do not greatly affect the resulting temperature.
  Thus we do not expect our choice of $T_{X}-r_{\delta{}}$ relation to be a dominant systematic in our calculated luminosities or temperatures.
}.
All centroids are calculated using the \emph{dmstat} CIAO tool.
  All ACIS-I source and background regions are constrained to lie within ACIS-I CCDs only; all ACIS-S source and background regions are constrained to lie within the CCD on which their center lies.
  This prevents any difficulties arising from having a region span multiple CCDs with different response characteristics.

\matcha{} additionally determines \tx{} and \lx{} values for a core-excised \rfh{} aperture, the calculation of which is presented in this section.
However due to the noisy nature of this data for faint clusters, we choose not to present scaling relations for the core-excised \rfh{} aperture in \autoref{sec:results}, instead leaving this analysis as a possibility for a later work.

The main steps in the \matcha{} cluster analysis are as follows:

\begin{figure*}[htbp!]
  \centering
  \includegraphics[width=\textwidth]{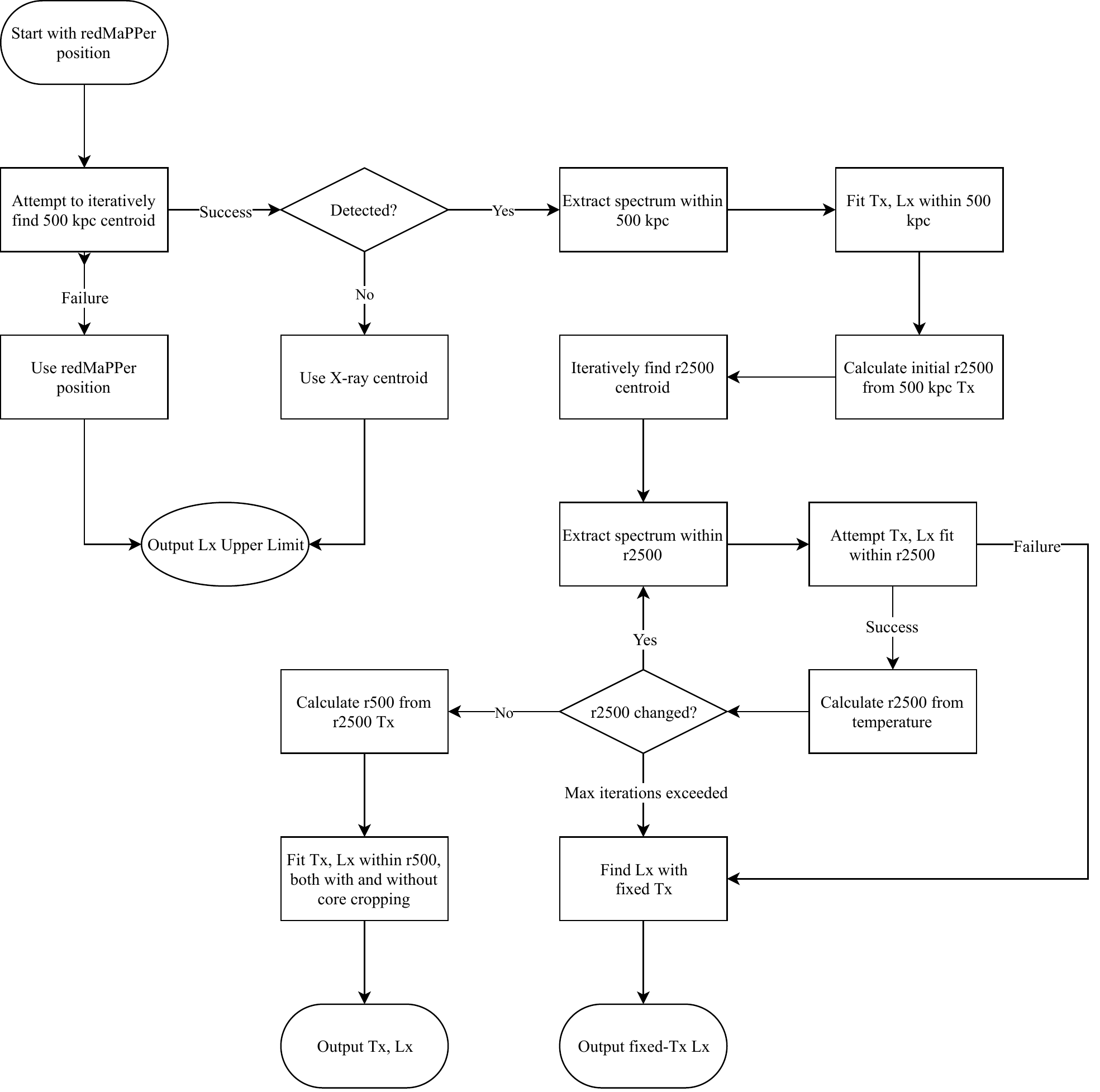}
  \caption{The \matcha{} analysis process.\label{fig:matcha}}
\end{figure*}
\begin{enumerate}
\item\label{itm:centroid} \matcha{} iteratively centers a region with a 500 kpc radius, using the \redmapper{} position as the initial center.
  The corresponding angular separation for this 500 kpc radius is calculated by dividing 500 kpc by the angular diameter distance to the cluster (given the \redmapper{} redshift).
  In each iteration, the new center is the X-ray centroid within the previous 500 kpc region.
  Iteration stops when the new center is within 15 kpc of the old center; the new center is chosen.
  This iterative nature of this process allows us to find centroids which lie more than 500 kpc from the \redmapper{} cluster position, so long as there is sufficient cluster emission within 500 kpc to point \matcha{} towards the centroid.
  If no stable center has been found after 20 iterations, \matcha{} aborts the attempt to find a center and marks the cluster as ``undetected''.
  \matcha{} then attempts to fit an \lx{} upper limit to this ``undetected'' cluster using the position from \redmapper{} and the calculated 500 kpc radius.
  See \autoref{subsec:lx-upper-lims}.
\item \matcha{} checks to see if the signal-to-noise ratio for the source region over the background region is greater than $5.0$.
  If so, the cluster is considered ``detected'', and \matcha{} continues the attempt to find \tx{} and \lx{}.
  If not, \matcha{} marks the cluster ``undetected'', and aborts the attempt to find \tx{} and \lx{}.
  In the latter case, \matcha{} attempts to assign the cluster an \lx{} upper limit using the converged position from~\ref{itm:centroid} and the calculated 500 kpc radius.
  See \autoref{subsec:lx-upper-lims}.
  For this target 500 kpc region, the background is taken as an annulus spanning 700 to 1000 kpc.

\item\label{itm:iter_start} \matcha{} uses the \emph{specextract} CIAO tool to extract a background-subtracted spectrum within the target region, centered on the converged centroid.
  Auxiliary response files are weighted by photon count if the source radius is less than 400 pixels.
  In the interest of efficiency, larger radii are not weighted; at larger radii the photon counts are low and weighting has little effect on the overall spectrum.

\item\label{itm:iter_end} \matcha{} fits a temperature to this spectrum, and calculates the unabsorbed luminosity in the soft-band (0.5--2.0 keV) as well as a bolometric (0.001--100 keV) luminosity.
  This fit is performed using \emph{XSPEC}, and assumes a galactic absorption hydrogen column density found using the \emph{nH} HEASOFT tool (this is a weighted average of the hydrogen densities found in \citet{Kalberla} and \citet{Dickey}).
  The metal abundance is fixed to $0.3 Z_{\odot}$, using the model from \citet{AndersGrevesse}.
  We find that the choice to fix the metal abundance is unimportant for clusters with $k T_X \gtrsim 3.0$ keV.
  For clusters with $k T_X \lesssim 3.0$ keV, we find that varying the abundance between $0.2 Z_\odot$ and $0.4 Z_\odot$ affects the fitted temperatures by $\approx 20\%$, which is usually less than our $1 \sigma$ statistical uncertainties.
  The spectral model used is \emph{XSPEC}'s $\mathrm{wabs} * \mathrm{mekal}$ model.
  Spectra are weighted by their aperture-correction factors (see~\autoref{subsec:aperture}).
\item\label{itm:r2500_centroid} \matcha{} repeats step~\ref{itm:centroid}, with the initial position being the 500-kpc centroid, and the radius being the calculated \rtfh{} radius.
  The converged position becomes our \rtfh{} position.
  If this new centroid does not converge within 20 iterations, the attempt to fit \lx{} and \tx{} is aborted.

\item\label{itm:r2500_fit} \matcha{} iteratively repeats steps~\ref{itm:iter_start}-\ref{itm:iter_end} to find the temperature and luminosities for the \rtfh{} region, stopping when the new \rtfh{} temperature is within $1 \sigma$ of the previous \rtfh{} temperature.
For our \rtfh{} regions, the background is taken as an annulus spanning $1.5 \cdot r_{2500}$ to $3.37 \cdot r_{2500}$.
(The latter number is approximately $1.5 \cdot{} r_{500}$, which is the outer limit of the \rfh{} background discussed in step~\ref{itm:r500_fit}.)
\item\label{itm:r500_fit} \matcha{} repeats steps~\ref{itm:centroid}-\ref{itm:iter_end}, using a region with the final \rtfh{} position as the initial center and \rfh{}, as estimated via the \rtfh{} temperature, as its radius.
  This gives a centroid, \lx{}, and \tx{} for \rfh{}.
  For this \rfh{} region, the background is taken as an annulus spanning $1.05 \cdot r_{500}$ to $1.5 \cdot r_{500}$.
\item \matcha{} repeats steps~\ref{itm:iter_start}-\ref{itm:iter_end}, using an annular region with the calculated \rfh{} as its outer radius, $0.15 \cdot r_{500}$ as its inner radius, and the \rfh{} position (from step~\ref{itm:r500_fit}) as its center.
  This gives \lx{} and \tx{} for a ``core cropped'' \rfh{} region.
  As with the non-core-cropped \rfh{} region, the background is taken as an annulus spanning $1.05 \cdot r_{500}$ to $1.5 \cdot r_{500}$.
\end{enumerate}

Note that in this section's description of the \matcha{} algorithm, all regions are taken as a Boolean AND with the \chandra{} field-of-view in order to avoid contaminating data with extraneous ``zeros'' from area outside the observation.
Additional steps are taken to account for this when the area of a region is required for a calculation; these steps are described in full in the next section of this paper (\autoref{subsec:aperture}).

For clusters with multiple observations, all fits described above are performed as a single simultaneous fit over all observations.

\subsection{Aperture Correction}%
\label{subsec:aperture}
In many observations, the entirety of the detectable cluster emission does not lie on the chip.
Furthermore point sources sometimes account for a significant portion of the cluster area, especially on non-aimpoint \chandra{} chips.
It is thus necessary to correct for area ``lost'' to chip edges and point sources.
To this end, we consider a series of equal-width annuli which cover the cluster source area.
We aim to use ten annuli, but if this would result in annuli with widths of less than 10 pixels, we instead use the maximum number of annuli that allows each annulus a width of at least 10 pixels.
For each annulus, we then take the photon count within the detector area (excluding areas marked as point sources), $N_{\mathrm{annulus,obs}}$, and multiply this count by the ratio of the ``full'' area of the annulus ($\pi(r_2^2-r_1^2)$, where $r_2$ is the outer annular radius and $r_1$ is the inner annular radius) to the exposed annular area $A_\mathrm{annulus}$.
\begin{equation}
N_{\mathrm{annulus,adj}} = N_{\mathrm{annulus,obs}} \cdot \frac{\pi\left(r_2^2-r_1^2\right)}{A_{\mathrm{annulus}}}
\end{equation}
The result, $N_{\mathrm{annulus,adj}}$ approximates the number of counts that we would have received within the annulus were there no point sources or chip edges, assuming that the flux is relatively constant around the annulus.
The sum of these adjusted counts is then compared with the total counts measured in the cluster source area ($N_{\mathrm{tot}}$).
This ratio gives an ``adjust factor'' $F_{\mathrm{adj}}$ for the missing area in each observation.
\begin{equation}
F_{\mathrm{adj}} = \frac{ \displaystyle\sum_{\mathrm{annuli}} N_{\mathrm{annulus,adj}} }{ N_{\mathrm{tot}} }
\end{equation}
We multiply our luminosities and upper limits by this factor.
For clusters with multiple observations, we correct individually before calculation of \tx{} or \lx{}.

We choose to perform this procedure because it maintains reasonable accuracy for faint clusters and because we do not want to make assumptions about the shape of the surface brightness profile.
This procedure may misestimate the correction for non-azimuthally-symmetric clusters.
Experimentation shows that errors in $F_{\mathrm{adj}}$ are no greater than 20\%, and are more typically 5\% for asymmetric clusters.
Thus, this procedure provides a good first-order estimate of the missing data, and we do not believe our choice of procedure to be a dominant source of uncertainty for any cluster.

\subsection{\texorpdfstring{\lx{} Without \tx{}}{Lx Without Tx}}%
\label{subsec:lx-no-tx}%
\label{subsec:lx-upper-lims}
If \matcha{} cannot fit an \rtfh{} \tx{} to a detected cluster, \matcha{} will still attempt to calculate the cluster's luminosity with an assumed temperature of 3.0 keV and \rtfh{} of 500 kpc.
This assumed value is chosen because it is typical for clusters without a \tx{} value: in \autoref{fig:lx-lambda-r2500} it may be seen that these clusters typically have a richness of 30--40; this corresponds to a temperature of approximately 3 keV in \autoref{fig:tx-lambda-r2500}.
As with the luminosities fitted alongside \tx{}, these luminosities are aperture-corrected.
The main source of uncertainty in this \lx{} measurement is from the unknown \tx{}.
Using an assumed \tx{} affects our calculated luminosity: a too-low \tx{} gives a too-high luminosity and a too-high \tx{} gives a too-low luminosity.
Additionally, assuming a 500 kpc radius (instead of using an \rtfh{} radius given by a \tx{}-\rtfh{} relation) means that we oversample lower-mass clusters and undersample higher-mass clusters.
To estimate the contribution of this \tx{} uncertainty to our \lx{} uncertainty, we use PIMMS\footnote{%
  https://heasarc.gsfc.nasa.gov/docs/software/tools/pimms.html.
  In general, PIMMS is not appropriate for detailed analysis (see http://cxc.harvard.edu/ciao/why/pimms.html), and it is not employed here for the actual determination of cluster luminosities.
  We employ it simply as a gauge of the typical size of systematic differences in luminosity when changing the assumed temperature.
  The errors in flux using PIMMS are typically much less than other sources of uncertainty and well within the generous systematic uncertainty in \lx{} we allow for.
}
to estimate the change in flux for a fixed count rate and a varying spectrum, and a $\beta$-model to estimate the change in flux between a 500 kiloparsec aperture and the temperature-determined \rtfh{} aperture.
We find that the effects of an uncertain \tx{} on the assumed flux and on the assumed radius partially cancel each other.
At a temperature of 1 keV we underestimate \lx{} due to the temperature by a factor of $1.2--2.0$ depending on the observing cycle and overestimate \lx{} due to the radius by a factor of $0.4$-$0.8$.
At a temperature of 12 keV we overestimate \lx{} due to the temperature by a factor of $0.69--0.84$ depending on observing cycle and underestimate \lx{} due to the radius by a factor of $1.0$--$1.3$.
The range of 1--12 keV is chosen to span the typical range of temperatures for X-ray detected clusters.
The net effect is roughly negligible for high \tx{}, and for low \tx{} we tend to slightly overestimate \lx{}.
We believe that this error is significantly less than the statistical uncertainty in our scaling relations' fitted slope and scatter (see \autoref{sec:results}).
However, to compensate for the potential systematic uncertainty from using an assumed \tx{} and \rtfh{}, we manually increase our error bars for every \lx{} value which comes from an assumed \tx{}.
The new error bars are taken to be $\left(0.5 \cdot{} L_X\right)$--$\left(2.0 \cdot{} L_X\right)$ plus the statistical errors.
This factor of two was chosen as a conservative error estimate which encompasses the majority of potential \lx{} changes.
We find that this choice has negligible effect on our derived scaling relations.
In principle, this method may be improved by using an \lx{}-\tx{} relation to generate a new \tx{}, and then using the above process to generate a new \lx{} from this \tx{}, continuing to convergence.
However, this is beyond the scope of this paper, and such an extension is left as potential future work.

For ``undetected'' clusters, an \lx{} upper limit may be placed by assuming that all emission received from the cluster location is background, and then calculating the flux that would be $3\sigma$ above this background.
Here we consider emission from the area within 500 kpc of the X-ray centroid determined in step~\ref{itm:centroid} of \autoref{subsec:finding-tx}.
If no such position can be found, we use the \redmapper{} position.
We then predict a model flux by assuming a 3.0 keV temperature and using the same $\mathrm{wabs} * \mathrm{mekal}$ model as in \autoref{subsec:finding-tx}.
An upper limit flux $\Phi_{\max{}, n_{\sigma}}$ is then given by
\begin{equation}
  \Phi_{\max{}, n_{\sigma}} = \Phi_{\mathrm{model}} \cdot{} \frac{ n_{\sigma}\sqrt{N_\mathrm{obs}} }{ N_{\mathrm{model}} }
\end{equation}
where $\Phi_{\mathrm{model}}$ is the fitted flux, $n_{\sigma}$ is the desired confidence level (in units of standard deviation), $N_\mathrm{obs}$ is the aperture-corrected observed number of counts, and $N_{\mathrm{model}}$ is the product of the exposure time and the model count rate.
These counts are not background subtracted, because by definition the source region for an undetected cluster is indistinguishable from background.
Here, we multiply the observed flux from the non-detection by the ratio of (the count rate that we would have detected the cluster with confidence $n_{\sigma}\sigma$) to (the count rate that we observed).
Typical values of $N_\mathrm{obs}$ for undetected clusters are a few hundred photons, with the middle 50\% of undetected clusters having between 148 and 642 counts.
Through this method, we place a $3\sigma$ \lx{} upper limit (within a 500 kpc aperture) on each ``undetected'' cluster.

\subsection{Peak Finding}%
\label{subsec:peak-finding}
In addition to finding the X-ray centroid, which is a useful measure of a cluster's center for spectral fitting, we explore using a cluster's most luminous X-ray region as an alternative centering measure which is better matched to the \redmapper{} central galaxy (see \autoref{subsec:centering}).
Simply taking the brightest pixel does not work as a reliable cluster center; more care must be taken in determining the X-ray peak.
This is both because observations can be quite noisy, and because we would like to avoid picking the peak of a small substructure of the galaxy cluster which happens to be X-ray bright over a more significant substructure which happens to be slightly dimmer.
Additionally, we may have cut out the X-ray peak when we cut out X-ray point sources, as there is occasionally an active galactic nucleus in the most luminous region of a galaxy cluster.
To deal with these problems, we smooth the binned X-ray image (with point sources removed) via convolution with a 2D Gaussian of 50 kpc radius.
We then take the X-ray peak to be the brightest pixel of this smoothed image that is within 500 kpc of the X-ray's 500-kpc-aperture centroid (see \autoref{subsec:finding-tx}).
As before, this 500 kpc radius is proper distance and is calculated using the \redmapper{} redshift.
We then check the results of the peak-finding visually (see \autoref{subsec:post-pipeline}), looking for cases where the brightest cluster peak lies outside of our initial 500 kpc search.
This occurs a only small fraction of the time, specifically for five clusters in this sample.
In these cases we manually re-run the above analysis with a larger peak search radius, chosen to include the actual peak.

\subsection{Post-Pipeline Analysis and Cleaning}%
\label{subsec:post-pipeline}
After running \matcha{} to get \tx{} and \lx{} (or \lx{} upper limits) for each cluster, we further examine the detected clusters to ensure a clean sample.
First, we compare the output cluster catalog to the known galaxy clusters in the \href{https://ned.ipac.caltech.edu/}{NASA/IPAC Extragalactic Database (NED)}\footnote{The NASA/IPAC Extragalactic Database (NED) is operated by the Jet Propulsion Laboratory, California Institute of Technology, under contract with the National Aeronautics and Space Administration}, to find any instances where our moving centroid (see \autoref{subsec:finding-tx} \autoref{itm:centroid}) causes the X-ray analysis to choose a bright, nearby X-ray cluster instead of a separate, foreground or background cluster detected by \redmapper{}.
We then manually examine each used observation, flagging both potential problems and interesting attributes.

``Potential problems'' include clusters whose X-ray centroids are not located on a cluster substructure (see e.g. \autoref{fig:centering} (a)), clusters which are too close to an outer chip edge to be considered reliable, clusters whose sole observation is in a non-imaging mode\footnote{A \chandra{} image may be generated even if the observation is in a non-imaging mode.}, clusters which are ``mismatched'' (as in our above NED search), and clusters whose background or source spectra are significantly contaminated by a separate nearby cluster.
Additionally, for \rfh{} regions we find a handful of clusters for which we cannot measure a reliable background because \rfh{} is approximately the angular size of the observation{(s)}; we flag these clusters as being too close to a chip edge and treat them as we treat our other clusters affected by proximity to chip edges.

``Interesting attributes'' include merging or disturbed clusters, clusters where the \redmapper{}-assigned center does not lie near an X-ray peak, and ``serendipitous'' clusters---clusters which are not the aimpoint of the \chandra{} observation and which are thus more free from selection bias (see \autoref{subsec:selection-effects}).
Our criterion for marking a cluster as serendipitous is that in each of its observations the cluster either lies on a non-aimpoint \chandra{} chip, or shares the observation with a cluster which is clearly the aimpoint cluster.
See \autoref{fig:bad-xray} in \autoref{sec:example-images} for examples of these common X-ray phenomena.

Because we only use undetected clusters as upper limits in our \lx{}-$\lambda{}$ scaling relations, we do not examine them in as-great a depth.
For these clusters, we only flag proximity to an outer chip edge and non-imaging-mode X-ray observations.

\begin{figure}[htbp!]
  \centering{}%
  \includegraphics[width=0.5\textwidth{}]{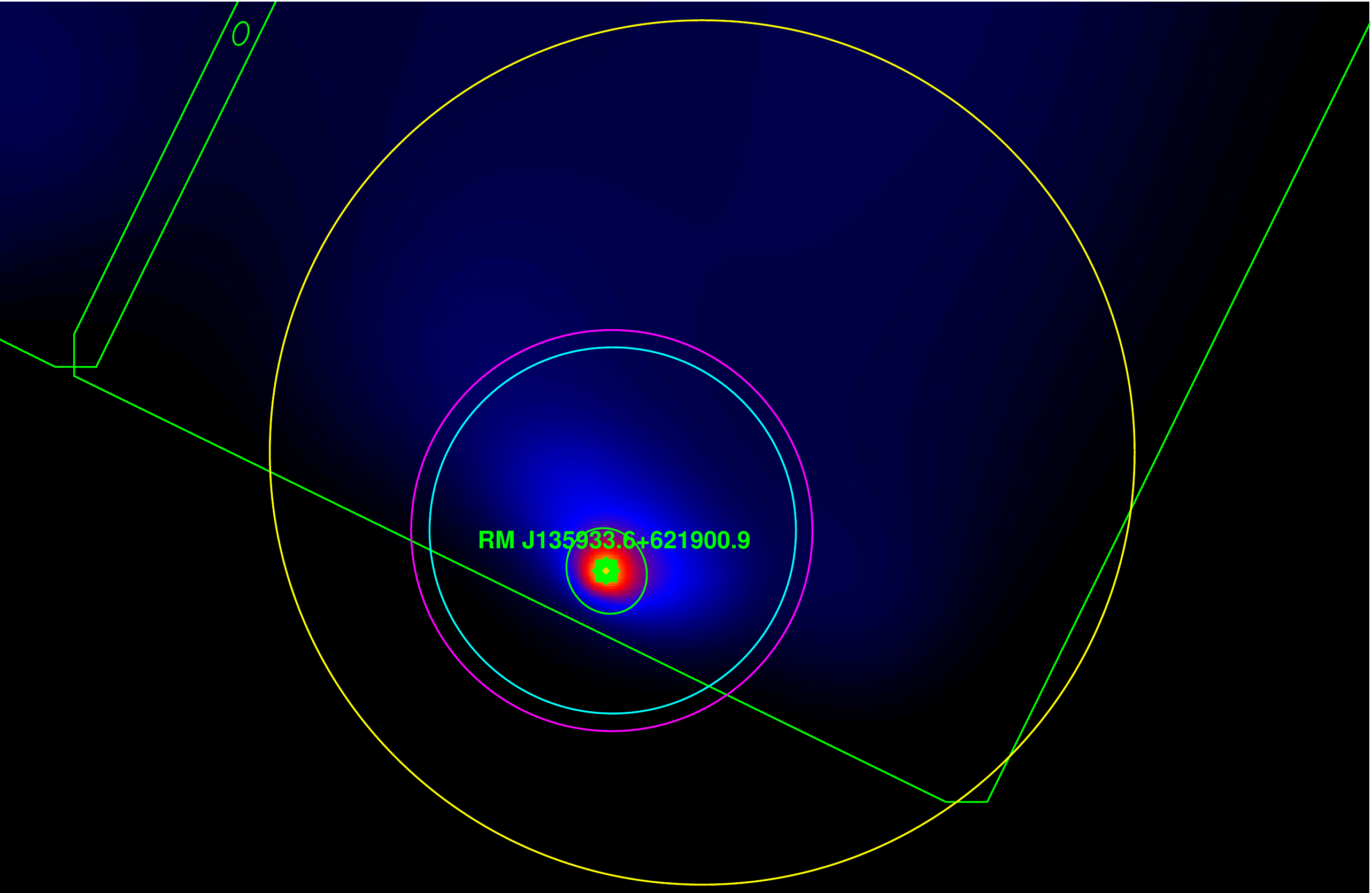}%
  \caption{%
    RM J135933.6+621900.9 (\memmatch{} 972, $z = 0.34$), ObsID 7714, ACIS-I detector.
    This cluster is very close to the chip edge, and thus we cannot determine accurate centering information for it.
    However, we still capture enough cluster emission to get accurate \lx{} and \tx{} values.
    \colorinfosingle{}\label{fig:cat-741}
  }
\end{figure}

We then use these flags to make cuts to our scaling-relation and centering data sets.
When fitting mass-proxy\textendash{}richness relations and when comparing \redmapper{} centers to X-ray peaks, we remove clusters for which proximity to the chip edge is deemed an issue, clusters whose sole observation is in a non-imaging mode, and ``mismatched'' clusters.
When comparing \redmapper{} centers to X-ray centroids (but not peaks), we remove the above cases, and additionally remove clusters for which the X-ray centroid does not lie on a major X-ray substructure.
This is because \redmapper{} is not expected to produce a center that agrees with the X-ray centroid in these cases (see \autoref{subsec:centering} for discussion).
Note that for each cluster we separately decide whether chip edge proximity is a problem for centering and whether it is a problem for each radius's \lx{} and \tx{}.
For example, in RM J135933.6+621900.9 (\memmatch{} 972, see \autoref{fig:cat-741}), we have an example of a cluster whose proximity to the chip edge is a problem for centering but not for scaling relations, because the proximity to the chip edge causes the centroid to move significantly yet we could capture enough of the cluster emission to determine \tx{} and \lx{} accurately.

Using this system of flagging, we are able to give \redmapper{} centering feedback directly to the \redmapper{} team.
See \autoref{subsec:centering} for more information on our follow-up on \redmapper{} centering.

\subsection{Mispercolations}%
\label{subsec:mispercolations}
Sometimes, when there are two or more separate physical clusters near one another, or when \redmapper{} has incorrectly split a single massive halo into two-or-more separate clusters in its catalog, \redmapper{} assigns a large richness to the smaller system and a small richness to the larger system.
We call this problem ``mispercolation'', as it is a failure of \redmapper{}'s ``percolation'' step (see \autoref{sec:cluster-selection}).
In our data, we correct these mispercolations by manually assigning the brightest halo's centroids, radii, \tx{}, and \lx{} values to the richest \redmapper{} halo.
We then remove the other \redmapper{} cluster entirely.
Effectively, this is equivalent to treating the two halos as a single halo with a very large centering error, which makes intuitive sense because mispercolation is a \redmapper{} centering issue.
Additionally, this approach acts as a compromise between removing mispercolated halos altogether, which artificially removes richness scatter and miscentering information, and keeping the halos untouched, which leads to extreme outliers in the scaling relations (because very hot or massive clusters are associated with low-richness entries in the \redmapper{} catalog).
For a full treatment of the effects of \redmapper{} miscentering on scaling relations, see \citet{ZhangCentering}.

\begin{figure*}[htbp]
  \centering
  \gridline{%
    \fig{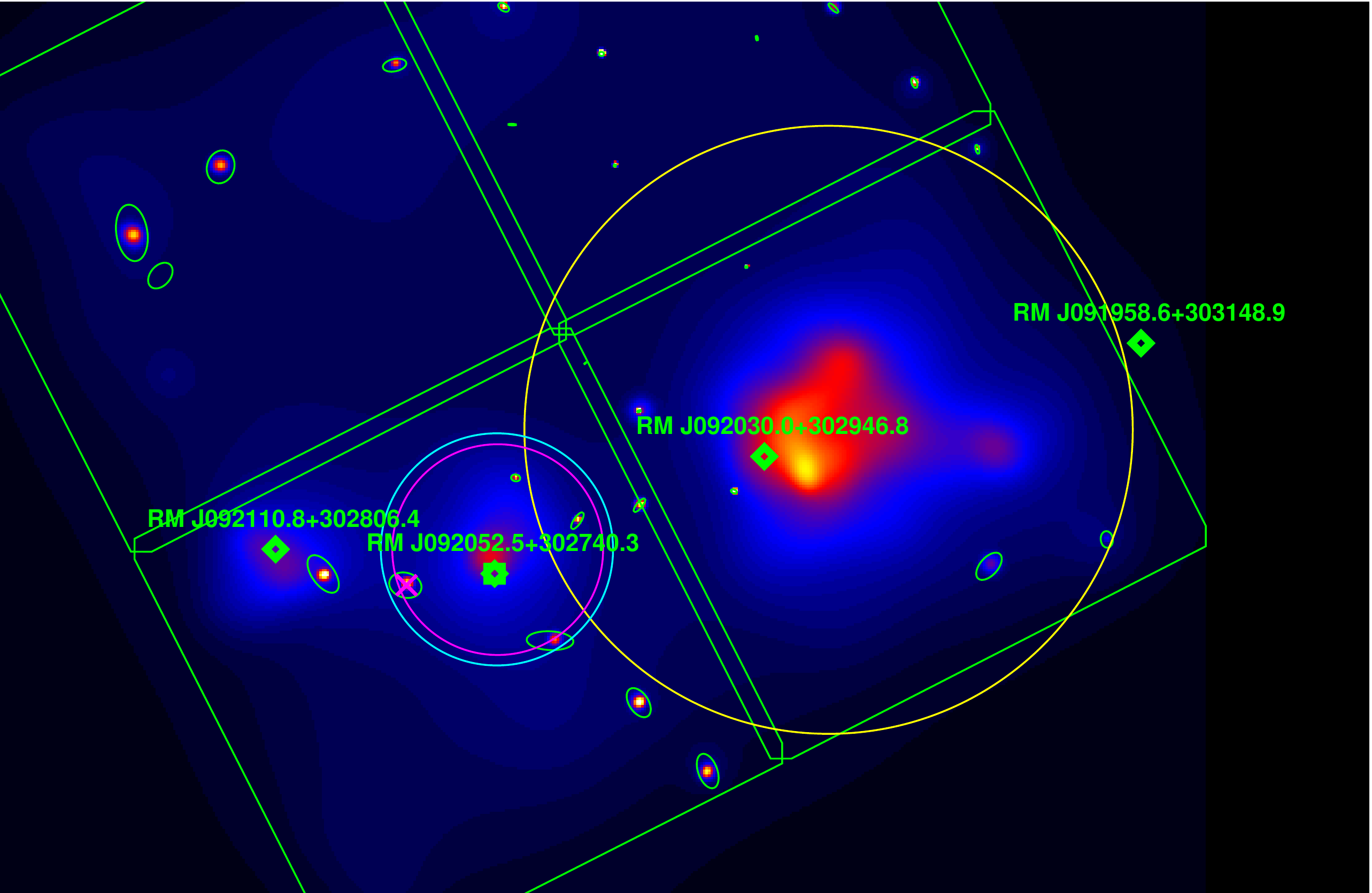}{0.5\textwidth}{(a)}
    \fig{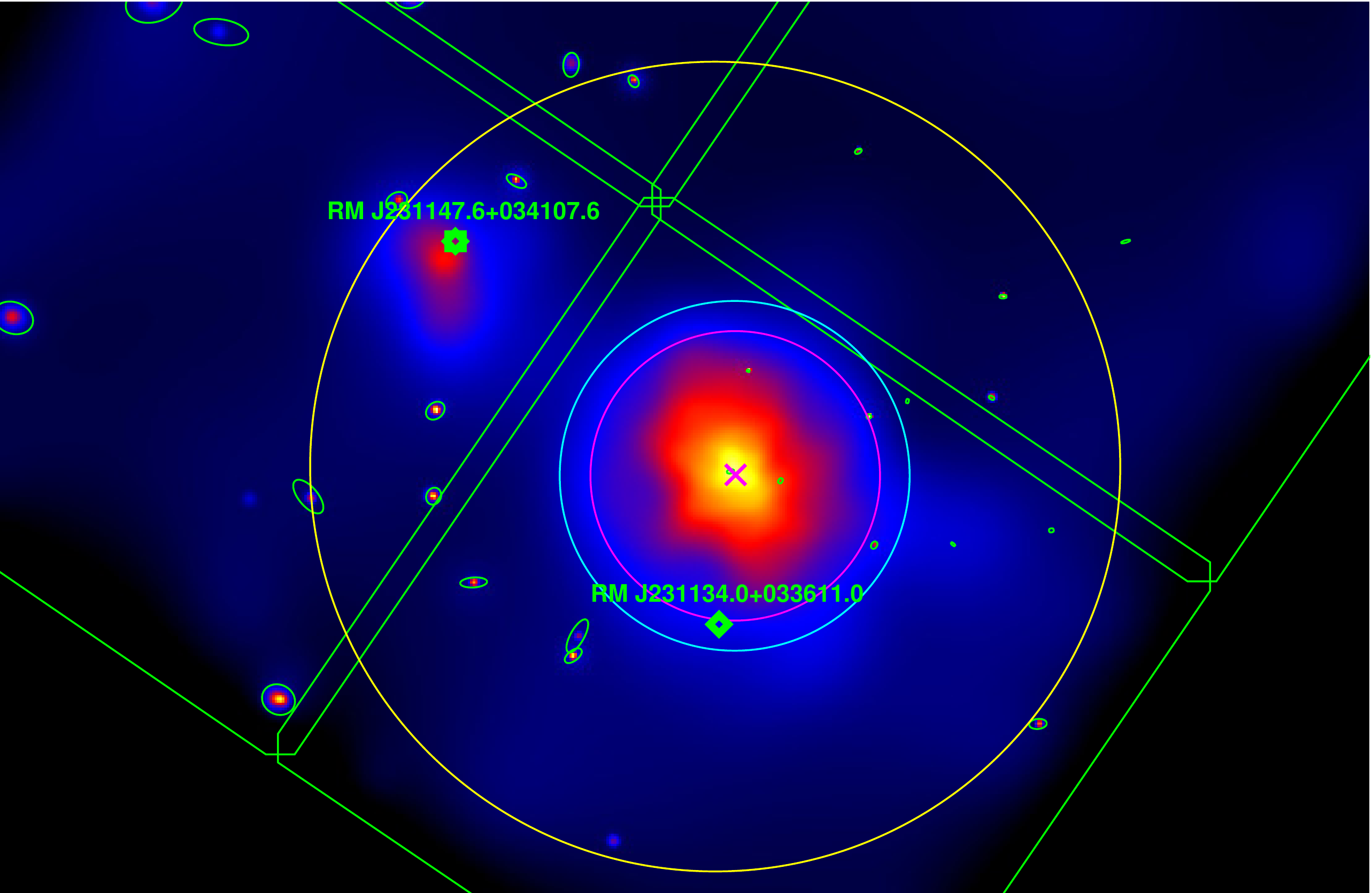}{0.5\textwidth}{(b)}
  }%
  \caption{%
    (a) RM J092052.5+302740.3 (\memmatch{} 23, $z = 0.29$) and RM J092030.0+302946.8 (\memmatch{} 21, $z = 0.31$), ObsID 534, ACIS-I detector.
    Here, \redmapper{} splits this merging cluster into two separate clusters.
    \redmapper{} then assigns a richness of 129 to the smaller subcluster (left, RM J092052.5+302740.3) and a richness of 39 to the larger subcluster (right, RM J092030.0+302946.8).
    In our analysis, we discard the latter and take this to be a single cluster: RM J092052.5+302740.3.
    We then manually assign this cluster the X-ray peak, \rtfh{} centroid, \rtfh{} radius, \rtfh{} \tx{}, and \rtfh{} \lx{} from RM J092030.0+302946.8.
    We determine the \rfh{} information to be acceptable without modification.
    This figure shows the final modified regions.
    (b) RM J231147.6+034107.6 (\memmatch{} 34, $z = 0.30$) and RM J231134.0+033611.0 (\memmatch{} 41, $z = 0.30$), ObsID 11730, ACIS-I detector.
    Here, \redmapper{} splits this merging cluster into two separate clusters.
    \redmapper{} assigns a richness of 166 to the smaller subcluster (left, RM J231147.6+034107.6) and a richness of 20 to the larger subcluster (right, RM J231134.0+033611.0).
    In our analysis, we discard the latter and take this to be a single cluster with the X-ray peak, \rtfh{} centroid, \rtfh{} radius, \rtfh{} \tx{}, \rtfh{} \lx{}, \rfh{} centroid, \rfh{} radius, \rfh{} \tx{}, and \rfh{} \lx{} from RM J231134.0+033611.0 (the less rich halo), but with the richness and \redmapper{} ID from RM J231147.6+034107.6 (the richer halo).
    This figure shows the final modified regions.
    \colorinfo{}\label{fig:mispercolation-a}
  }
\end{figure*}

\begin{figure*}[htbp]
  \centering
  \gridline{%
    \fig{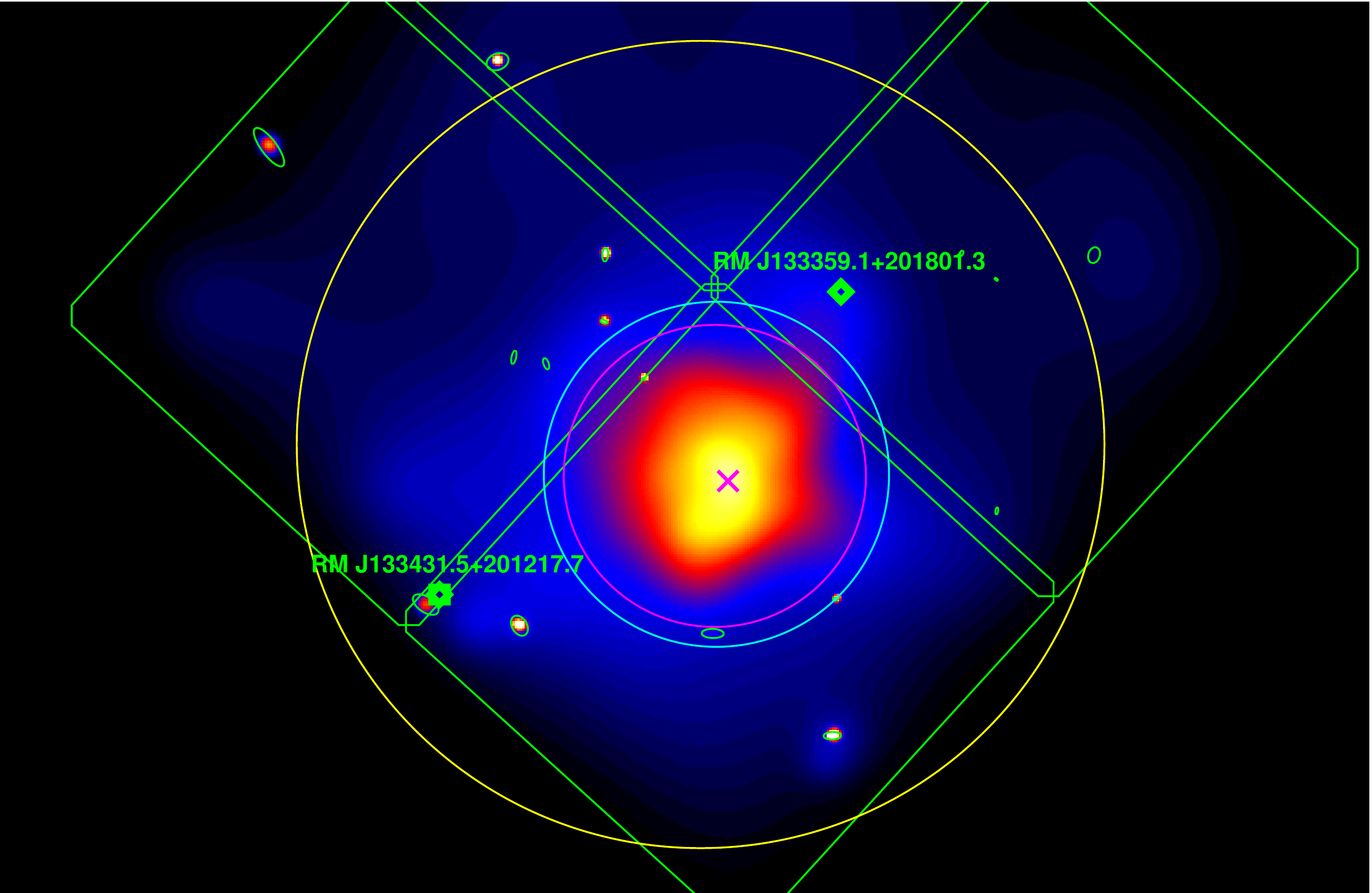}{0.5\textwidth}{(a)}
    \fig{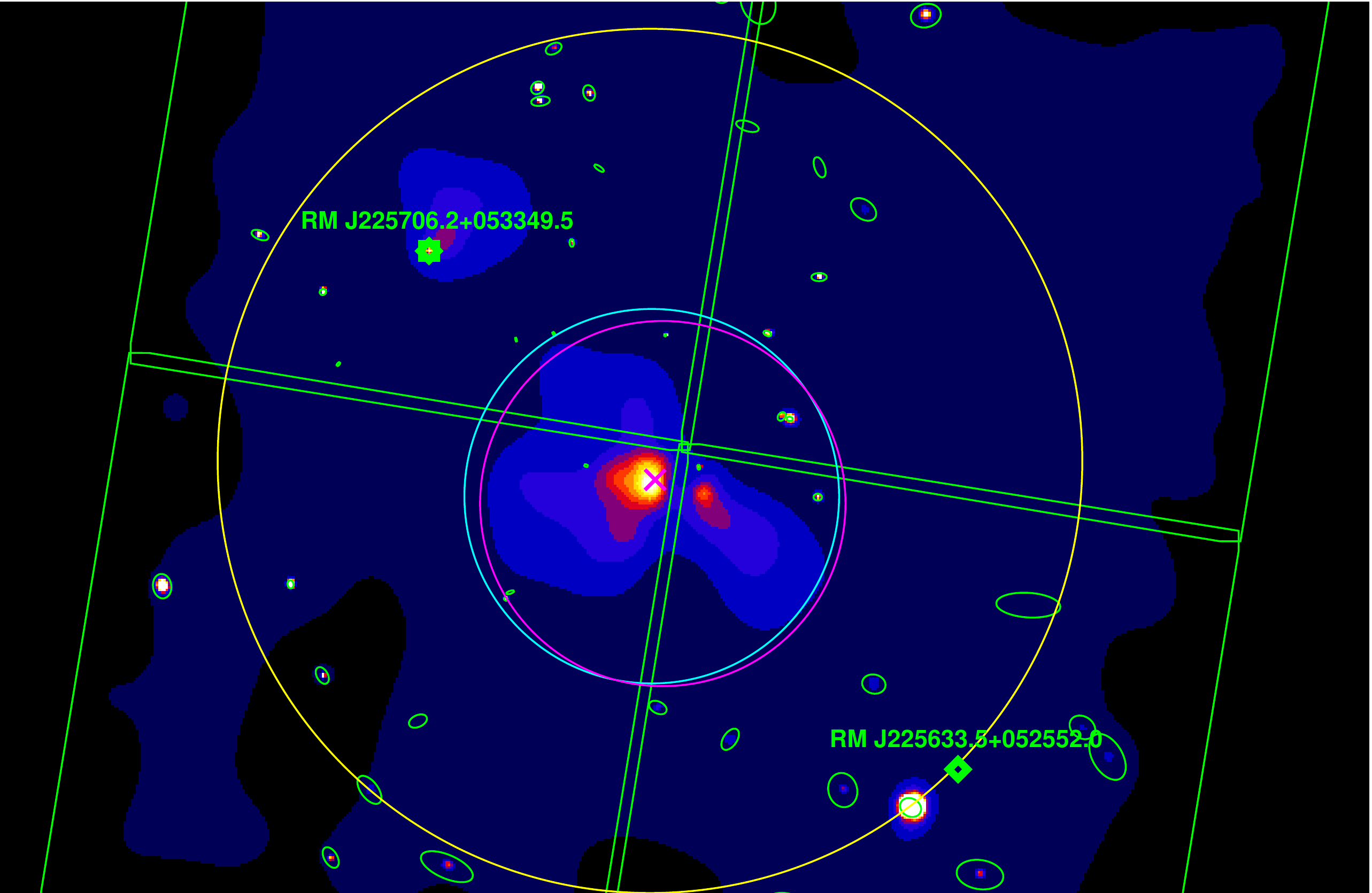}{0.5\textwidth}{(b)}
  }
  \caption{%
    (a) RM J133431.5+201217.7 (\memmatch{} 25, $z = 0.17$) and RM J133359.1+201801.3 (\memmatch{} 24, $z = 0.17$), ObsID 17159, ACIS-I detector.
    Here, \redmapper{} splits this single, relaxed cluster into two separate clusters.
    \redmapper{} divides the richness between the two clusters, assigning a richness of 73 to RM J133431.5+201217.7 (left) and a richness of 26 to RM J133359.1+201801.3 (right).
    In our analysis, we discard the latter and use the former as-is.
    This figure shows the regions for RM J133431.5+201217.7.
    (b) RM J225706.2+053349.5 (\memmatch{} 236, $z = 0.18$) and RM J225633.5+052552.0 (\memmatch{} 164, $z = 0.16$), ObsID 12248, ACIS-I detector.
    Here, \redmapper{} splits this merging cluster into two separate clusters.
    \redmapper{} assigns a richness of 70 to a fragment of the cluster (left, RM J225706.2+053349.5) and a richness of 23 to a second fragment (right, RM J225633.5+052552.0).
    Both of these are offset from the X-ray cluster.
    In our analysis, we discard the latter and take this to be a single cluster: RM J225706.2+053349.5.
    We then manually assign this cluster the X-ray peak, \rtfh{} centroid, \rtfh{} radius, \rtfh{} \tx{},\rtfh{} \lx{}, \rfh{} centroid, \rfh{} radius, \rfh{} \tx{}, and \rfh{} \lx{} from RM J225633.5+052552.0.
    This figure shows the final modified regions.
    \colorinfo{}\label{fig:mispercolation-b}
  }
\end{figure*}

In the \redmapper{} \sdss{} DR8 sample, we identify four cases of mispercolation.
Images of each mispercolated cluster are presented in Figures~\ref{fig:mispercolation-a}--\ref{fig:mispercolation-b} along with a brief discussion of how we handle each individual case.
The cases are summarized in \autoref{tab:mispercolated}.

\begin{deluxetable*}{rrr|l}
  \tablecolumns{6}
  \tablecaption{Summary of Mispercolation Handling\label{tab:mispercolated}}
  \tablehead{\colhead{ID} & \colhead{$\lambda$} & \colhead{$z$} & \colhead{Action Taken}}
  \startdata{}
    21  & 38.7  & 0.31 & Remove from data. \\
    23  & 128.7 & 0.29 & Replace \rtfh{} centroid, radius, \lx{}, and \tx{} with that of \#21. Keep \rfh{} data as-is. \\
    \midrule{}%
    34  & 166.2 & 0.30 & Replace \rtfh{} and \rfh{} centroid, radius, \lx{}, and \tx{} with that of \#41. \\
    41  & 20.0  & 0.30 & Remove from data. \\
    \midrule{}%
    25  & 73.4  & 0.17 & Keep as-is. \\
    24  & 26.5  & 0.17 & Remove from data. \\
    \midrule{}%
    236 & 69.8  & 0.18 & Replace \rtfh{} and \rfh{} centroid, radius, \lx{}, and \tx{} with that of \#164. \\
    164 & 22.7  & 0.16 & Remove from data. \\
  \enddata{}
  \tablecomments{Here the ``ID'' column gives the \memmatch{} from the \redmapper{} catalog}
\end{deluxetable*}

\section{Results}%
\label{sec:results}
We analyze 863 \redmapper{} clusters which fall within archival \chandra{} observations.
Of these 863 clusters, we successfully clean 850 clusters (as described in \autoref{subsec:data-prep}).
Of these 850 clusters, 447 are considered ``detected'', and 403 are considered ``undetected''.
We then manually review each of these clusters as described in \autoref{subsec:post-pipeline}, removing 39 of the 447 detected clusters.
(Information for clusters removed in review is still available in the table described in \autoref{sec:matcha-data}.)
After removing these problematic clusters, we find \rtfh{} temperatures for 235 clusters of the 408 remaining detected clusters.
We find \rtfh{} luminosities for each of these 235 clusters via the method described in \autoref{subsec:finding-tx}.
Out of the 235 clusters for which we find an \rtfh{} luminosity and temperature, we additionally find an \rfh{} luminosity and temperature for 190 clusters.
For 172 of the 173 valid detected clusters with no \rtfh{} temperature, we successfully estimate \rtfh{} luminosities via the method described in~\autoref{subsec:lx-no-tx}\footnote{%
  The remaining cluster is Abell 1795, which has a massive 88 \chandra{} observations.
  Analyzing this many simultaneous observations with \emph{XSPEC} triggers \matcha{}'s internal
  time limits for its subprocesses, and \emph{XSPEC} is terminated before it can produce anything useful.
  Time limits are used for all \emph{HEASOFT} and \emph{CIAO} tools (given their propensity to hang), and are generously set to five hours by default.
}.
We place $3\sigma$ \lx{} upper limits on all 403 ``undetected'' clusters.
We identify 89 of the 408 detected clusters as serendipitous (see \autoref{subsec:post-pipeline}), and fit \rtfh{} temperatures to 29 of these.

All luminosities quoted in this section are rest-frame, and are soft-band (0.5--2.0 keV) unless otherwise noted.
We consider bolometric luminosities (0.001--100 keV) only for the purpose of comparison with scaling relations from the literature.

\subsection{X-ray Observable\textendash{}Richness Scaling Relations}%
\label{subsec:lambda-scaling-relations}
For the regression analysis, we employ the hierarchical Bayesian model proposed in \citet{Kelly}.
This method uses a Gaussian mixture model to estimate the distribution of the independent variable.
We choose this method because it provides an unbiased estimation of the scaling parameters for data with correlated and heteroscedastic measurement uncertainties and accounts for the effect of censored data and correlated and heteroscedastic measurement uncertainties.
To compute the joint posterior distribution of the model parameters, we run a Gibbs sampler algorithm proposed in \citet{Kelly}.
The marginalized estimate of the model parameters are summarized in Tables~\ref{tab:r2500-scaling-relations}--\ref{tab:r500-scaling-relations}, and select relations are highlighted below.
Our derived relations are of the form $\ln\left(y\right) = \alpha\ln\left(\lambda/70\right) + \beta$, where $\lambda$ is the cluster richness.

\begin{deluxetable*}{lrrrr}
  \tablecaption{\rtfh{} X-Ray Observable Scaling with Richness\label{tab:r2500-scaling-relations}}
  \tablehead{\colhead{Relation} & \colhead{$\beta$} & \colhead{$\alpha$} & \colhead{$\sigmaintr$} & \colhead{Figure}}
  \startdata{}%
  $T_X-\lambda$ (all redshift) & $1.82 \pm 0.02$ & $0.54 \pm 0.04$ & $0.26 \pm 0.02$ & \autoref{fig:tx-lambda-r2500} (a) \\
  $T_X-\lambda$ ($0.1 < z < 0.35$) & $1.85 \pm 0.03$ & $0.52 \pm 0.05$ & $0.27 \pm 0.02$ & \autoref{fig:tx-lambda-r2500} (b) \\
  \midrule{}%
  $L_X-\lambda$ (all redshift, w/o upper limits) & $-0.08 \pm 0.05$ & $1.37 \pm 0.08$ & $0.84 \pm 0.03$ & \autoref{fig:lx-lambda-r2500} (a) \\
  $L_X-\lambda$ ($0.1 < z < 0.35$, w/o upper limits) & $-0.08 \pm 0.07$ & $1.31 \pm 0.12$ & $0.92 \pm 0.05$ & \autoref{fig:lx-lambda-r2500} (b) \\
  $L_X-\lambda$ ($0.1 < z < 0.35$, w/o upper limits, fit \tx) & $0.02 \pm 0.09$ & $1.11 \pm 0.16$ & $0.99 \pm 0.06$ & \nodata{} \\
  $L_X-\lambda$ ($0.1 < z < 0.35$, w/ upper limits) & $-0.26 \pm 0.08$ & $1.78 \pm 0.12$ & $1.04 \pm 0.06$ & \autoref{fig:lx-lambda-r2500} (c) \\
  \enddata{}
  \tablecomments{%
    Relations are of the form $\ln\left(y\right) = \alpha\ln\left(\lambda/70\right) + \beta$, where $\lambda$ is the cluster richness.
    \lx{} is normalized by $E\left(z\right)$ and has units of $10^{44}$ ergs/s; \tx{} has units of keV.
    $\sigmaintr$ is the standard deviation of the intrinsic scatter in this relation.
    The $L_X-\lambda$ (all redshift without upper limits), $L_X-\lambda$ ($0.1 < z < 0.35$, w/o upper limits, fit \tx), and $T_X-\lambda$ ($0.1 < z < 0.35$) relations have scatter distributions which are slightly asymmetric, with a longer tail in the large-scatter direction.
    Uncertainties are listed as their $1\sigma{}$ values.
    The \lx{} relation labeled ``fit \tx{}'' contains only \lx{} values which were calculated alongside \tx{} (see \autoref{subsec:finding-tx}, c.f.\ \autoref{subsec:lx-no-tx}).
  }
\end{deluxetable*}

\begin{deluxetable*}{lrrrr}
  \tablecaption{\rfh{} X-Ray Observable Scaling with Richness\label{tab:r500-scaling-relations}}
  \tablehead{\colhead{Relation} & \colhead{$\beta$} & \colhead{$\alpha$} & \colhead{$\sigmaintr$} & \colhead{Figure}}
  \startdata{}%
  $T_X-\lambda$ (all redshift) & $1.83 \pm 0.03$ & $0.54 \pm 0.05$ & $0.28 \pm 0.02$ & \autoref{fig:tx-lambda-r500} (a) \\
  $T_X-\lambda$ ($0.1 < z < 0.35$) & $1.86 \pm 0.03$ & $0.51 \pm 0.05$ & $0.24 \pm 0.03$ & \autoref{fig:tx-lambda-r500} (b) \\
  \midrule{}%
  $L_X-\lambda$ (all redshift, fit \tx) & $0.46 \pm 0.06$ & $1.07 \pm 0.11$ & $0.77 \pm 0.04$ & \autoref{fig:lx-lambda-r500} (a) \\
  $L_X-\lambda$ ($0.1 < z < 0.35$, fit \tx) & $0.44 \pm 0.09$ & $1.02 \pm 0.15$ & $0.84 \pm 0.06$ & \autoref{fig:lx-lambda-r500} (b) \\
  \enddata{}
  \tablecomments{%
    Relations are of the form $\ln\left(y\right) = \alpha\ln\lambda/70 + \beta$, where $\lambda$ is the cluster richness.
    \lx{} is normalized by $E\left(z\right)$ and has units of $10^{44}$ ergs/s; \tx{} has units of keV.
    $\sigmaintr$ is the standard deviation of the intrinsic scatter in this relation.
    Uncertainties are listed as their $1\sigma{}$ values.
    The \lx{} relations contain only \lx{} values which were calculated alongside \tx{} (see \autoref{subsec:finding-tx}, c.f.\ \autoref{subsec:lx-no-tx}).
    The scatter distributions for each of these relations are slightly asymmetric, with longer tails in the large-scatter direction.
  }
\end{deluxetable*}

In the presented relations, we primarily focus on data within the redshift range $0.1 < z < 0.35$.
This redshift range is chosen because it selects the best possible data from \redmapper{}~\citep{redmapperI}.
At $z < 0.1$, \redmapper{} centering degrades due to an increased fraction of poorly measured central galaxies and observations flagged for processing issues.
At $z > 0.35$, \redmapper{}'s scatter in both richness and redshift are significantly increased by the 4000 \r{A} break transitioning \sdss{} bands, and by \sdss{}'s magnitude limit.
See \citet{redmapperI} for more details on these effects.
We choose to limit our manual follow-up of undetected clusters to this $0.1 < z < 0.35$ range due to their sheer number.
We thus only present upper-limit luminosities for this redshift range.

For $T_X-\lambda$ in the $0.1 < z < 0.35$ range, \rtfh{} aperture, we derive
\begin{equation}
  \txrelation[r2500]{1.85 \pm{} 0.03}{0.52 \pm{} 0.05}
\end{equation}
with standard-deviation-of-intrinsic-scatter $\sigmaintr = 0.27 \pm 0.02$.
We thus constrain $\sigmaintr$ within 7\%.
This does not differ significantly from our derived all-redshift \tx{}-$\lambda$ relation in slope, intercept, or $\sigmaintr$.

We now compare our \tx{}-$\lambda$ relation with those presented in two previous \redmapper{} papers: \citet{redmapperII} and \citet{redmapperSV}.
For the former, we compare to their data from the ACCEPT cluster catalog~\citep{Accept}.
This sample is a collection of galaxy clusters with deep, pointed Chandra observations.
\citet{redmapperII} utilize the temperature and gas density profiles from \citet{Accept} to calculate spectroscopic-like average temperatures which are core-excised at 150 kpc when possible given the radial range probed.
The total sample used by \citet{redmapperII} is 56 clusters.
Despite the differences in the X-ray analysis, our \tx{}-$\lambda$ relation is consistent with theirs.
Our relation's intercept agrees well with this ACCEPT relation, which (after normalization to our choice of pivots) is listed as $1.852 \pm 0.032$.
Their derived slope of $0.407 \pm 0.066$ is shallower than ours by $1.2 \sigma$ considering our $0.1<z<0.35$ sample, but largely consistent.
However, they derive a lower $\sigmaintr$ of $0.196 \pm 0.021$.
As the ACCEPT sample is biased toward X-ray bright clusters, it is perhaps not surprising that these clusters might undersample the scatter of an optically-selected cluster sample.
We further discuss selection effects in \autoref{subsec:selection-effects}.

\citet{redmapperSV} instead combines data for 14 clusters from the \matcha{} pipeline and 14 clusters from a similar \xmm{} pipeline~\citep{XCS}, both using non-core-excised temperatures within \rtfh{}.
\citet{redmapperSV} calculates a slope of $0.61 \pm 0.09$ and an intercept of $1.52 \pm 0.07$, with $\sigmaintr = 0.28^{+0.07}_{-0.05}$.
These data have been normalized to our pivots, and we have used the relation quoted in \citet{redmapperSV} to convert the \xmm{} temperatures to equivalent \chandra{} temperatures.
Our slopes are in statistical agreement, with their slope differing from ours by $0.9 \sigma$ in the opposite direction of \citet{redmapperII}.
In this case our $\sigmaintr$ agrees nicely as well.
Their intercept here appears to disagree, however without more information on the uncertainty in their \chandra{}-to-\xmm{} conversion it is difficult to determine the degree of disagreement.
It is reassuring that our slope and scatter agree with \citet{redmapperSV}, given that they use the \matcha{} pipeline (in conjunction with a similar pipeline for \xmm{} data) to supply the X-ray data for their scaling relations.

\begin{figure*}[htbp!]
  \centering%
  \tx{}-$\lambda$ Scaling Relations, \rtfh{} Aperture
  \gridline{%
    \fig{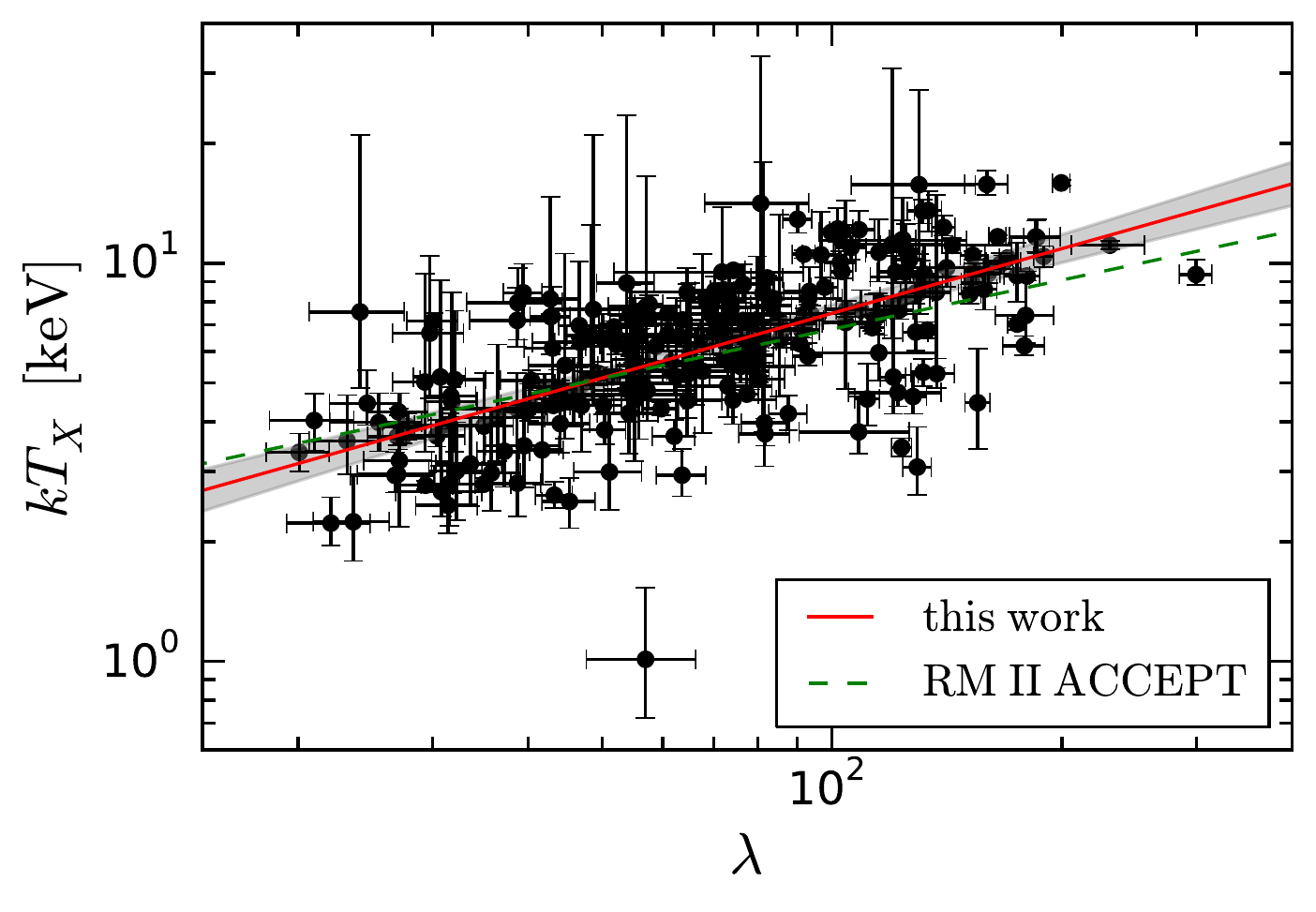}{0.5\textwidth}{(a)}
    \fig{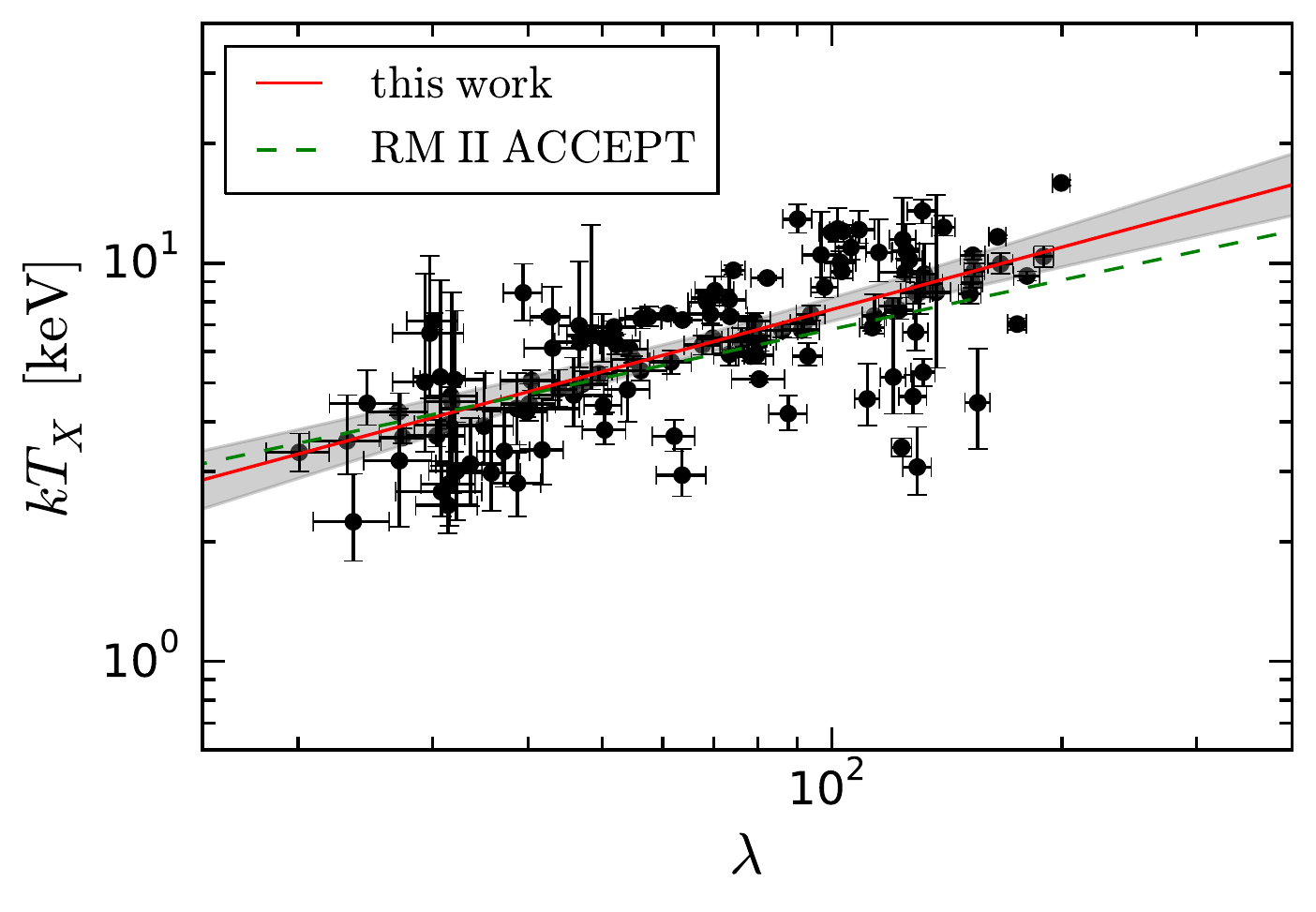}{0.5\textwidth}{(b)}
  }
  \caption{%
    (a)
    \rtfh{} \tx{}-$\lambda$, all redshift.
    (b)
    \centering{}\rtfh{} \tx{}-$\lambda$, $0.1 < z < 0.35$.
    The black dots and associated error bars are the data produced by \matcha{} (see \autoref{sec:matcha}).
    The red lines are the best-fits given in \autoref{tab:r2500-scaling-relations}.
    The surrounding gray areas are the $2\sigma{}$ uncertainties in the fits.
    The dotted green lines are the ACCEPT \tx{}-lambda scaling relations from \citet{redmapperII}.
    For discussion of the outliers in these plots, see \autoref{subsec:outliers}.\label{fig:tx-lambda-r2500}
  }
\end{figure*}

Due to the low total exposure times of many clusters within our sample and the resulting high uncertainty in our core-excised temperatures, we choose not to present core-excised \rfh{} relations here.
Plots of our \rtfh{} \tx{}-$\lambda$ data may be found in \autoref{fig:tx-lambda-r2500}.

\begin{figure*}[htbp!]
  \centering%
  \tx{}-$\lambda$ Scaling Relations, \rfh{} Aperture
  \gridline{%
    \fig{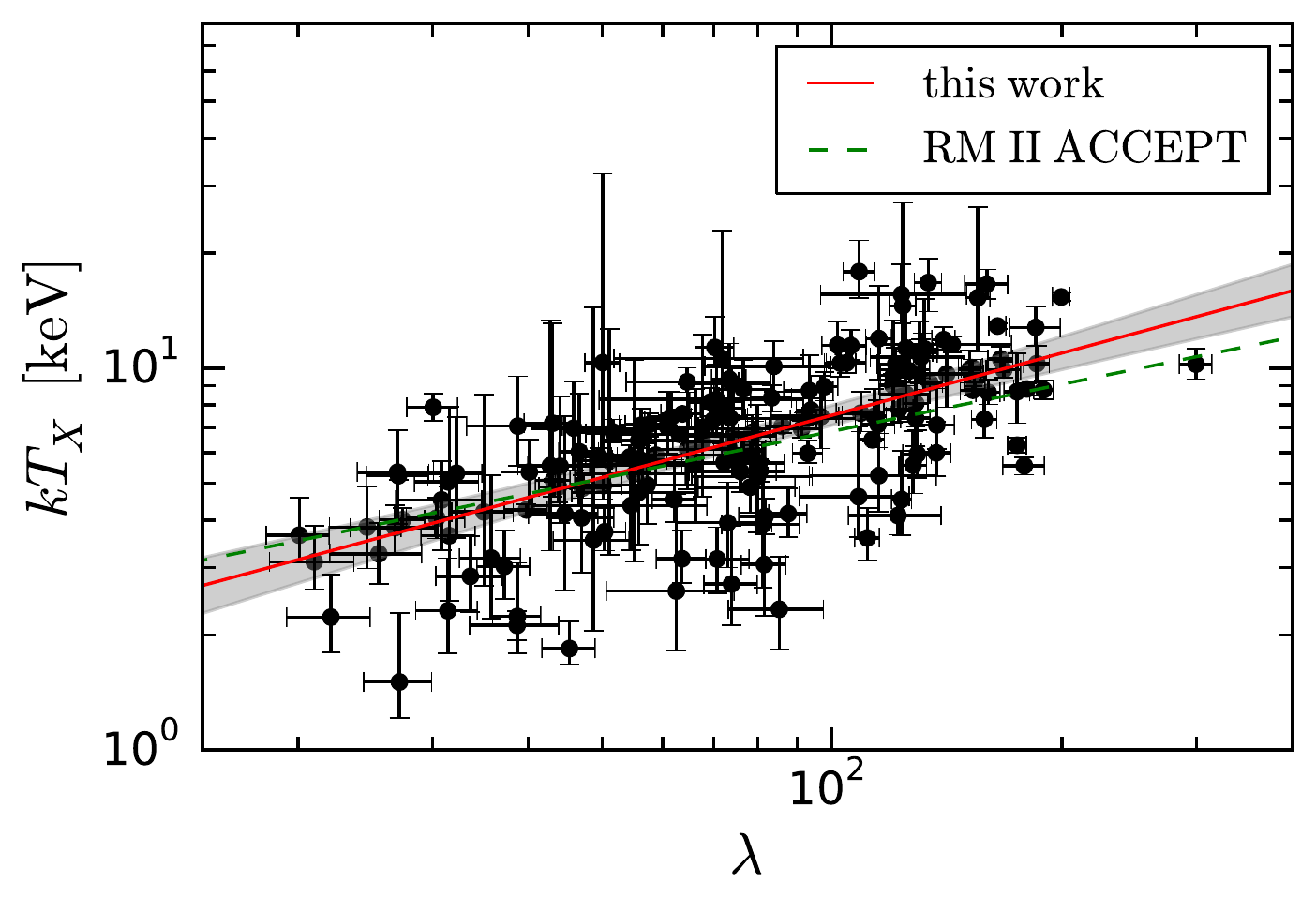}{0.5\textwidth}{(a)}
    \fig{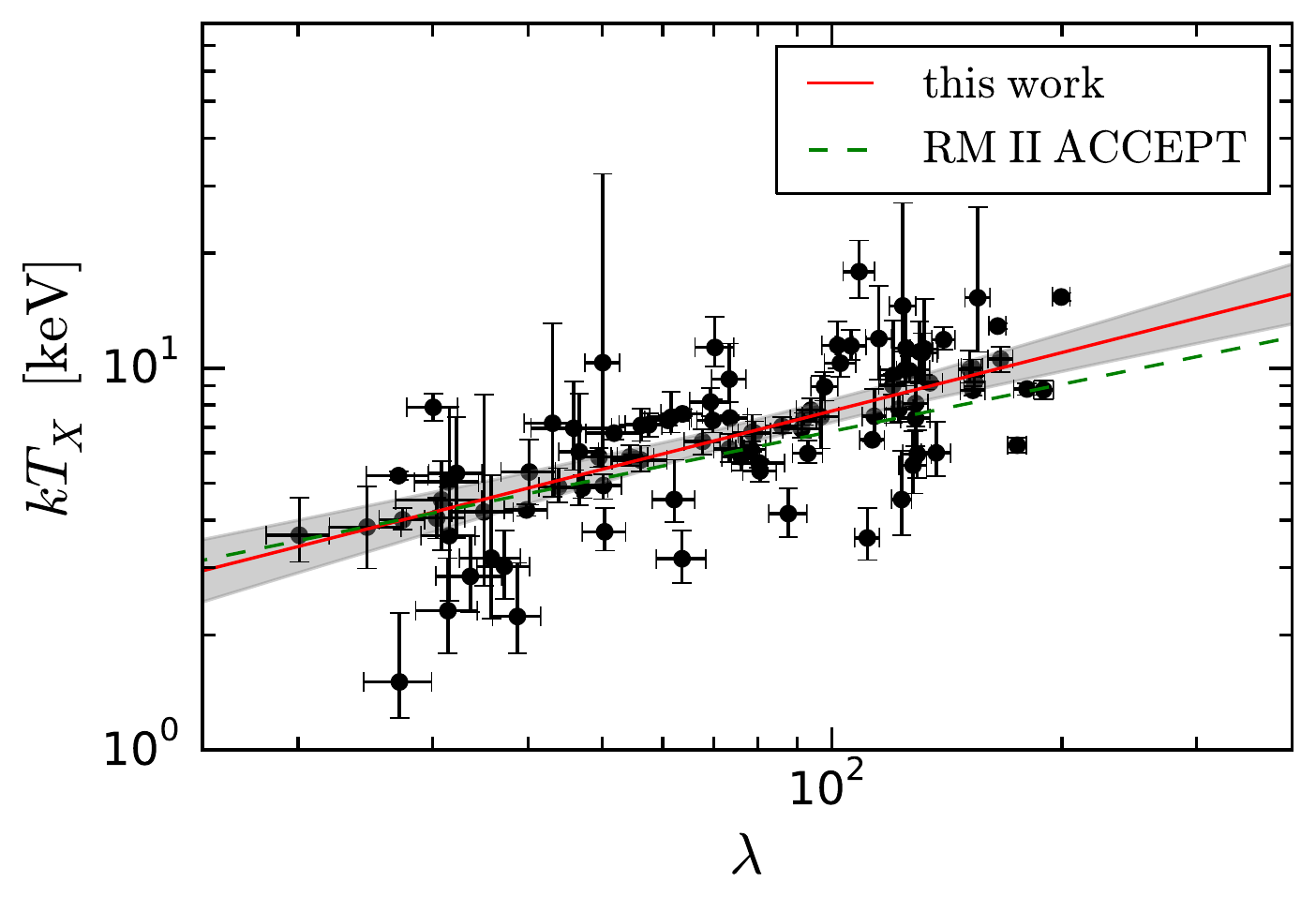}{0.5\textwidth}{(b)}
  }
  \caption{%
    (a)
    \rfh{} \tx{}-$\lambda$, all redshift.
    (b)
    \centering{}\rfh{} \tx{}-$\lambda$, $0.1 < z < 0.35$.
    The black dots and associated error bars are the data produced by \matcha{} (see \autoref{sec:matcha}).
    The red lines are the best-fits given in \autoref{tab:r500-scaling-relations}.
    The surrounding gray areas are the $2\sigma{}$ uncertainties in the fits.
    The dotted green lines are the ACCEPT \tx{}-lambda scaling relations from \citet{redmapperII}.%
    For discussion of the outliers in these plots, see \autoref{subsec:outliers}.\label{fig:tx-lambda-r500}
  }
\end{figure*}

We find that our slope, intercept, and $\sigmaintr$ do not change significantly when we increase the considered aperture from \rtfh{} to \rfh{}.
Plots of this \rfh{} \tx{}-$\lambda$ data may be found in \autoref{fig:tx-lambda-r500}.

\begin{figure*}[htbp!]
  \centering%
  \lx{}-$\lambda$ Scaling Relations, \rtfh{} Aperture
  \gridline{%
    \fig{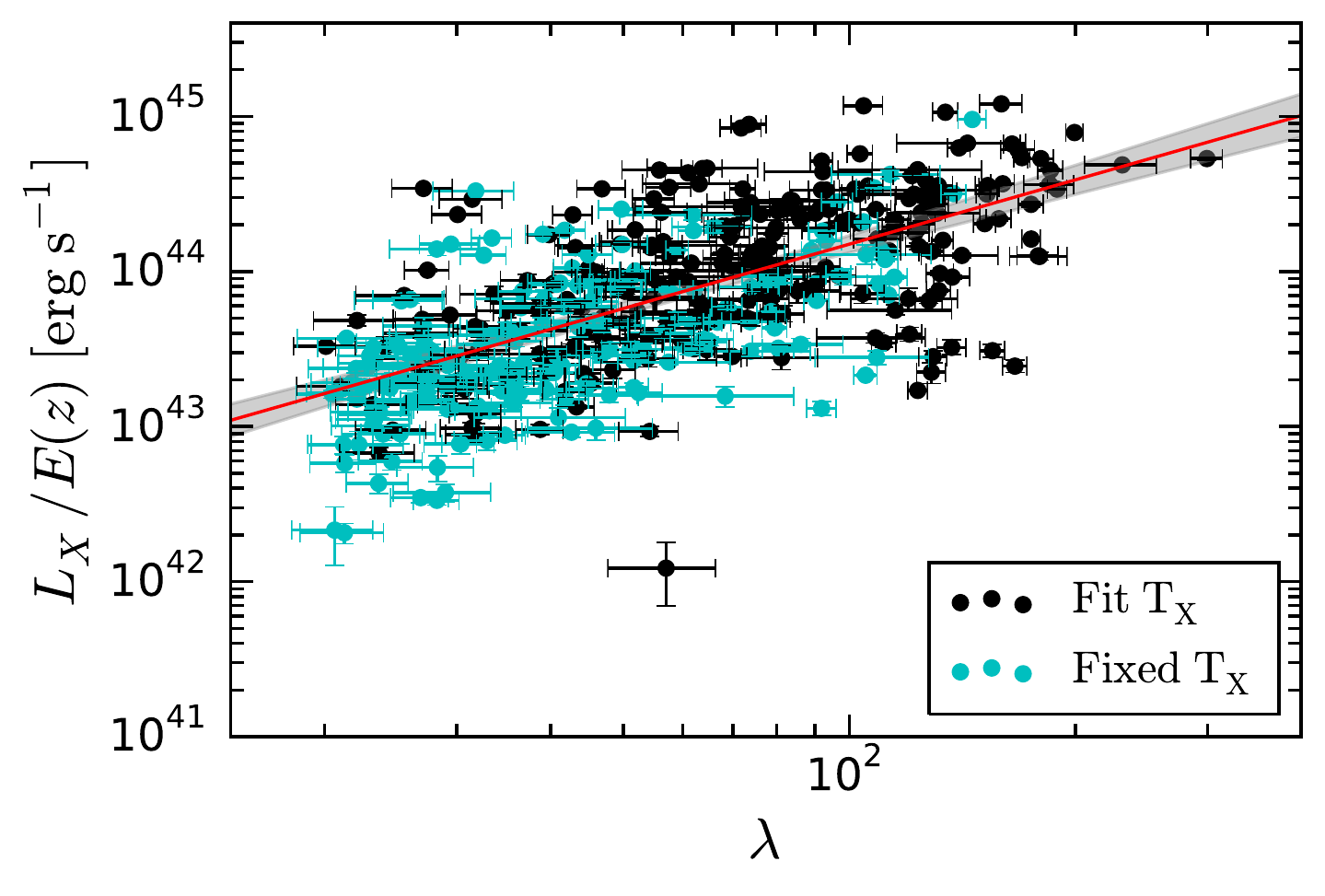}{0.5\textwidth}{(a)}
    \fig{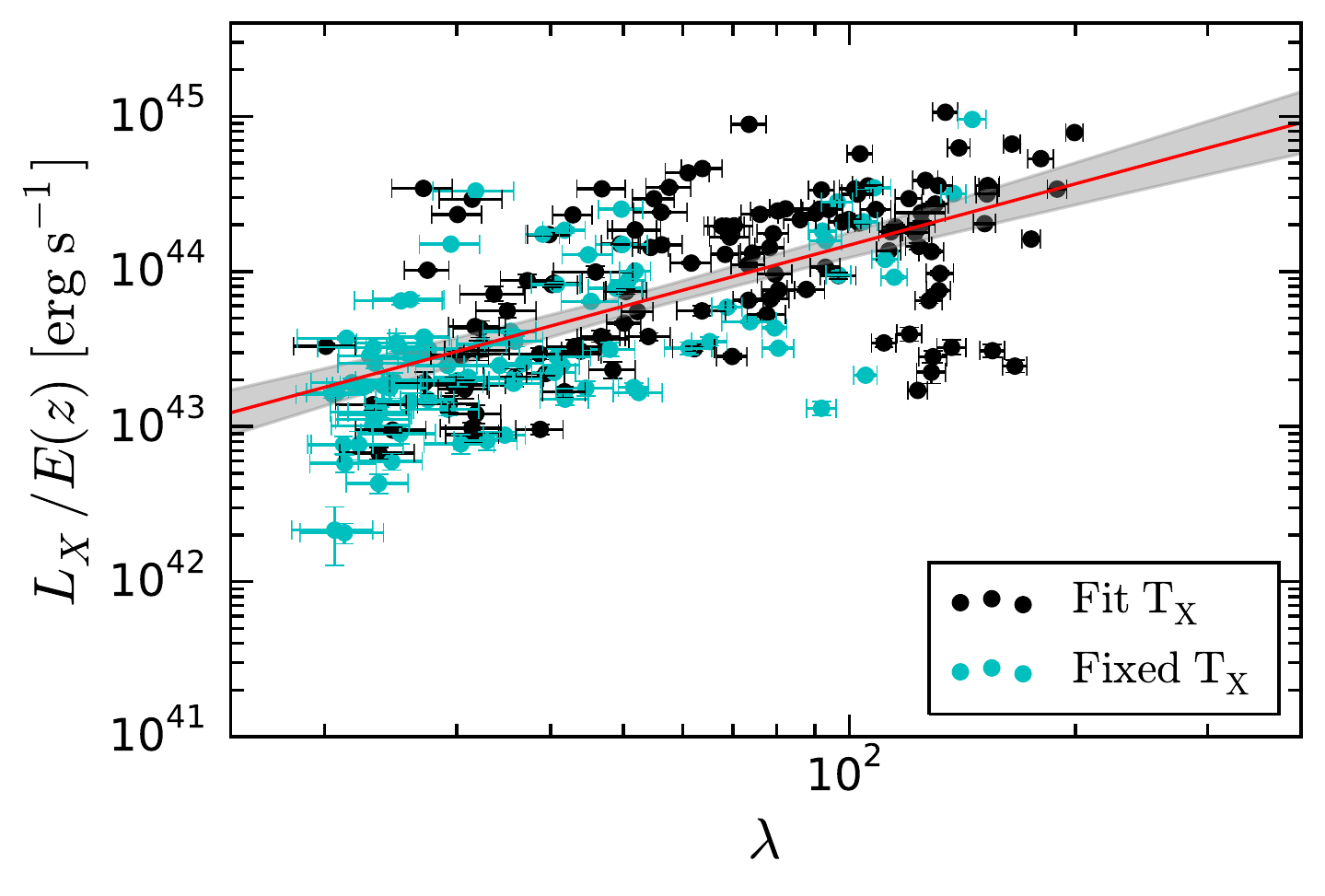}{0.5\textwidth}{(b)}
  }
  \gridline{%
    \fig{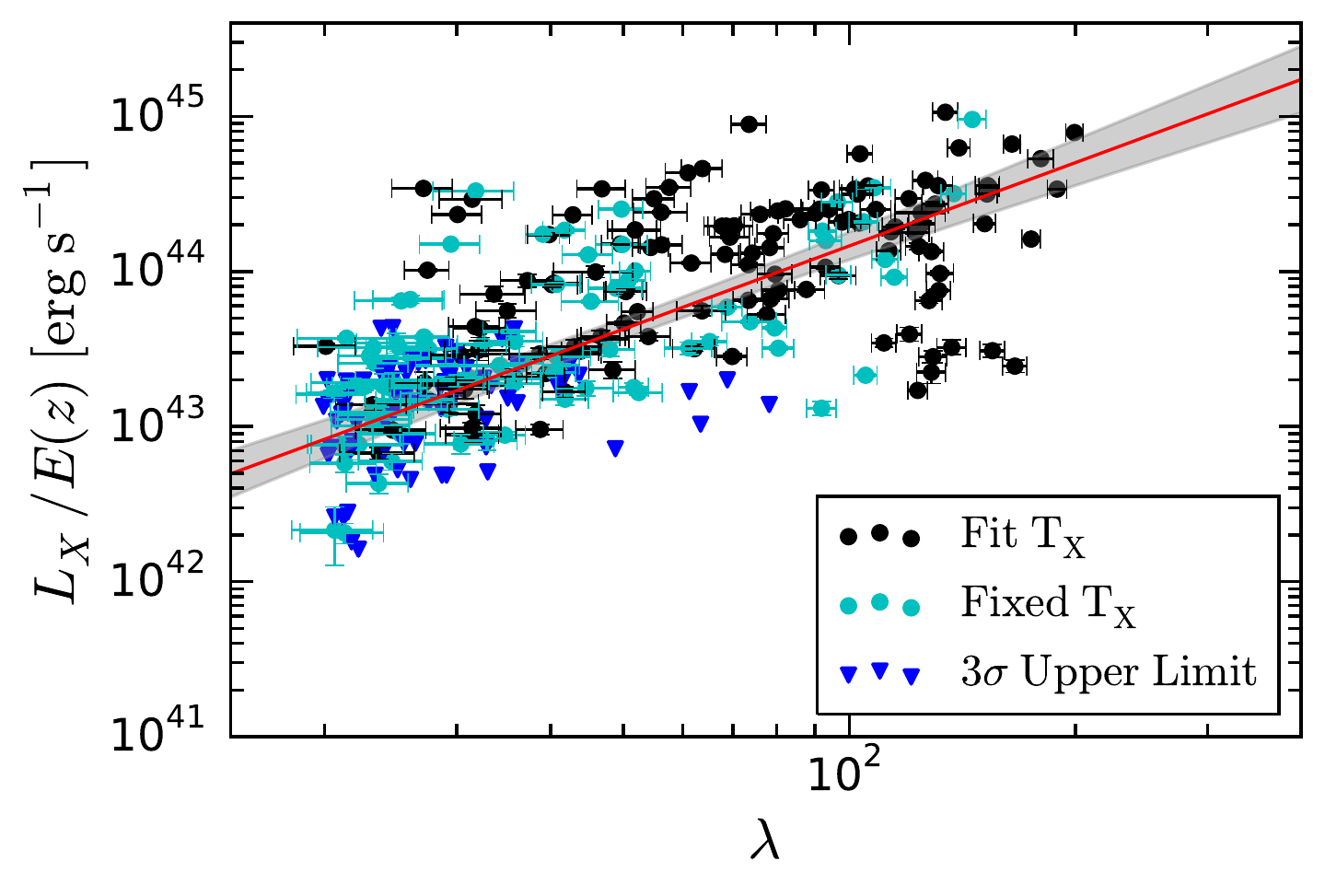}{0.5\textwidth}{(c)}
  }
  \caption{%
    (a)
    \rtfh{} \lx{}-$\lambda$, all redshift, no upper limits.
    (b)
    \rtfh{} \lx{}-$\lambda$, $0.1 < z < 0.35$, no upper limits.
    (c)
    \rtfh{} \lx{}-$\lambda$, $0.1 < z < 0.35$, with upper limits.
    ``Fit \tx'' (black dots) are clusters which are detected and which have their \lx{} fit along with their measured \tx{}, as described in \autoref{subsec:finding-tx}.
    ``Fixed \tx{}'' (cyan dots) are detected clusters which have their luminosities fit with an assumed rather than fit \tx{}, as described in \autoref{subsec:lx-no-tx}.
    ``\lx{} $3\sigma$ upper limit'' (blue triangles) are $3 \sigma$ upper-limit luminosities for undetected clusters, as described in \autoref{subsec:lx-upper-lims}.
    The red lines are the best-fits given in \autoref{tab:r2500-scaling-relations}.
    The surrounding gray areas are the $2\sigma$ uncertainties in the fits.
    The notable outlier in (a) is RM J115807.3+554459.4 (\memmatch{} 13419, $z = 0.50$); it is discussed in \autoref{subsec:outliers}.\label{fig:lx-lambda-r2500}
  }
\end{figure*}

Our best-fit to the \rtfh{} \lx{}-richness relation in the $0.1 < z < 0.35$ range, without taking into account non-detections, is $\lxrelation[r2500,detected]{-0.08 \pm{} 0.07}{1.31 \pm{} 0.12}$ with $\sigmaintr = 0.92 \pm 0.05$.
As with \tx{}-$\lambda$, this is not significantly different from the same relation including all redshifts.
This suggests that our decision to include \lx{} upper limits solely in the $0.1 < z < 0.35$ range does not affect our result significantly, however the greater volume of data would help us constrain the \lx{}-$\lambda$ $\sigmaintr$ to a greater degree of certainty.
When we include luminosity upper limits, we find
\begin{equation}
  \lxrelation[r2500]{-0.26 \pm 0.08}{1.78 \pm 0.12}
\end{equation}
with $\sigmaintr = 1.04 \pm 0.06$.
This is a significant increase in the slope, a significant decrease in the intercept, and a slight increase in $\sigmaintr$.
This steepening of the \lx{}-$\lambda$ relation is expected because in the low-$\lambda$ regime we only detect clusters on the high-\lx{} side of the scatter.
These three \lx{}-$\lambda$ relations may be found in \autoref{fig:lx-lambda-r2500}.
Additionally, we find that the presence-or-lack of fixed-\tx{} \lx{} values (see \autoref{subsec:lx-no-tx}) does not significantly affect $\sigmaintr$, and has only minor effect on our slope and intercept.
For more information on the effects of selection on X-ray scaling relations, see e.g.\ \citet{MantzII} and \autoref{subsec:selection-effects}.

\begin{figure*}[htbp!]
  \centering%
  \lx{}-$\lambda$ Scaling Relations, \rfh{} Aperture
  \gridline{%
    \fig{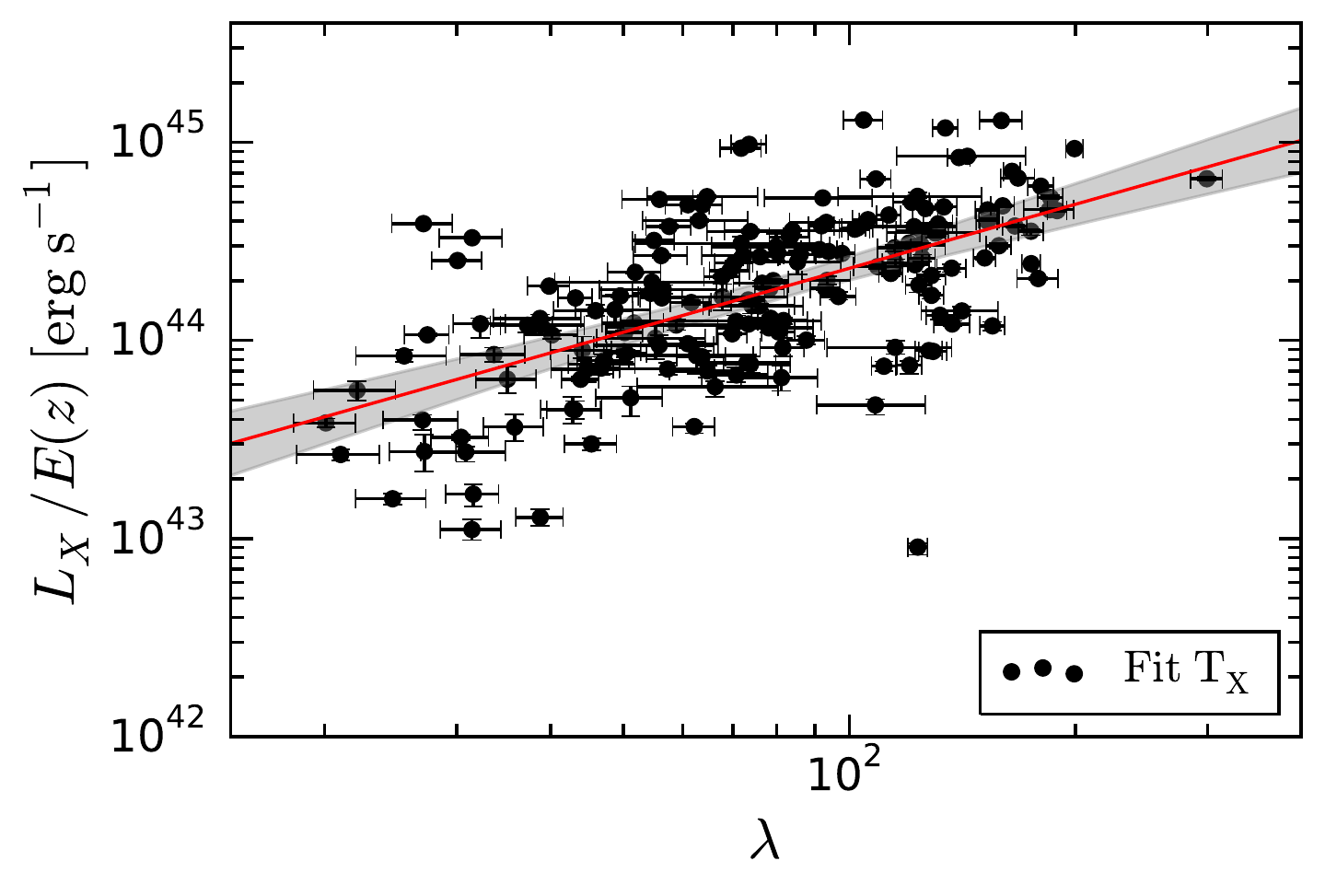}{0.5\textwidth}{(a)}
    \fig{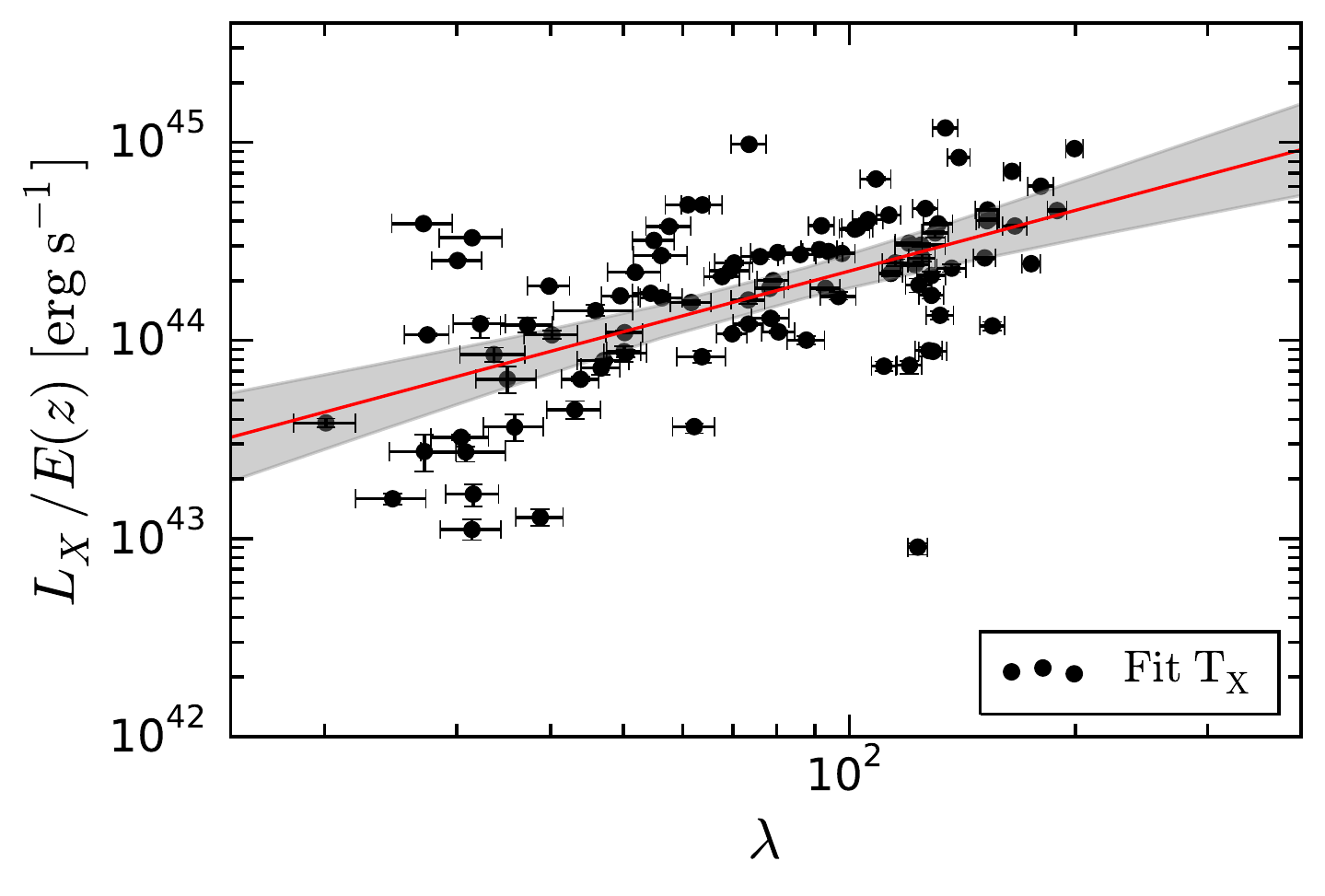}{0.5\textwidth}{(b)}
  }
  \caption{%
    (a)
    \rfh{} \lx{}-$\lambda$, all redshift.
    (b)
    \rfh{} \lx{}-$\lambda$, $0.1 < z < 0.35$.
    ``Fit \tx{}'' (black dots) are clusters which are detected and which have their \lx{} fit along with their \tx{}, as described in \autoref{subsec:finding-tx}.
    The red lines are the best-fits given in \autoref{tab:r500-scaling-relations}.
    The surrounding gray areas are the $2\sigma$ uncertainties in the fits.
    The striking outlier here is RM J004629.3+202804.8 (\memmatch{} 15, $z = 0.10$); see \autoref{subsec:outliers} for more information.\label{fig:lx-lambda-r500}
  }
\end{figure*}

We find that our relation has an increased intercept as we widen the considered aperture from \rtfh{} to \rfh{}.
$\sigmaintr$ decreases slightly (from $0.99 \pm 0.06$ to $0.84 \pm 0.06$), and the slope does not change significantly.
The \rfh{} relations may be found in \autoref{fig:lx-lambda-r500}.

\subsection{X-Ray\textendash{}X-Ray Scaling Relations}%
\label{subsec:x-ray-scaling-relations}
In addition to our \tx{}-$\lambda$ and \lx-$\lambda$ scaling relations, we derive \lx{}-\tx{} and \tx{}-\lx{} scaling relations within $0.1 < z < 0.35$.
As before, we use the Bayesian fitting method presented in \citet{Kelly}.
Our resulting relations are discussed below in the form $\ln\left(y\right) = \alpha\ln\left(x/\mathrm{pivot}\right) + \beta$, and are presented in \autoref{tab:x-ray-scaling-relations}.

\begin{deluxetable*}{lrrrr}
  \tablecaption{X-Ray\textendash{}X-Ray Scaling Relations}
  \tablehead{\colhead{Relation} & \colhead{$\beta$} & \colhead{$\alpha$} & \colhead{$\sigmaintr$} & \colhead{Figure}}
  \startdata{}%
  \rtfh{} $L_X-T_X$ ($0.1 < z < 0.35$) & $-0.05 \pm 0.08$ & $2.07 \pm 0.20$ & $0.81 \pm 0.06$ & \autoref{fig:lx-tx} (a) \\
  \rtfh{} $L_X-T_X$ ($0.1 < z < 0.35$, bolometric) & $1.08 \pm 0.08$ & $2.49 \pm 0.19$ & $0.80 \pm 0.06$ & \autoref{fig:lx-tx} (b) \\
  \midrule{}%
  \rfh{} $L_X-T_X$ ($0.1 < z < 0.35$) & $0.39 \pm 0.07$ & $2.08 \pm 0.18$ & $0.58 \pm 0.06$ & \autoref{fig:lx-tx} (c) \\
  \rfh{} $L_X-T_X$ ($0.1 < z < 0.35$, bolometric) & $1.54 \pm 0.07$ & $2.53 \pm 0.18$ & $0.54 \pm 0.06$ & \autoref{fig:lx-tx} (d) \\
  \midrule{}%
  \rtfh{} $T_X-L_X$ ($0.1 < z < 0.35$) & $1.89 \pm 0.03$ & $0.26 \pm 0.03$ & $0.28 \pm 0.02$ & \autoref{fig:tx-lx} (a) \\
  \midrule{}%
  \rfh{} $T_X-L_X$ ($0.1 < z < 0.35$) & $1.78 \pm 0.03$ & $0.32 \pm 0.03$ & $0.23 \pm 0.02$ & \autoref{fig:tx-lx} (b) \\
  \enddata{}
  \tablecomments{%
    Relations are of the form $\ln\left(y\right) = \alpha\ln\left(x/\mathrm{pivot}\right) + \beta$.
    \lx{} is normalized by $E\left(z\right)$ and has units of $10^{44}$ ergs/s; \tx{} has units of keV.
    When the independent variable, $L_{X}/E\left(z\right)$ has pivot $0.3 \cdot{} 10^{44}$ and \tx{} has pivot 2.0 keV.
    $\sigmaintr$ is the standard deviation of the intrinsic scatter in this relation.
    Uncertainties are listed as their $1\sigma{}$ values.\label{tab:x-ray-scaling-relations}.
    The scatter distributions for each of these relations are slightly asymmetric, with longer tails in the large-scatter direction.
  }
\end{deluxetable*}

\begin{figure*}[htbp!]
  \centering%
  \lx{}-\tx{} Relations
  \gridline{%
    \fig{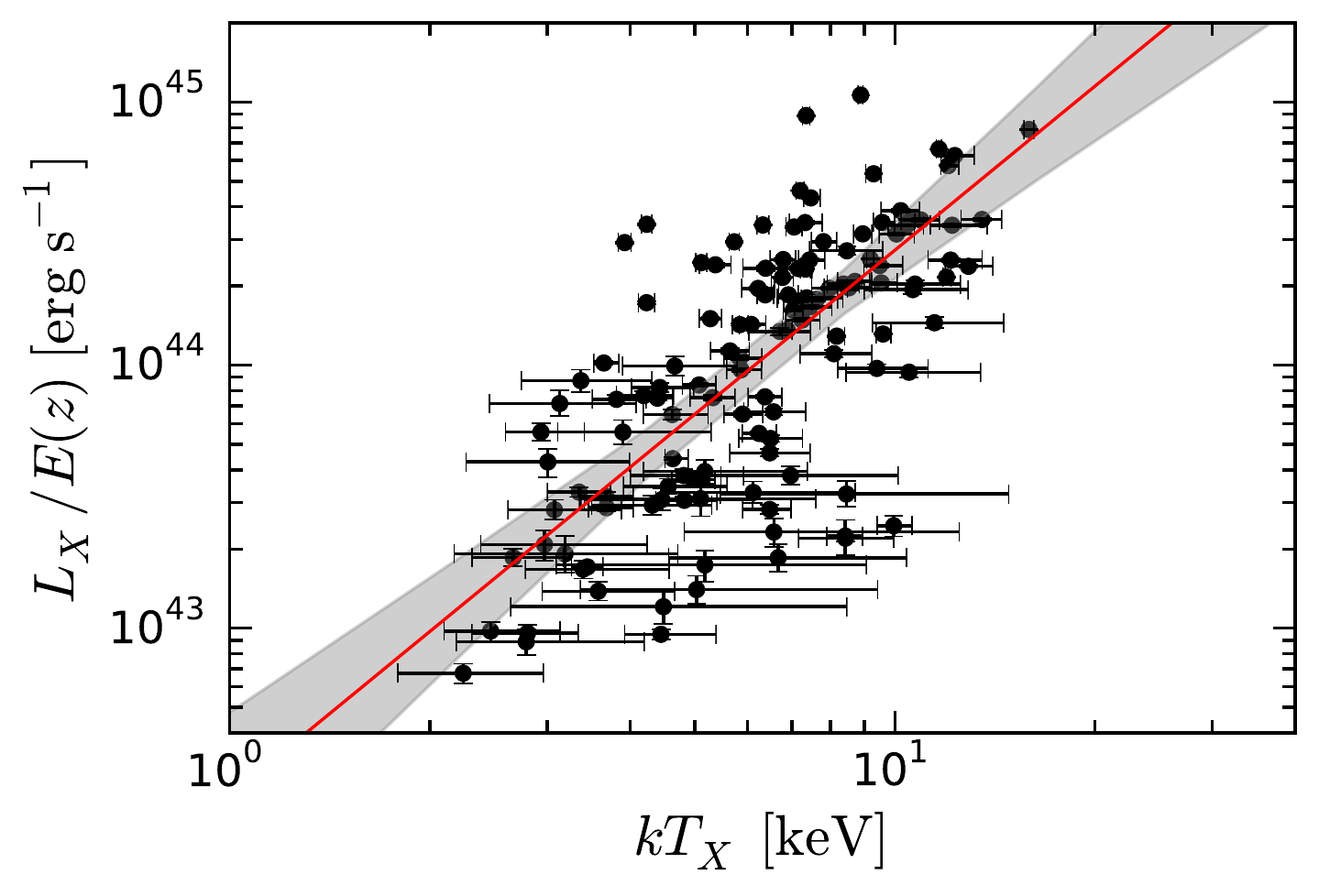}{0.5\textwidth}{(a)}
    \fig{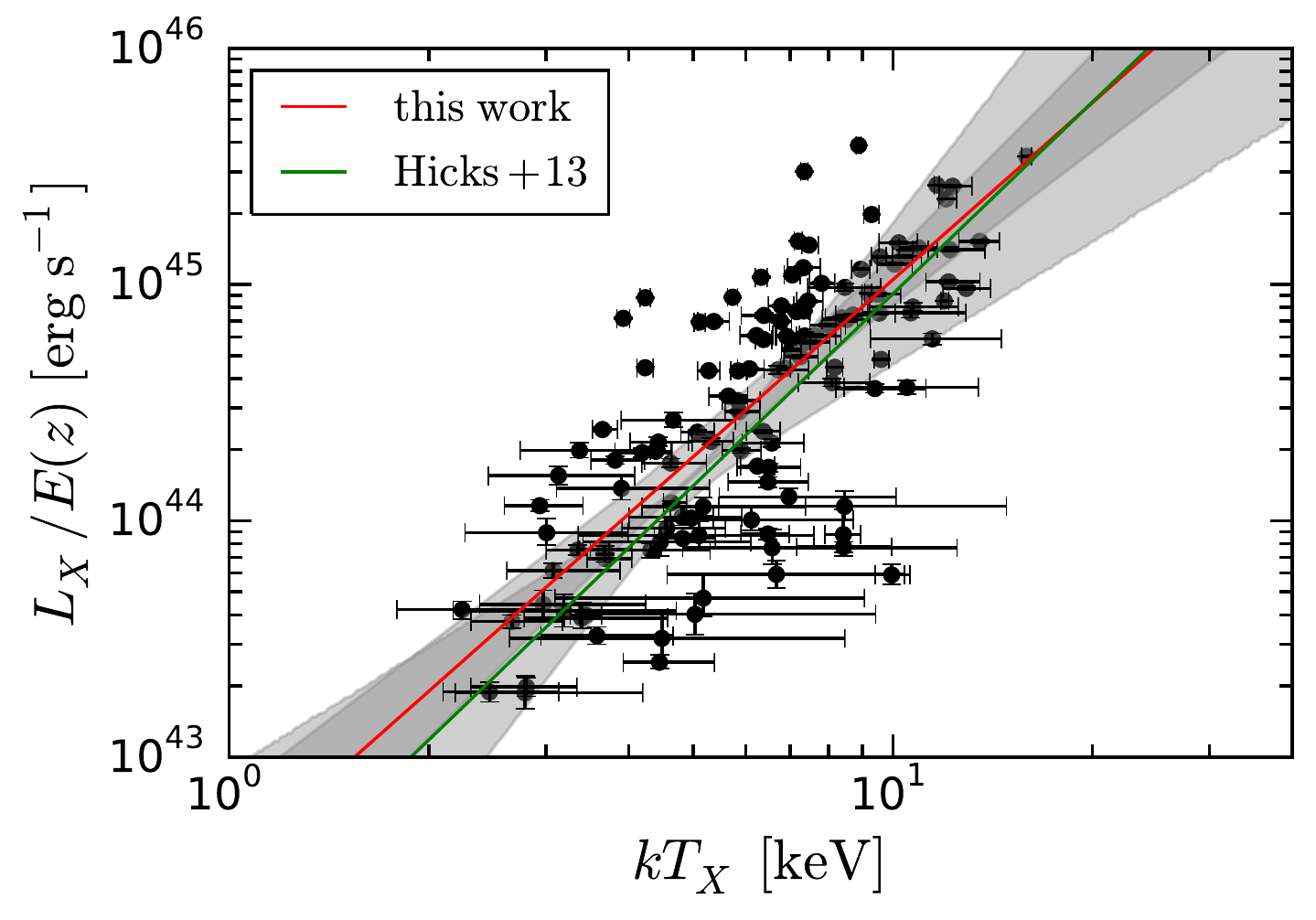}{0.5\textwidth}{(b)}
  }
  \gridline{%
    \fig{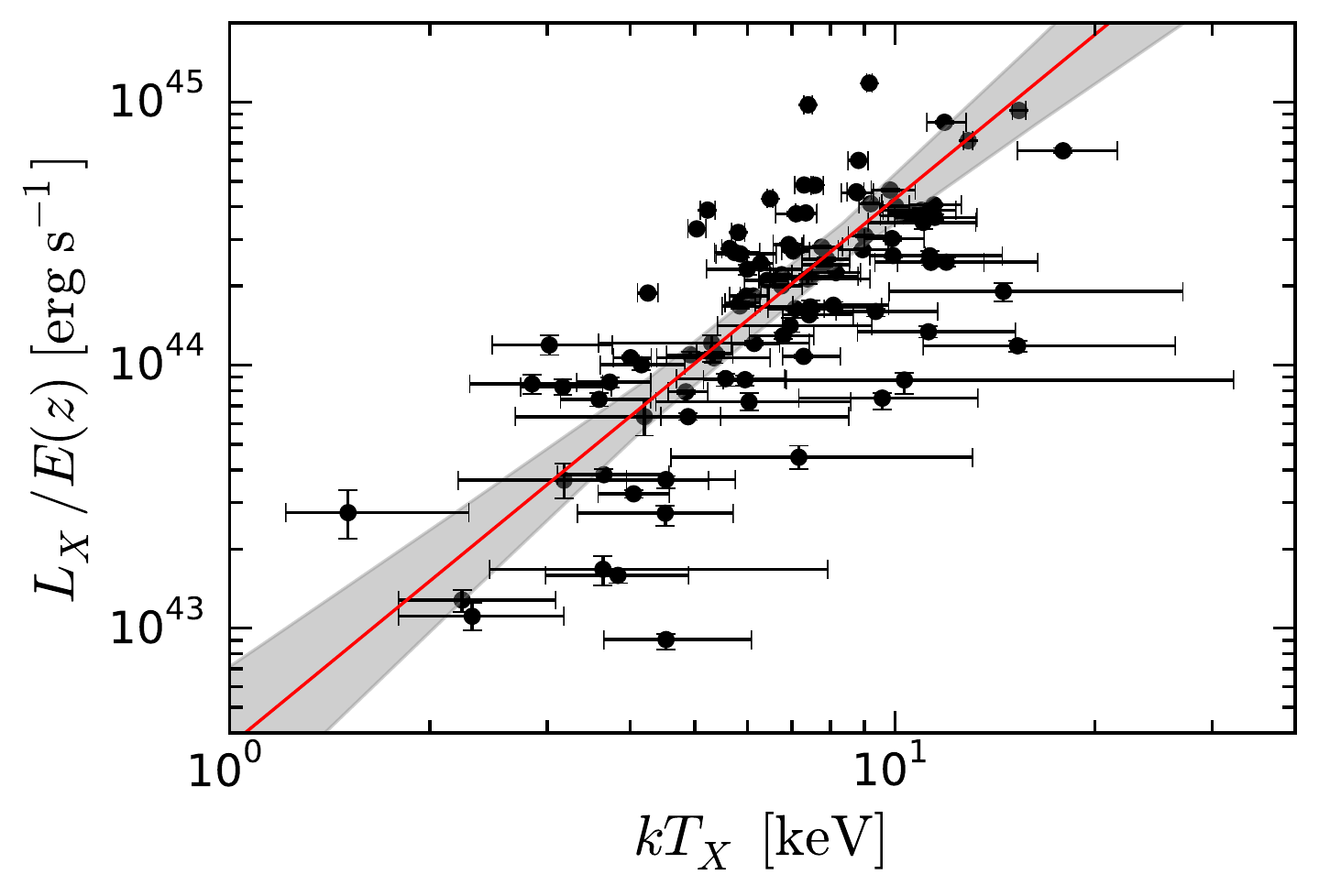}{0.5\textwidth}{(c)}
    \fig{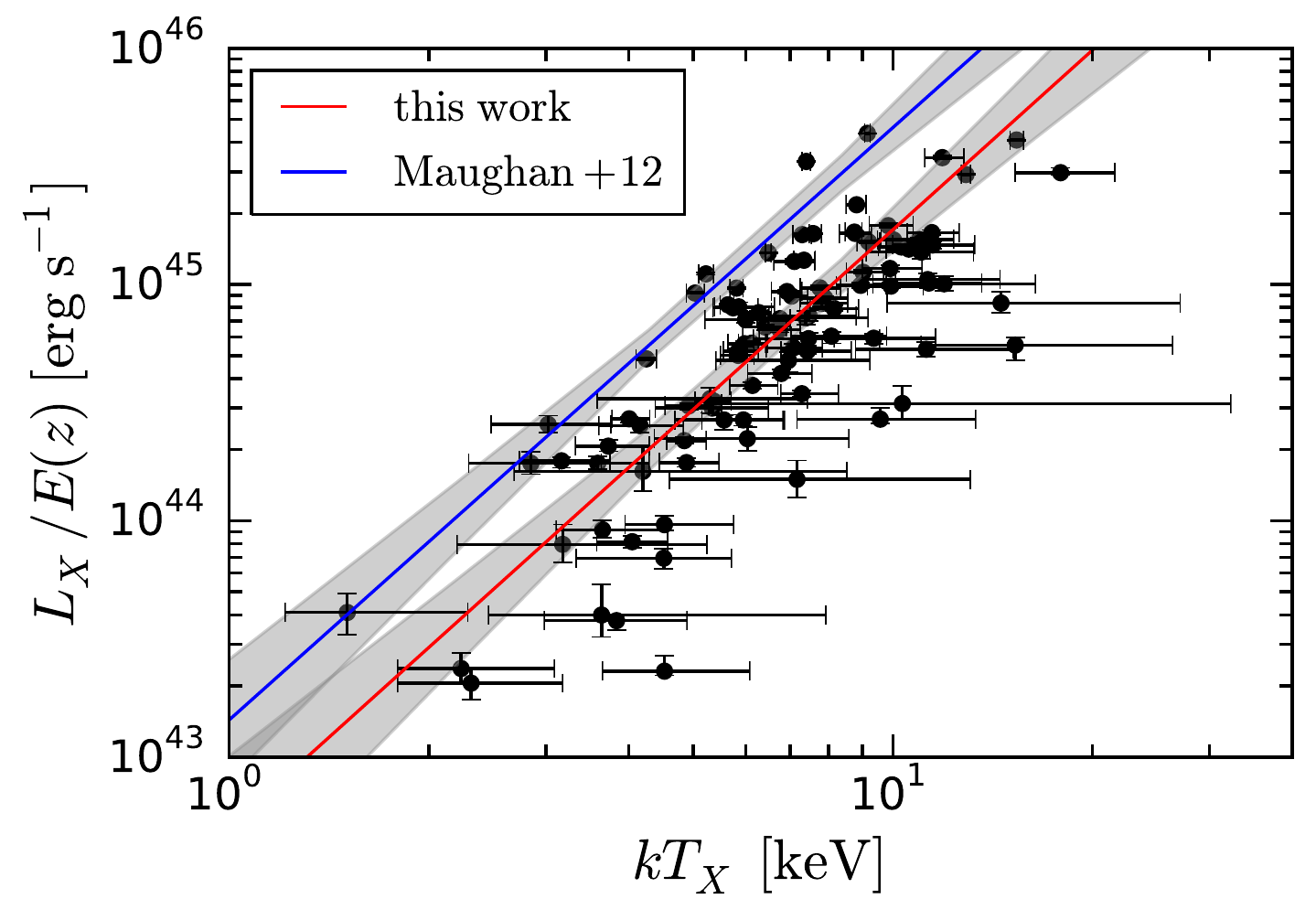}{0.5\textwidth}{(d)}
  }
  \caption{%
    (a)
    \rtfh{} soft-band \lx{}-\tx{}, $0.1 < z < 0.35$.
    (b)
    \rtfh{} bolometric \lx{}-\tx{}, $0.1 < z < 0.35$.
    (c)
    \rfh{} soft-band \lx{}-\tx{}, $0.1 < z < 0.35$.
    (d)
    \rfh{} bolometric \lx{}-\tx{}, $0.1 < z < 0.35$.
    In each figure, the \lx{} and \tx{} obtained by \matcha{} are shown in black.
    The red line represents the best-fit given in \autoref{tab:r2500-scaling-relations}.
    The surrounding gray area is the $2\sigma{}$ uncertainty in our fit.
    In figure (b), the green line represents the best-fit given in \citet{Hicks13}.
    In figure (d), the blue line represents our best-fit to the data given in \citet{Maughan12}.
    In figures (b) and (d), we mirror \citet{Hicks13}'s and \citet{Maughan12}'s choice to use bolometric luminosities.\label{fig:lx-tx}
  }
\end{figure*}

For \lx{}-\tx{} in \rtfh{}, we derive
\begin{equation}
  \lxtxrelation[r2500]{-0.05 \pm{} 0.08}{2.07 \pm{} 0.20}
\end{equation}
with $\sigmaintr = 0.81 \pm 0.06$.
Here, there are not many particularly comparable relations from the literature: most papers either choose to excise cluster cores, measure their luminosities in a different band, or use differing instruments (which are known to have an offset when compared with \chandra{}).
After a literature search, we find that the best comparison for \rtfh{} is \citet{Hicks13}, and for \rfh{} is \citet{Maughan12}.
The comparison with \citet{Hicks13} is straightforwards in both method of analysis and cluster selection, and we find that our \rtfh{} bolometric luminosities agree well with their listed \lx{}-\tx{} slope of $2.7 \pm{} 0.5$.
See \autoref{fig:lx-tx} (b) for a visual comparison.

As we increase our aperture from \rtfh{} to \rfh{}, we find that the our bolometric \lx{}-\tx{} intercept increases to $1.54 \pm 0.07$, our slope steepens to $2.54 \pm 0.18$, and our $\sigmaintr$ decreases to $0.55 \pm 0.05$.
This differs significantly from \citet{Maughan12}, which lists a slope of $3.63 \pm 0.27$ for this relation.
However, \citet{Maughan12} uses a different regression method, specifically the BCES orthogonal regression \citep{BCES}, compared to the Bayesian regression method of \citet{Kelly} we employ.
Fitting their data with our methodology, we find 2.55$\pm$0.06, 2.51$\pm$0.17, and 0.58$\pm$0.05 for the slope, intercept, and scatter respectively.
Both the slope and scatter are in very good agreement with our results.
There is, however, an offset in the normalization (see \autoref{fig:lx-tx} (d) for a visual comparison).
Comparing clusters which appear in both samples, we find that on average our \tx{} values are 28\% higher than those found by \citet{Maughan12}.
This offset is likely due to a combination of changes in the Chandra calibration, which include significant updates to the contamination model, and our use of the Cash statistic instead of $\chi^2$ in the spectral fitting.
We use CALDB version 4.6.7 compared to \citet{Maughan12}'s use of CALDB 4.3.0.
\citet{Humphrey09} find that \tx{} values fit using $\chi^2$ are biased low by $\approx$10\% even for well-sampled clusters while the Cash statistic is relatively unbiased.
Up to 20\% changes in \tx{} have also been found when employing different CALDB versions \citep{Reese10, Giles17}.
In particular, \citet{Giles17} finds hydrostatic masses 29\% higher for their analysis using CALDB 4.5.9 and a Cash statistic when fitting \tx{} compared to using CALDB 4.3.1 and $\chi^2$, with 20\% originating from the CALDB change and the rest from the fit statistic.
This difference is quite similar to what we find.

\begin{figure*}[htbp!]
  \centering%
  \tx{}-\lx{} Relations
  \gridline{%
    \fig{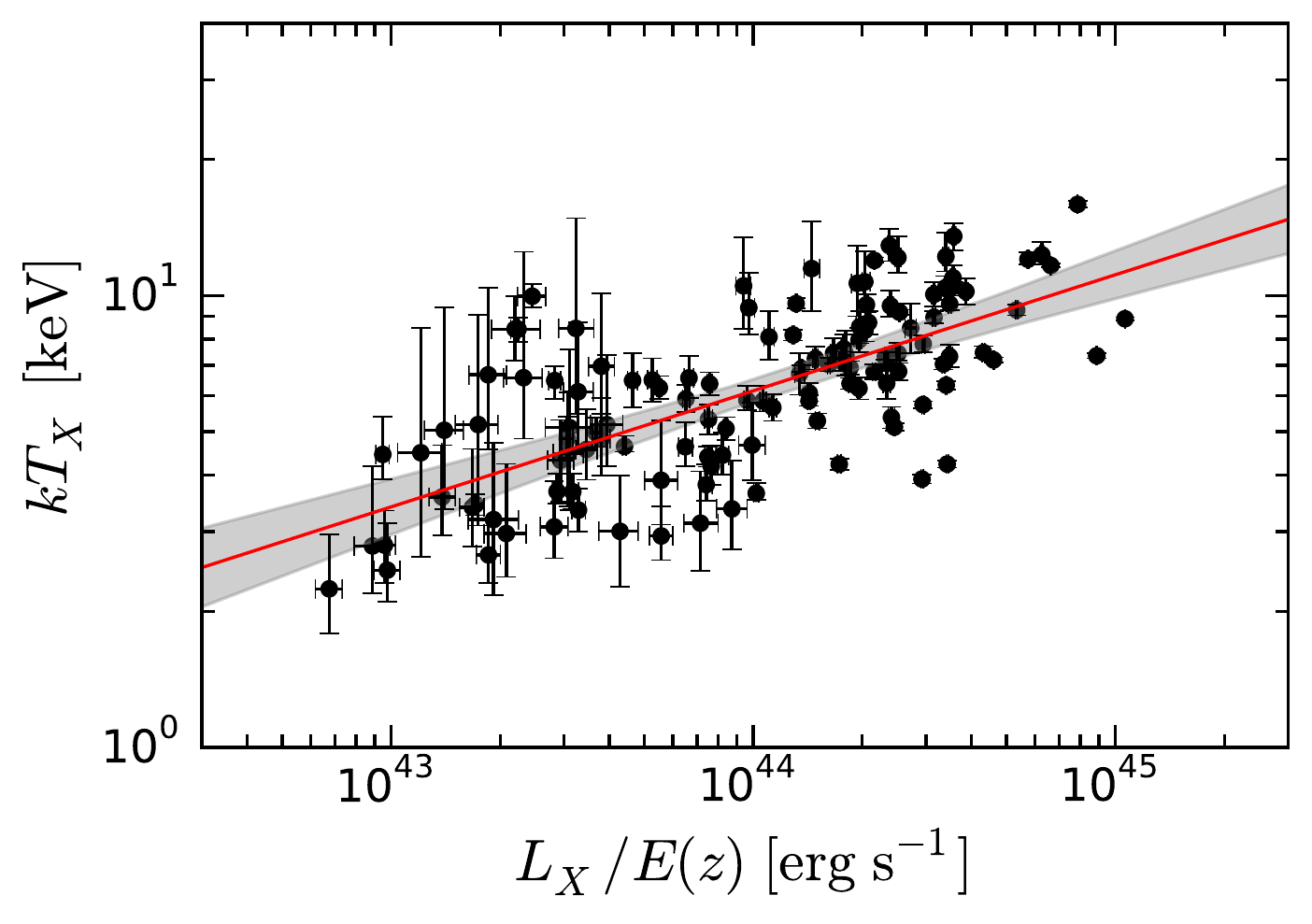}{0.5\textwidth}{(a)}
    \fig{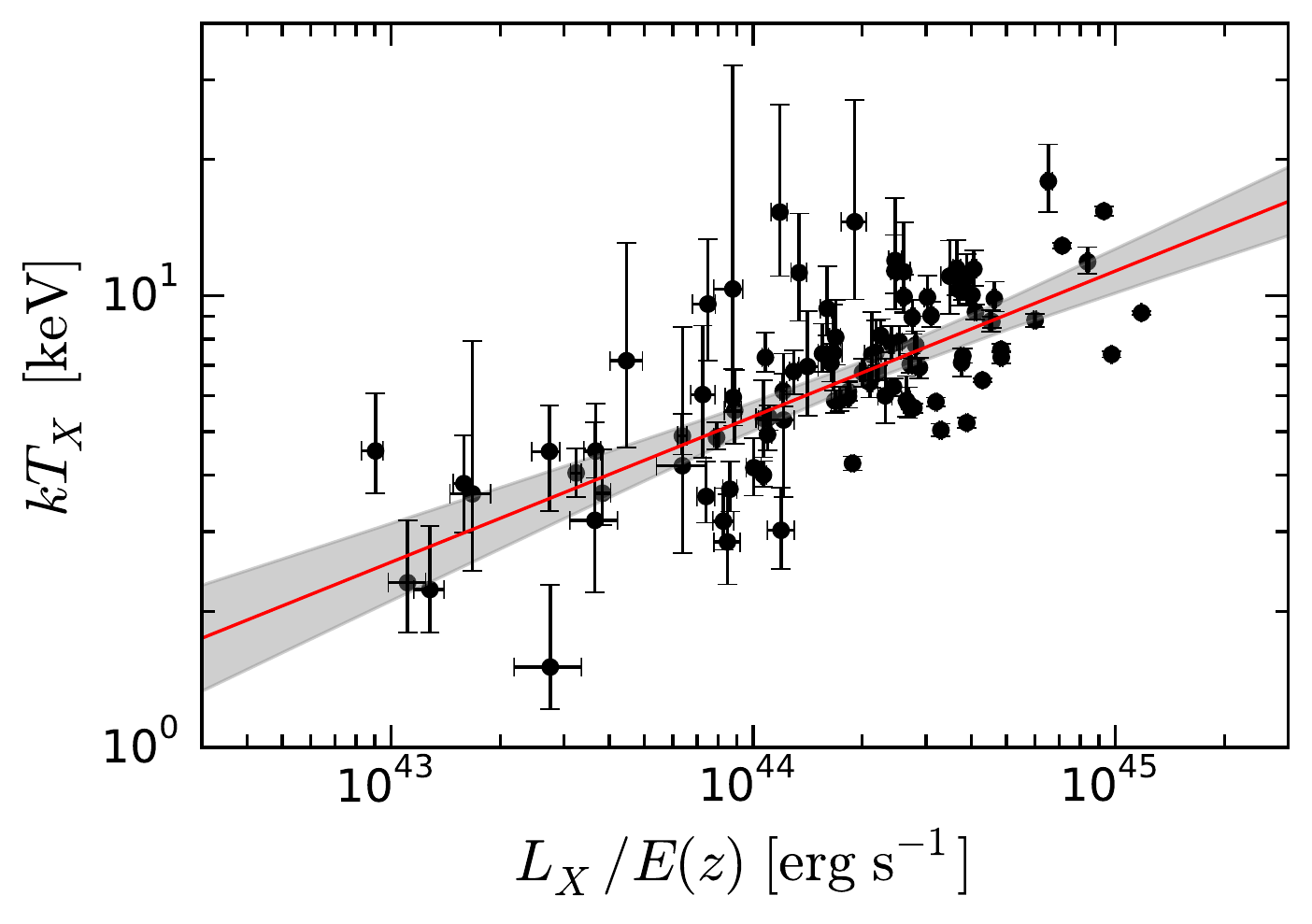}{0.5\textwidth}{(b)}
  }
  \caption{%
    (a)
    \rtfh{} soft-band \tx{}-\lx{}, $0.1 < z < 0.35$.
    (b)
    \rfh{} soft-band \tx{}-\lx{}, $0.1 < z < 0.35$.
    In each figure, the \tx{} and \lx{} obtained by \matcha{} are shown in black.
    The red line represents the best-fit given in \autoref{tab:r500-scaling-relations}.
    The surrounding gray area is the $2\sigma{}$ uncertainty in our fit.\label{fig:tx-lx}
  }
\end{figure*}

Because the selection effect of \lx{} in our X-ray data is much stronger than that of \tx{} (see \autoref{subsec:selection-effects}), it is desirable to examine the reverse relation with \lx{} as the dependent variable.
We derive a soft-band \tx{}-\lx{} relation within $0.1 < z < 0.35$ of
\begin{equation}
  \txlxrelation{1.89 \pm 0.03}{0.26 \pm 0.03}
\end{equation}
with $\sigmaintr = 0.28 \pm 0.02$.
At \rfh{}, the intercept drops to $1.78 \pm 0.03$, the slope increases to $0.32 \pm 0.03$, and the scatter drops to $0.23 \pm 0.02$.
These results are shown in \autoref{fig:tx-lx}.

\subsection{Scaling Relation Outliers}%
\label{subsec:outliers}
As can be seen in Figures~\ref{fig:tx-lambda-r2500}--\ref{fig:lx-lambda-r500}, there are a number of clusters that seem to have high richnesses for their X-ray properties.
These clusters tend to be low-\tx{} clusters showing evidence of projection effects: \redmapper{} has added correlated foreground and/or background halos to these clusters, increasing their richness significantly.
For more information on projection effects in \redmapper{}, see e.g.\ \citet{redmapperIV} and \citet{Costanzi18}.

\begin{figure*}[htbp!]
  \centering%
  \includegraphics[width=0.5\textwidth]{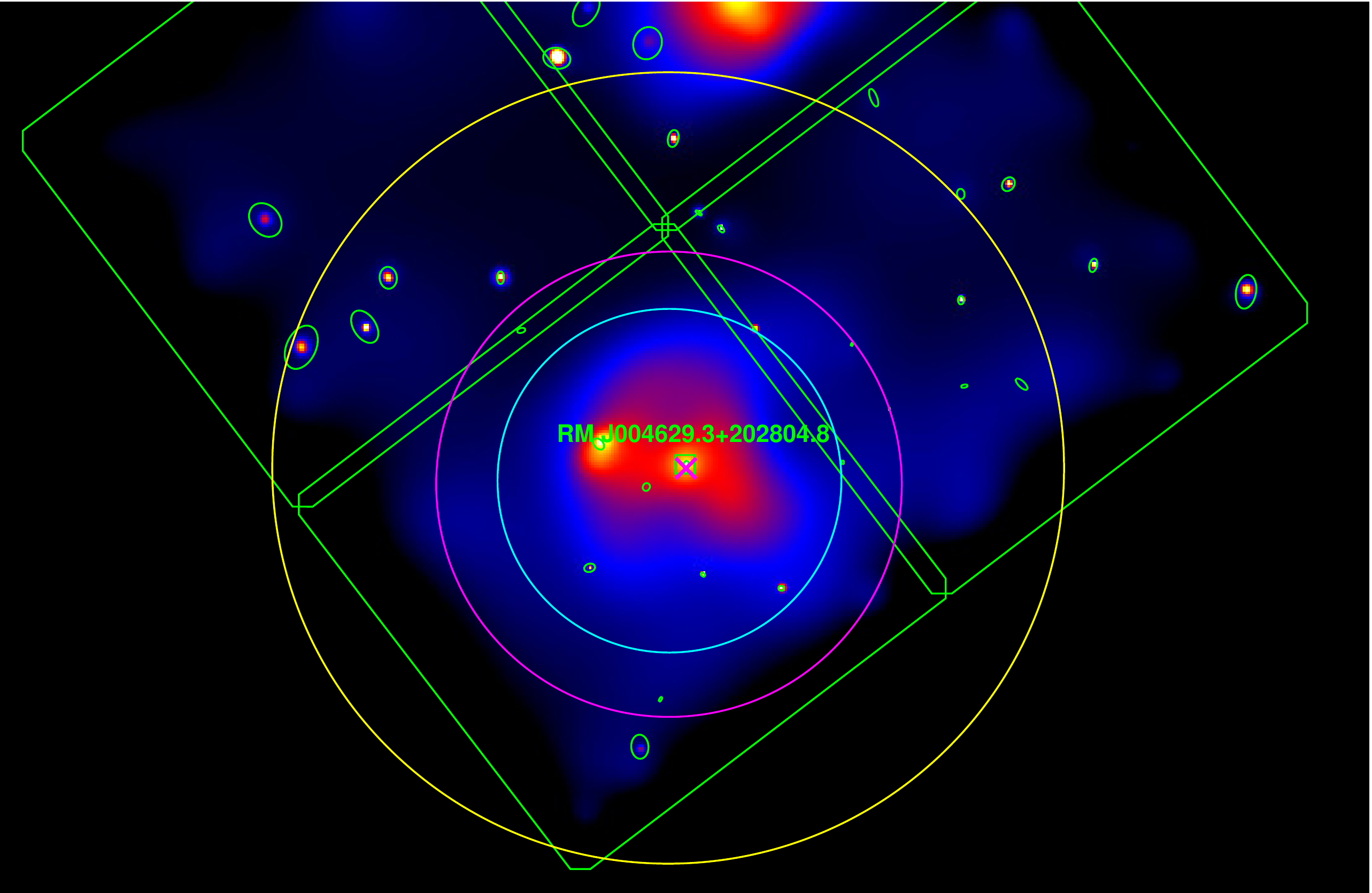}
  \caption{%
    RM J004629.3+202804.8 (\memmatch{} 15, $z = 0.10$), ObsID 11876, ACIS-I detector.
    This ``galaxy cluster'' is actually three separate halos which lie near to one another when projected along line-of-sight (the third halo is in a separate observation, to the south).
    \redmapper{} incorrectly counts this as a single cluster with a very large richness of 123.4.
    Because this sort of error is intrinsic to the \redmapper{} algorithm, we keep this cluster as-is for our analysis.
    \colorinfosingle{}\label{fig:supercluster-merge}
  }
\end{figure*}

An example of this is RM J004629.3+202804.8 (see \autoref{fig:supercluster-merge}), which is actually a supercluster composed of three separate galaxy clusters.
\redmapper{} merges these separate galaxy clusters into one single large cluster with a very large richness.
This is the striking low-\lx{}, high-richness outlier in \autoref{fig:lx-lambda-r500}.

Another issue is mispercolation (discussed in \autoref{subsec:mispercolations}), in which \redmapper{} incorrectly divides a single large halo into two or more separate ``clusters''.
We handle these on a case-by-case basis, with a typical decision being to flag the smaller ``galaxy-cluster'' as being masked by the larger, true galaxy cluster and thus discarded from our analysis.
See \autoref{subsec:post-pipeline} for details on the process of flagging galaxy clusters for potential problems, and \autoref{fig:flag-effects} (c) in \autoref{sec:flag-effects} for a plot of the locations of mispercolated clusters within our \rtfh{} \tx{}-$\lambda$ data.
Were these included, they would be outliers with high \tx{}/\lx{} and low $\lambda$, and would artificially flatten the slope and increase the scatter of the richness scaling relations.

Finally, there is the notable outlier in \autoref{fig:lx-lambda-r2500} (a): RM J115807.3+554459.4 (\memmatch{} 13419, $z = 0.50$).
After careful checking of the \matcha{} X-ray analysis, we believe that \redmapper{} has significantly overestimated the richness of this cluster.
This is likely due to an issue in \redmapper{}'s extrapolation of richness for high-z clusters (see \autoref{sec:cluster-selection}).

\subsection{Effects of Selection}%
\label{subsec:selection-effects}
Due to the archival nature of our sample, our results may exhibit bias due to selection.
Bigger, brighter clusters may have been more likely to be the object of \chandra{} observations than less-luminous clusters at the same redshift.
Were this effect equivalent to a simple flux cut on our data, we would see the effects of classic Malmquist bias: we would observe a decreased slope and scatter in our \lx{}-$\lambda$ relations when compared with unbiased data~\citep{MantzII}.
Indeed, we explore the effects of applying a flux cut to simulated \redmapper{} data and find that removing low-flux clusters flattens the slope of our resulting \lx-$\lambda$ relation.
In truth our selection function is much more complicated than a flux cut.
This is because observers will choose longer exposure times for dimmer clusters if those clusters are known prior to observation.
Thus more sophisticated modeling is needed, and such modeling is well beyond the scope of this paper.

There are a number of ways of probing the effects of our selection function on our fitted \lx{}-$\lambda$ relation.
When we include upper limits in \lx{}-$\lambda$, we see our slope increase.
The scatter does not increase significantly.
When we compare serendipitous and non-serendipitous clusters (see \autoref{subsec:post-pipeline}), we see that the former have a lower scatter than the latter.
Here the slope does not change significantly.
This change in scatter may be due to the fact that serendipitous clusters primarily lie within the low richness regime, where our sample is less complete (see \autoref{fig:flag-effects} (d)).
These aspects of our data make some intuitive sense: including upper limits helps take into account the fact that we are missing low-luminosity clusters, and serendipitous clusters should be less effected by observers' selection biases than clusters which were the targets of pointed observations.

We find that our \tx{}-$\lambda$ relations are significantly less susceptible to these selection effects than our \lx{}-$\lambda$ relations.
We do not find any significant effect on our fitted \rtfh{} \tx{}-$\lambda$ slope from limiting our sample to serendipitous clusters nor from our simulated flux cut.
Additionally, we have a complete sample of \rtfh{} \tx{} values for \redmapper{} clusters above $\lambda = 120$ and within $0.1 < z < 0.3$.
This complete sample exhibits a larger scatter in \rtfh{} \tx-$\lambda$ than our full catalog in the same redshift range, at $0.37 \pm 0.07$ vs.\ our full catalog's $\sigmaintr$ of $0.27 \pm 0.02$.
This effect may be due to the presence in this sample of an unusually high fraction of clusters with projection problems (see \autoref{subsec:outliers}) when compared with our full sample.
The complete sample's fitted slope has too large an uncertainty to draw conclusions there; its intercept is similarly uninstructive.

\subsection{Centering}%
\label{subsec:centering}
In order to understand the \redmapper{} miscentering function, we compare the \redmapper{} position with our X-ray centroids, which we calculate as described in \autoref{subsec:finding-tx} step~\ref{itm:centroid} and measure within \rtfh{}.
We additionally compare \redmapper{} positions with X-ray peak positions (see \autoref{subsec:peak-finding}).

\begin{figure*}[htbp]
  \centering%
  \gridline{%
    \fig{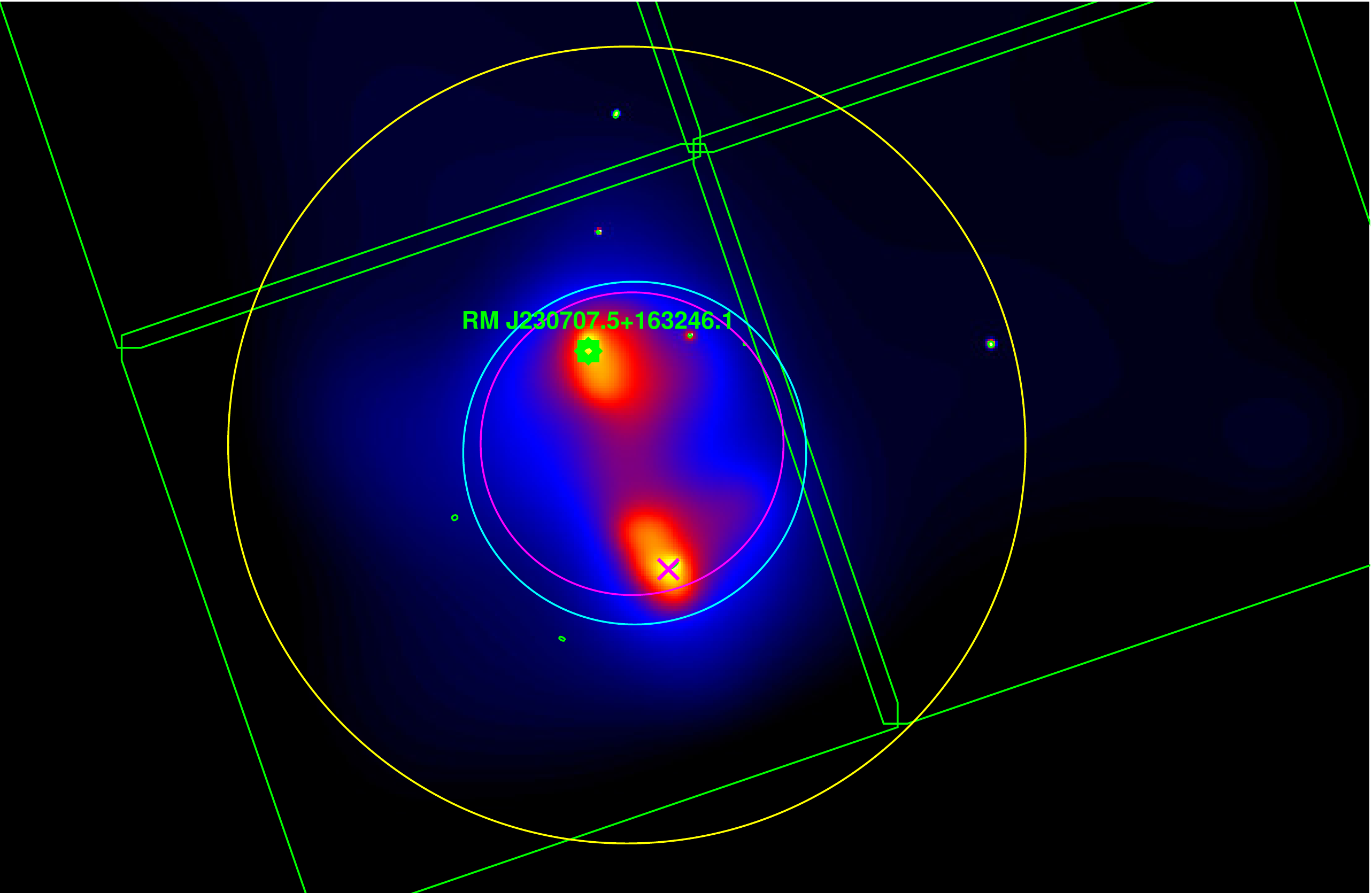}{0.5\textwidth}{(a)}
    \fig{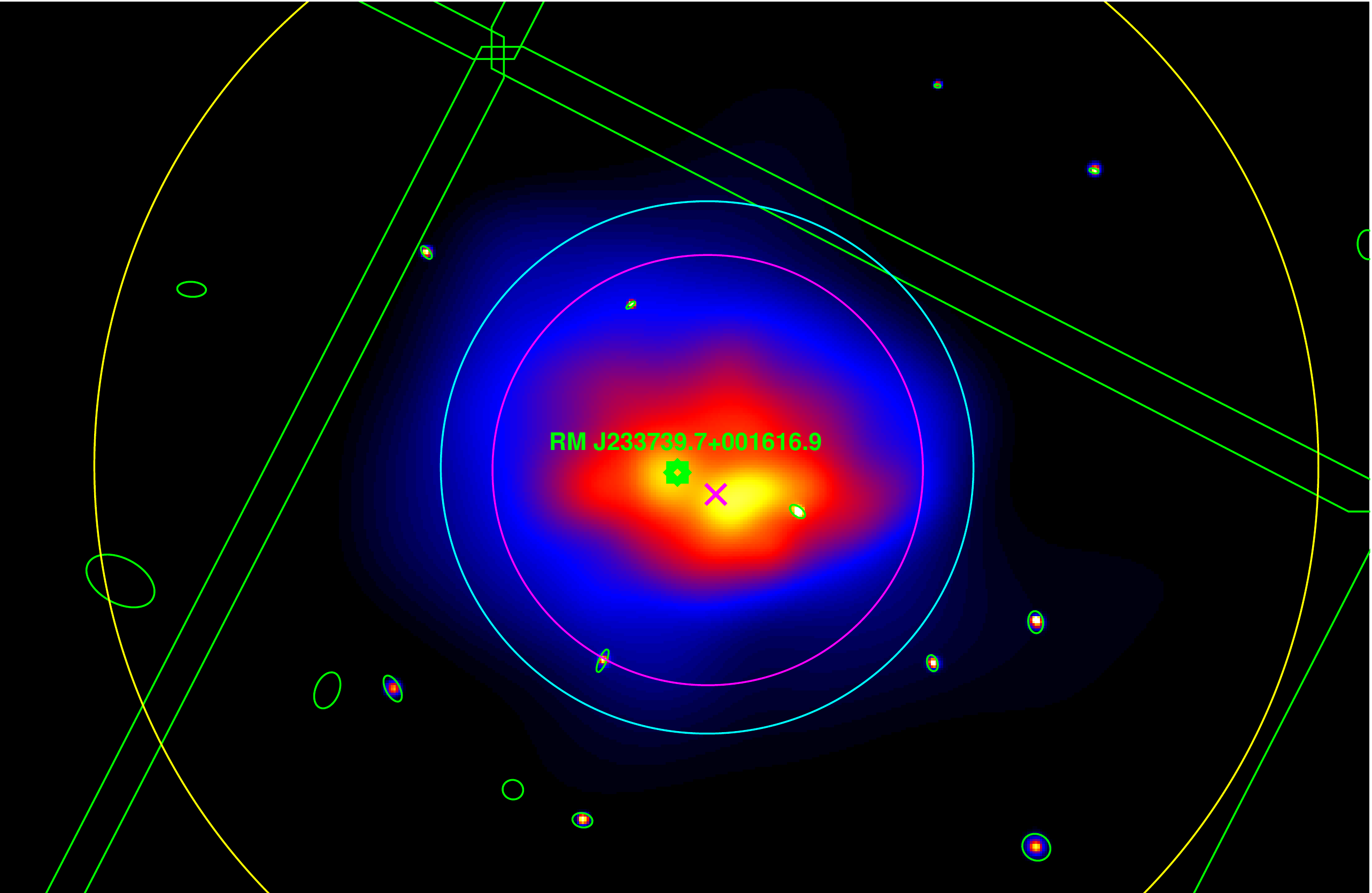}{0.5\textwidth}{(b)}
  }
  \gridline{%
    \fig{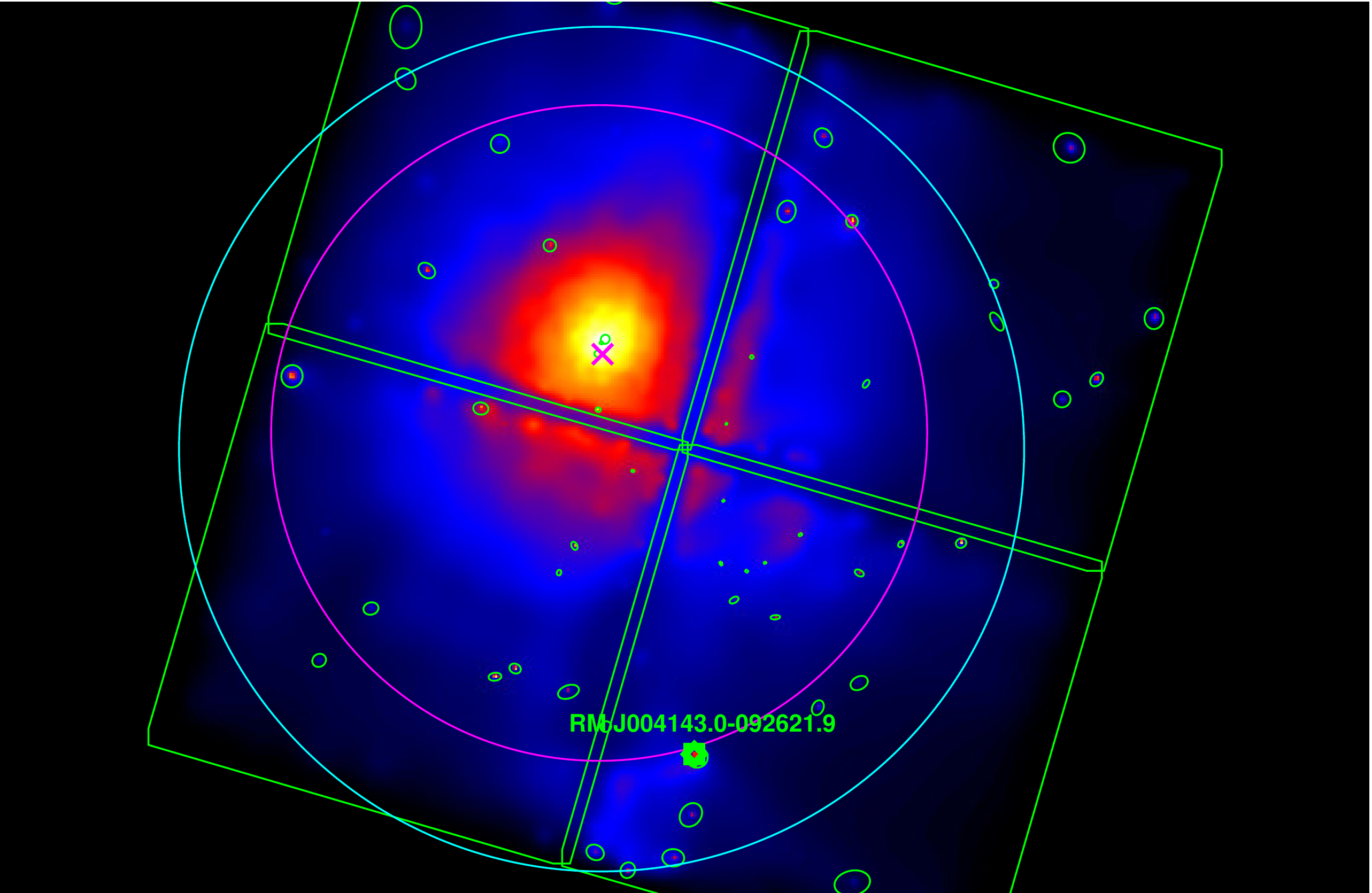}{0.5\textwidth}{(c)}
    \fig{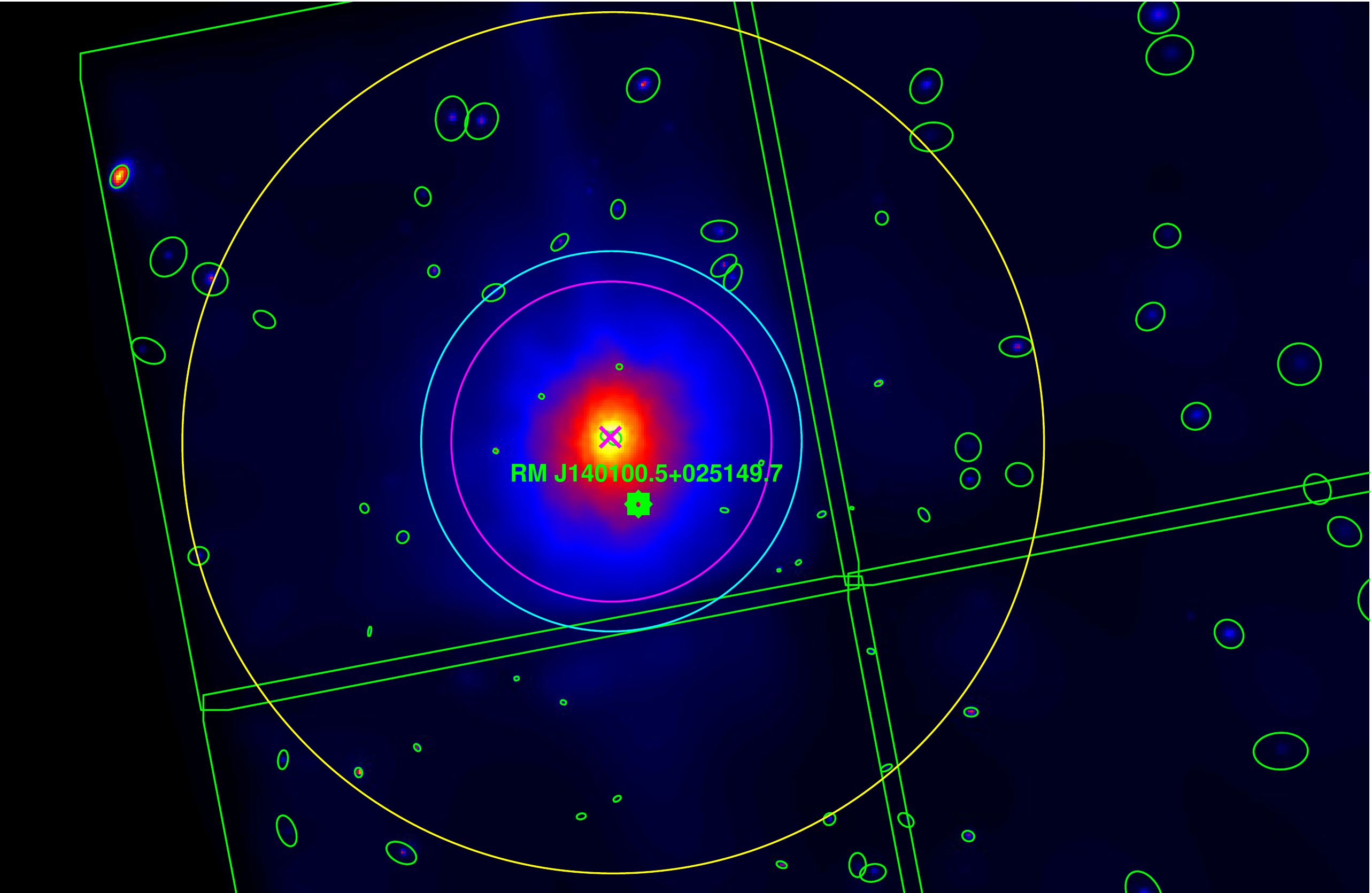}{0.5\textwidth}{(d)}
  }
  \caption{%
    (a) RM J230707.5+163246.1 (\memmatch{} 46, $z = 0.25$), ObsID 17157, ACIS-I detector.
    This cluster has two sub-halos of nearly the same size, so the centroid is located between the two sub-halos.
    However, the \redmapper{} center is at the central galaxy of upper halo.
    The X-ray peak is on the lower halo, so in this case \redmapper{}'s center agrees with neither X-ray centering measure.
    (b) RM J233739.7+001616.9 (\memmatch{} 68, $z = 0.30$), ObsID 11728, ACIS-I detector.
    This cluster has been disturbed in such a way that the \redmapper{} center does not agree well with the X-ray centroid or with the X-ray peak.
    (c) RM J004143.0$-$092621.9 (\memmatch{} 145, $z = 0.07$), ObsID 16264, ACIS-I detector.
    In this example, \redmapper{} chose to center on the small southern substructure (at $z = 0.07$) instead of the large structure which dominates the observation (at $z = 0.06$).
    (d) RM J140100.5+025149.7 (\memmatch{} 43, $z = 0.26$), ObsID 6880, ACIS-I detector.
    Here, the central galaxy has an active galactic nucleus (marked by a green ellipse surrounding the X-ray peak).
    This causes the central galaxy to appear blue.
    As a result, \redmapper{} ignores the true central galaxy and instead chooses an off-center galaxy as its most-likely central galaxy.
    \colorinfo{}\label{fig:centering-images}
  }
\end{figure*}

Centroids and peaks have differing merits as measures of galaxy cluster centers in X-ray.
Consider a merging cluster composed of two sub-halos of roughly the same size, each within \rtfh{} (e.g.\ \autoref{fig:centering-images} (a)), or a cluster which is a composed of a single ``lumpy'' halo (e.g.\ \autoref{fig:centering-images} (b)).
In these cases, the centroid will be located between subhalos, near the cluster's center of mass.
The peak will be located on one of the subhalos, along with the \redmapper{} center, which by definition is centered on a galaxy.
Indeed, this similarity of definition implies that the \redmapper{} center should more closely align with the X-ray peak than the X-ray centroid as a general trend, although as demonstrated by both the above clusters it also is possible that the \redmapper{} center will not be on the same subhalo as the X-ray peak and will thus instead be nearer to the X-ray centroid.

For practical use, the ``correct'' choice of centering measure depends on the purpose for which you are using the centers.
For example a weak lensing pipeline based on simulations would probably wish to choose a centering measure which aligns with the centers chosen by their simulated data, irrespective of whether that center is the center of mass, the densest point, or something else.

\begin{figure*}[htbp!]
  \centering%
  \includegraphics[width=0.5\textwidth]{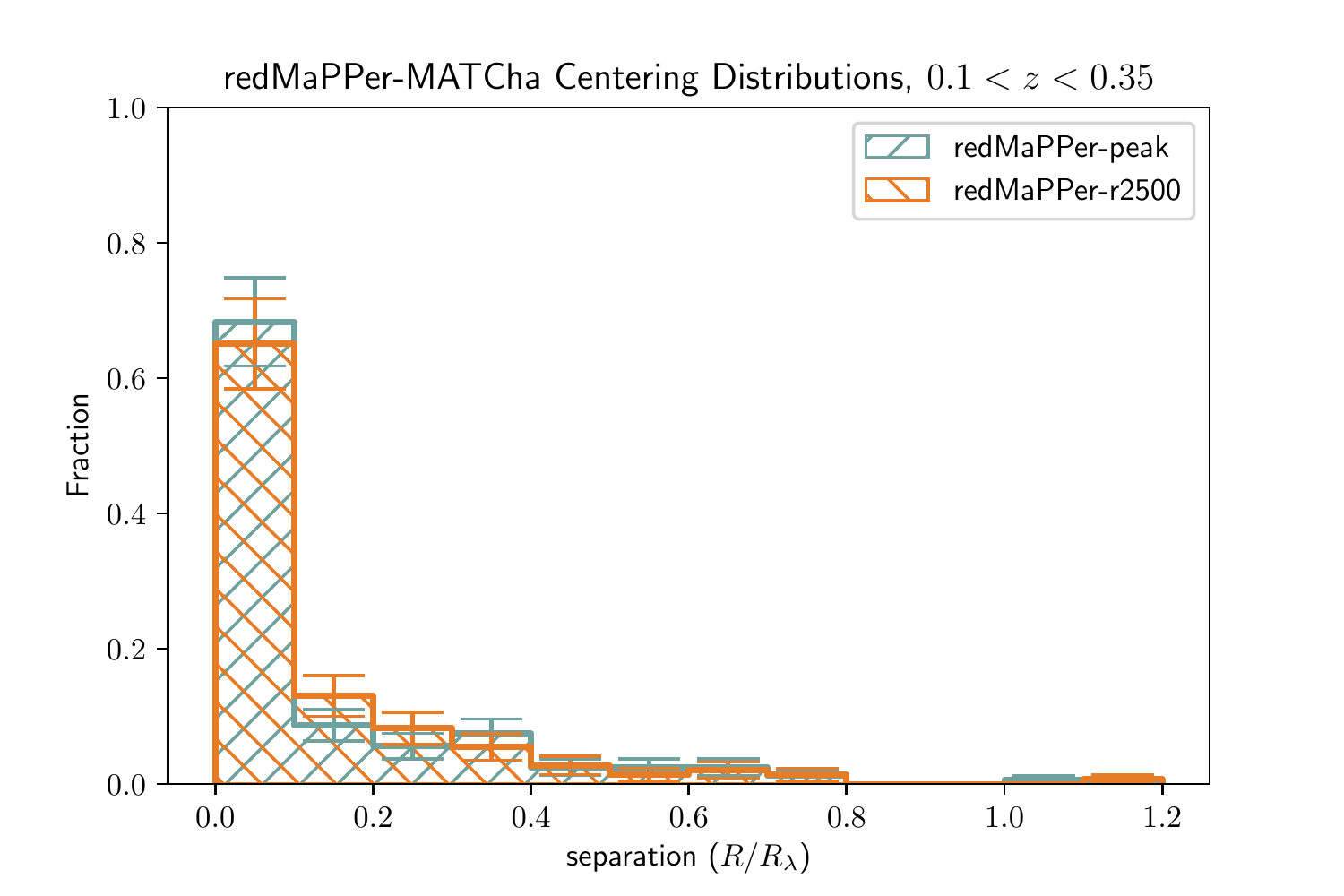}
  \caption{%
    Centering distributions.
    The histogram labeled ``\redmapper{}-peak'' shows the offset between the \redmapper{} central galaxy and the X-ray peak.
    The histogram labeled ``\redmapper{}-\rtfh{}'' shows the offset between the \redmapper{} central galaxy and the X-ray centroid within \rtfh{}.
    Of the two X-ray centering measures, the peak somewhat better agrees with the \redmapper{} definition of cluster center.\label{fig:centering}
  }
\end{figure*}

Comparisons of the \redmapper{} position, the \rtfh{} X-ray centroid, and the X-ray peak (after the sample cleaning discussed on \autoref{subsec:post-pipeline}) are shown in \autoref{fig:centering}.
We find that $68.3 \pm 6.5$\% of \redmapper{} BCGs are within 0.1 $R_\lambda$ of the peak and $65.1 \pm 6.7$\% are within the same distance of the centroid.
Here $R_\lambda$ is the richness-scaling radius measure defined in \citet{Rykoff12} and described in \autoref{sec:cluster-selection}.

In \autoref{fig:centering} it is clear that although many clusters are well-centered, there is a long tail to the \redmapper{} centering distribution.
In examining the clusters composing this tail, we identify the following major failure modes for \redmapper{} centering.

\begin{enumerate}
  \item \redmapper{} picks a central galaxy in a small cluster substructure, instead of in the main substructure.
  See \autoref{fig:centering-images} (c).

  \item \redmapper{} splits a single cluster into two separate clusters, or mis-assigns galaxies from one cluster to another nearby cluster.
  This can lead to choosing an incorrect center in the affected clusters and may cause \redmapper{} to assign wildly incorrect cluster richnesses.
  We call this problem ``mispercolation'' and discuss it in \autoref{subsec:mispercolations}.
  Figures~\ref{fig:mispercolation-a}--\ref{fig:mispercolation-b} examine the four cases of mispercolation that we observe in our data.

  \item The central galaxy is too blue and is thus ignored by \redmapper{}'s central galaxy selection algorithm.
  This leads \redmapper{} to choose an off-center galaxy instead.
  This problem often occurs when the central galaxy features an active galactic nucleus or significant ongoing star formation.
  See e.g.\ \autoref{fig:centering-images} (d).

  \item \redmapper{} misses the central galaxy due to masking caused by a bright star located along the line-of-sight, or due to data problems such as missing observations.
  See \citet{redmapperI} for more information.

\end{enumerate}

For more information on \redmapper{} centering, including data from \matcha{} and a comparison with X-ray centers from \xmm{}, see \citet{ZhangCentering}.

\section{Summary and Future Work}%
\label{sec:summary}
In this paper we introduce \matcha{}, a pipeline which is capable of performing parallel analysis of hundreds of galaxy clusters in archival \chandra{} data.
We run \matcha{} on the galaxy cluster catalog generated by \redmapper{}'s analysis of \sdss{} DR8 data.

Using this information we derive temperature-richness, luminosity-richness, and luminosity-temperature relations within \rtfh{} and \rfh{} apertures.
In particular, we find we find an \rtfh{} \tx{}-richness relation of $\txrelation{1.85 \pm{} 0.03}{0.52 \pm{} 0.05}$ and a standard-deviation-of-intrinsic-scatter of $0.27 \pm{} 0.02$ ($1 \sigma$) within $0.1 < z < 0.35$.
We also derive a number of other \tx{}-richness, \lx{}-richness, \lx{}-\tx{}, and \tx{}-\lx{} relations within \rtfh{} and \rfh{} apertures.
Our data offer improved constraints on $\sigmaintr$ when compared with similar prior work.
We find a slightly greater \tx{}-richness slope than that presented in \citet{redmapperII} ($1.4 \sigma$), and a much larger standard deviation of intrinsic scatter.
We find a similar $\sigmaintr$ to \citet{redmapperSV}, and here our slope is smaller than theirs by $0.9 \sigma$.
Finally, we find that our bolometric \lx{}-\tx{} relation's slope agrees well with \citet{Hicks13}, however we derive a much lower slope than \citet{Maughan12}.

We then measure the miscentering distribution in \redmapper{} by comparing the locations of \redmapper{}'s bright central galaxies with X-ray centroids and peaks measured by \matcha{}.
We find that $\approx 68\%$ of the clusters are well-centered.
We explore the tail of our centering distribution and identify failure modes of the \redmapper{} centering algorithm.

In addition to this current work, \matcha{} has already been used in large-scale X-ray analyses such as \citet{Bufanda}, in which \matcha{} is used to examine the AGN population of galaxy clusters; and \citet{redmapperSV}, in which \matcha{} is used to analyze DES Science Verification Data.
Further \matcha{} results on DES Year 1 and \sdss{} data will be presented in papers on \redmapper{} centering~\citep{ZhangCentering}, \redmapper{} scaling relations~\citep{FarahiScaling}, and cosmology results from both \redmapper{} \sdss{} DR8~\citep{SloanCosmology} and DES Y1~\citep{Y1Cosmology}.

\acknowledgments{}
\section*{Acknowledgments}
This material is based upon work supported by the U.S. Department of Energy, Office of Science, Office of High Energy Physics, under Award Numbers DE-SC0013541 and DE-SC0007093. KR and PG acknowledge support from the UK Science and Technology Facilities Council via grant ST/N504452/1.

Support for this work was provided by the National Aeronautics and Space Administration through Chandra Award Number AR4--15014X issued by the Chandra X-ray Center, which is operated by the Smithsonian Astrophysical Observatory for and on behalf of the National Aeronautics Space Administration under contract NAS8--03060.

Funding for SDSS-III has been provided by the Alfred P. Sloan Foundation, the Participating Institutions, the National Science Foundation, and the U.S. Department of Energy Office of Science. The SDSS-III web site is http://www.sdss3.org/.

SDSS-III is managed by the Astrophysical Research Consortium for the Participating Institutions of the SDSS-III Collaboration including the University of Arizona, the Brazilian Participation Group, Brookhaven National Laboratory, Carnegie Mellon University, University of Florida, the French Participation Group, the German Participation Group, Harvard University, the Instituto de Astrofisica de Canarias, the Michigan State/Notre Dame/JINA Participation Group, Johns Hopkins University, Lawrence Berkeley National Laboratory, Max Planck Institute for Astrophysics, Max Planck Institute for Extraterrestrial Physics, New Mexico State University, New York University, Ohio State University, Pennsylvania State University, University of Portsmouth, Princeton University, the Spanish Participation Group, University of Tokyo, University of Utah, Vanderbilt University, University of Virginia, University of Washington, and Yale University.

\appendix{}

\section{Example Images}%
\label{sec:example-images}

\begin{figure*}[h!]
  \centering
  \gridline{%
    \fig{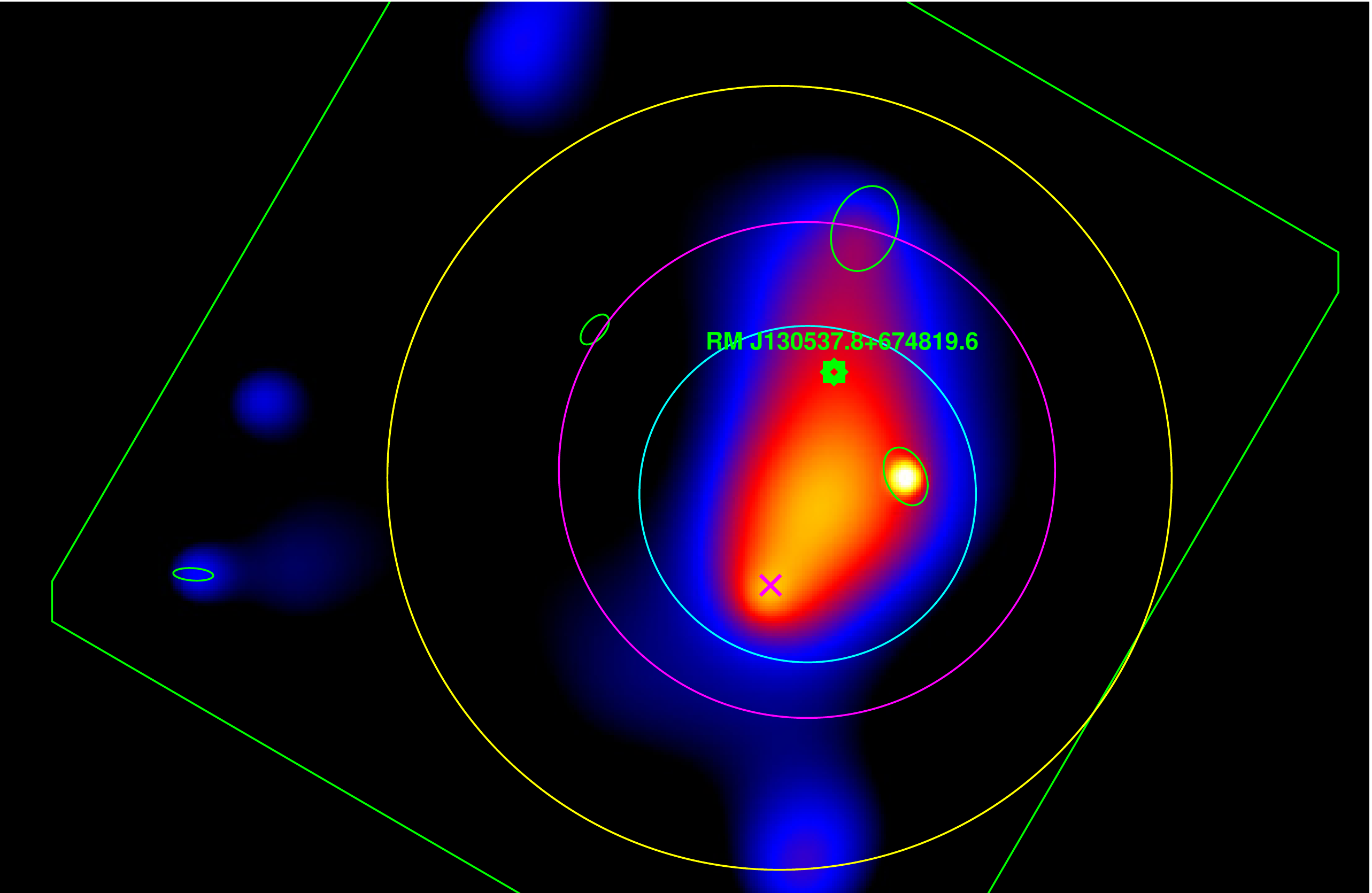}{0.5\textwidth}{(a)}
    \fig{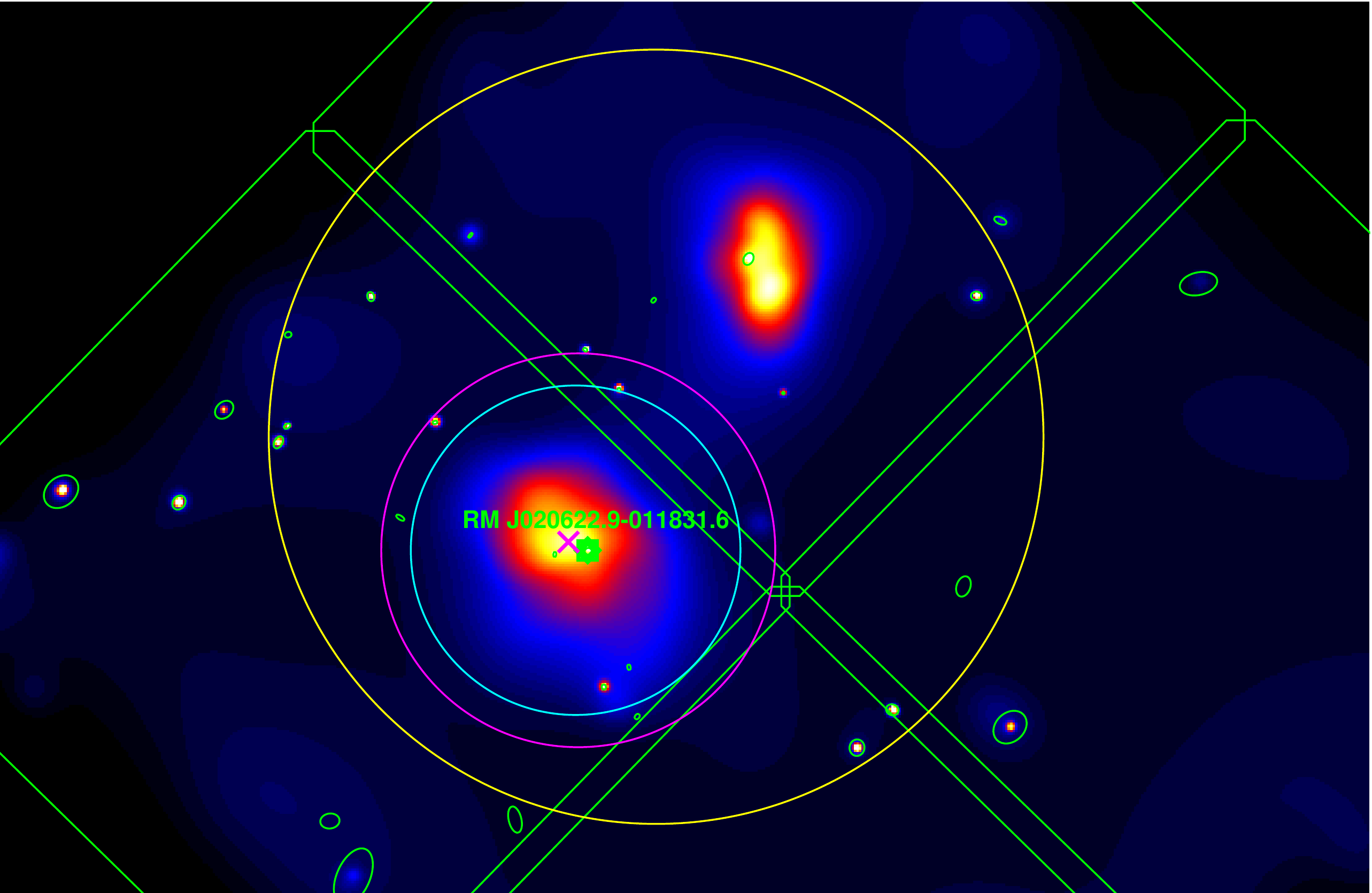}{0.5\textwidth}{(b)}
  }
  \gridline{%
    \fig{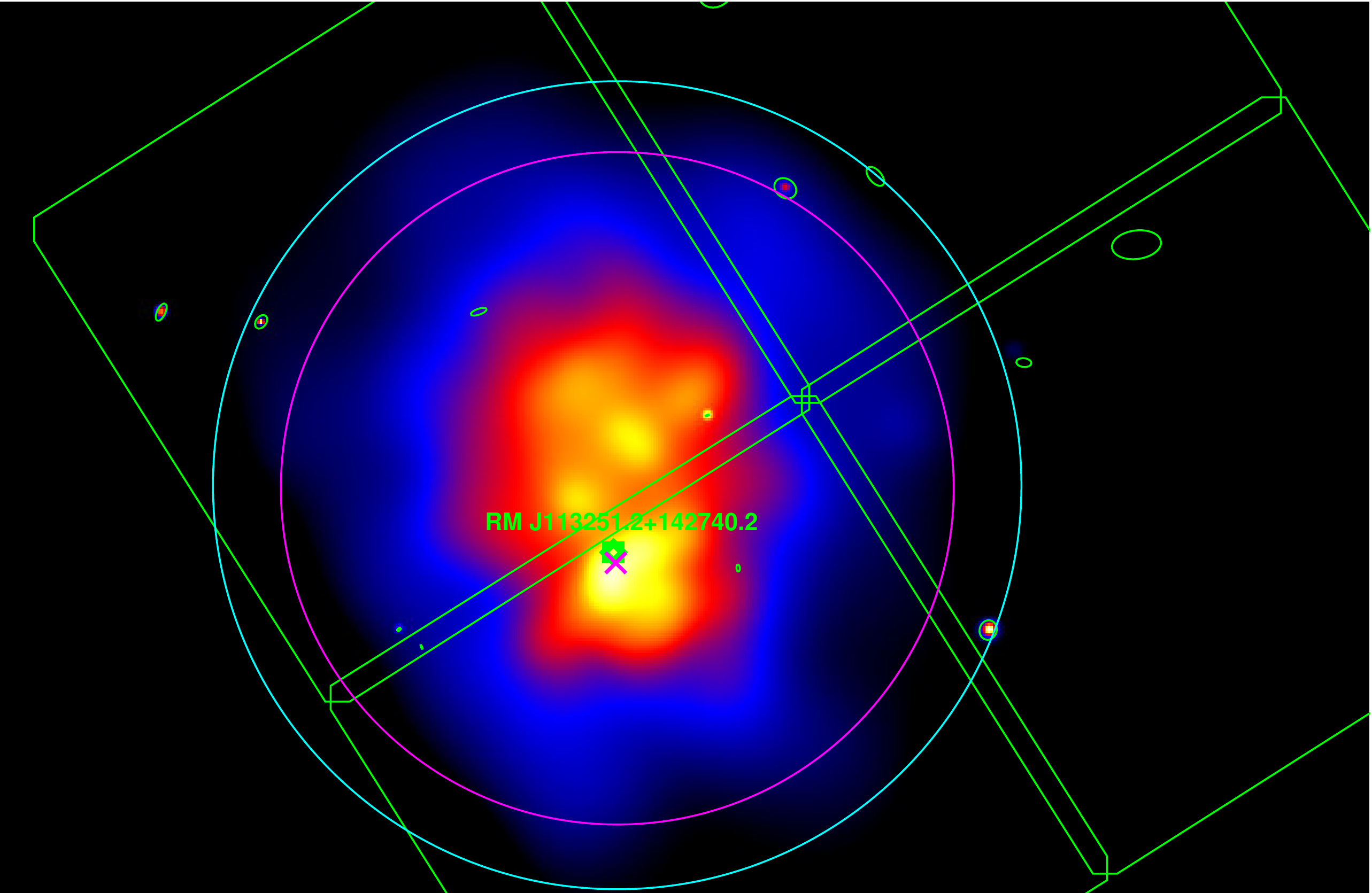}{0.5\textwidth}{(c)}
  }
  \caption{%
    (a) RM J130537.8+674819.6 (\memmatch{} 2598, $z = 0.23$), ObsID 10868, ACIS-S2 chip.
    This is an example of the output of \matcha{} for an asymmetric, and is also an example of a serendipitous cluster.
    (b) RM J020622.9$-$011831.6 (\memmatch{} 1382, $z = 0.19$), ObsID 16229, ACIS-I detector.
    This is an example of a cluster featuring substructure.
    In this case, \matcha{} analyzes the southern structure only until it reaches the \rfh{} analysis, at which point it moves the centroid to a point between the two clusters.
    (c) RM J113251.2+142740.2 (\memmatch{} 384, $z = 0.09$), ObsID 14387, ACIS-I detector.
    This is an example of a low-redshift cluster, which takes up much of the \chandra{} observation.
    \rfh{} does not fit on the observation, and is thus not analyzed.
    \matcha{} automatically handles the fact that \rtfh{} goes off the observation, see \autoref{subsec:aperture}.
    \colorinfo{}\label{fig:matcha-visual}
  }
\end{figure*}

\begin{figure*}[h!]
  \centering
  \gridline{%
    \fig{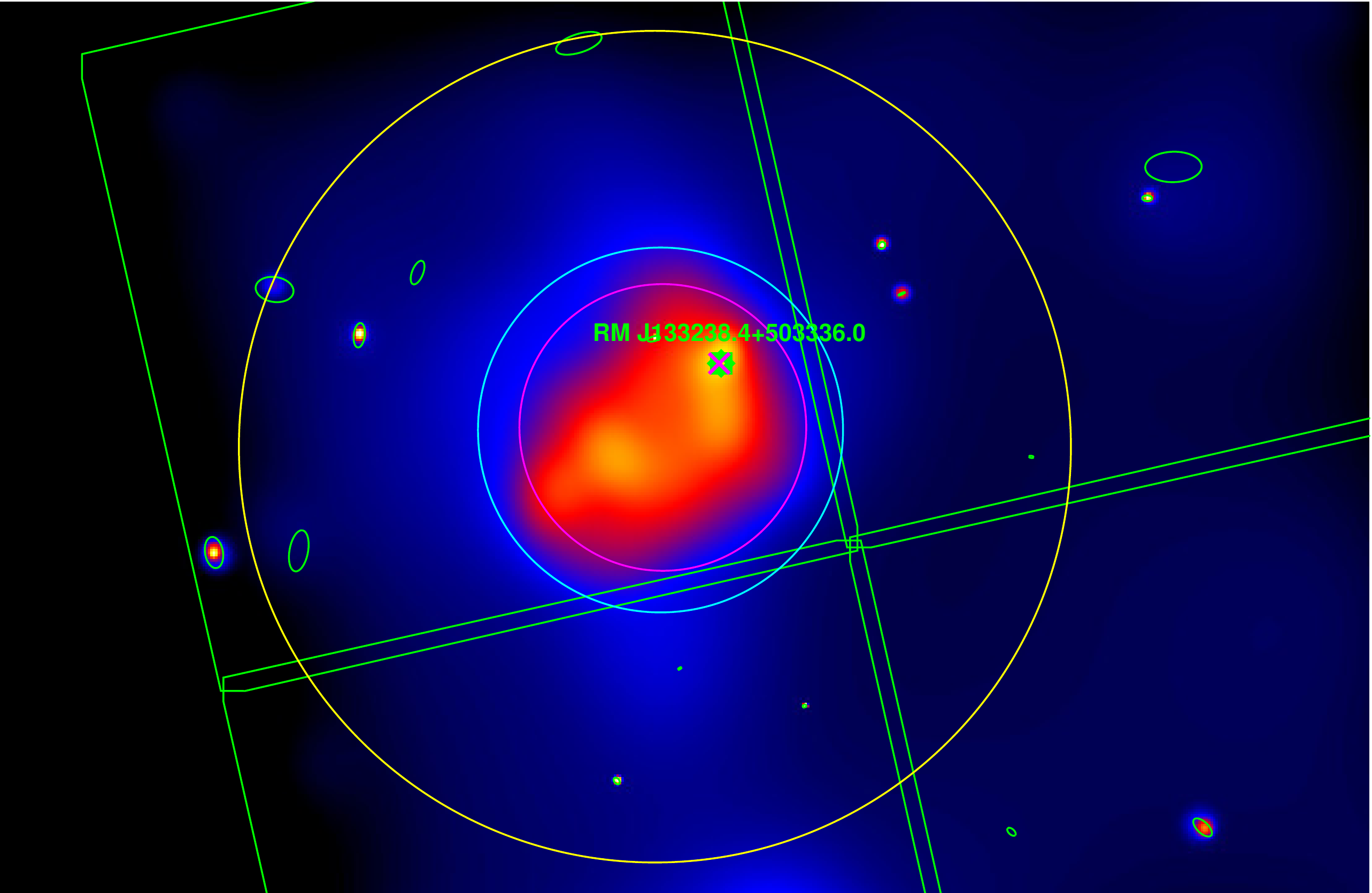}{0.5\textwidth}{(a)}
    \fig{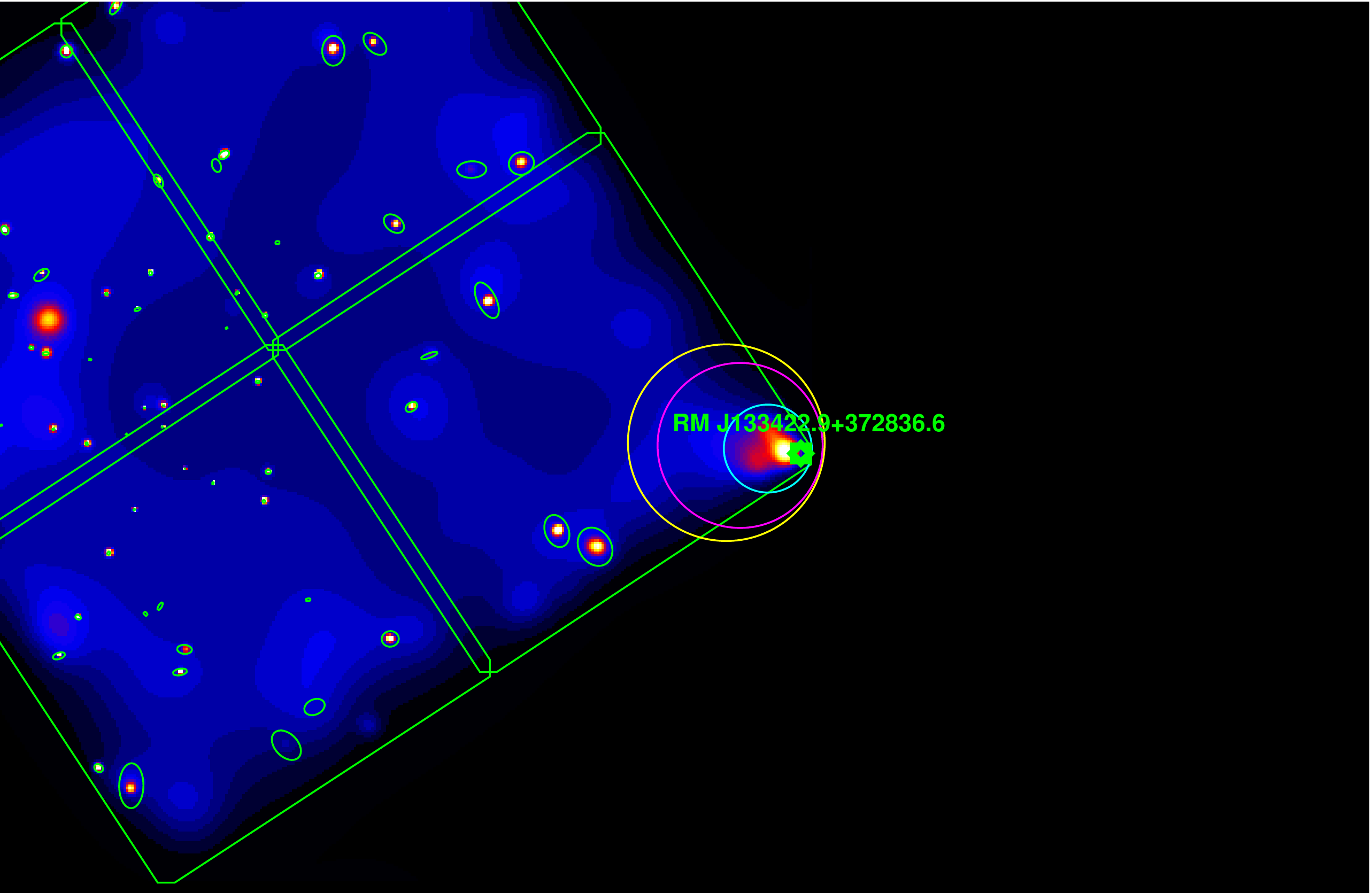}{0.5\textwidth}{(b)}
  }
  \gridline{%
    \fig{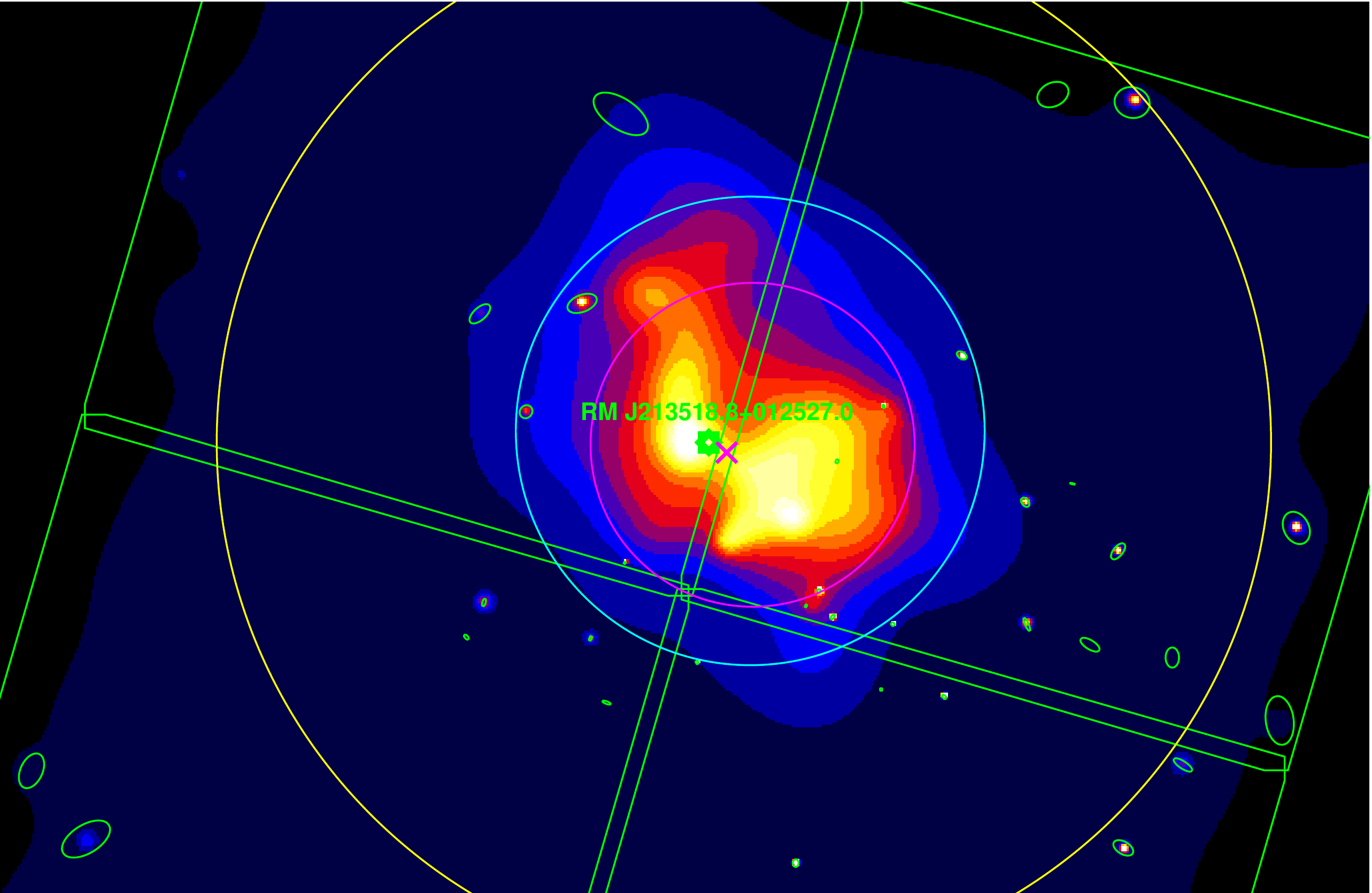}{0.5\textwidth}{(c)}
    \fig{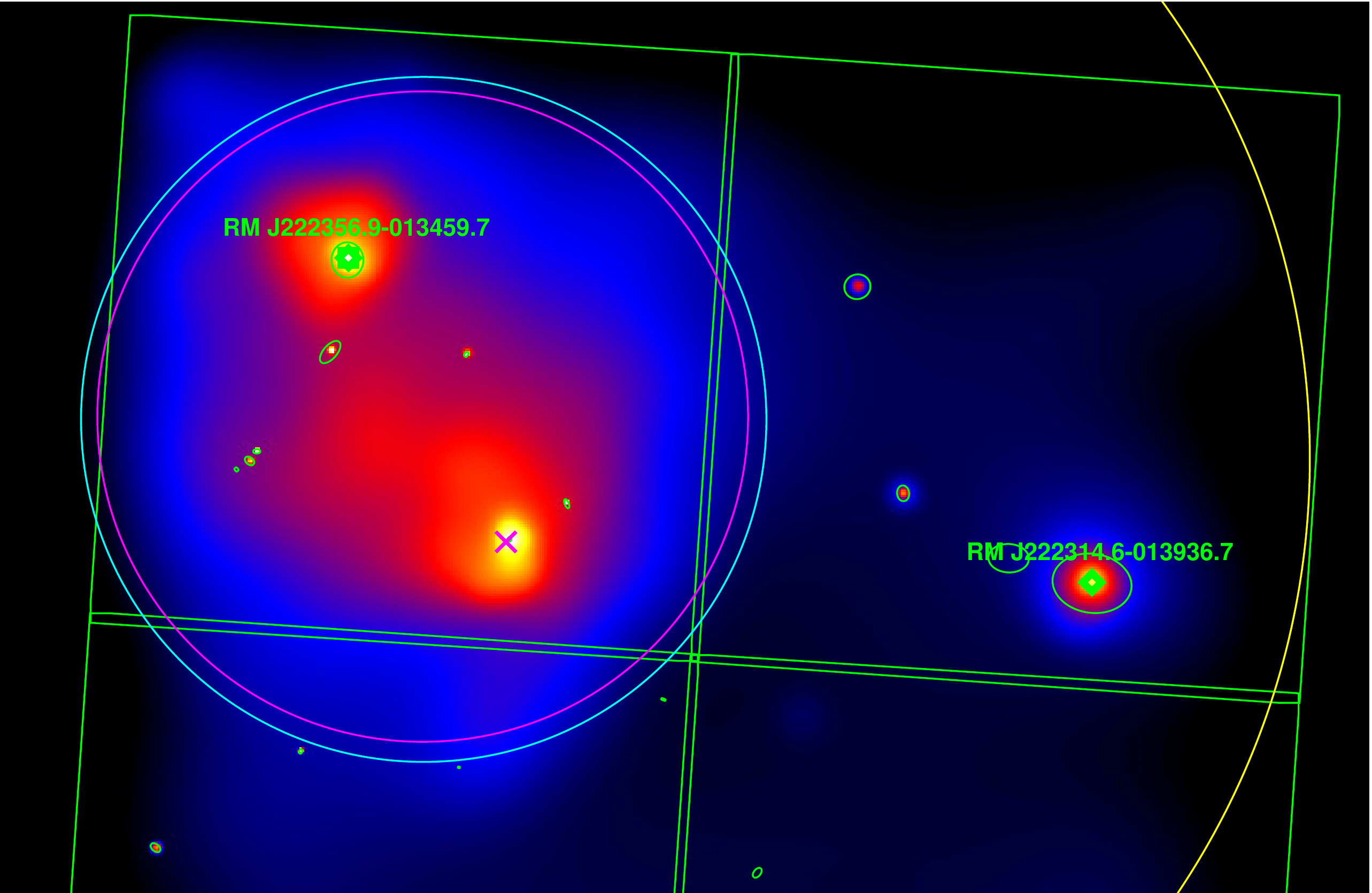}{0.5\textwidth}{(d)}
  }
  \caption{%
    (a) RM J133238.4+503336.0 (\memmatch{} 35, $z = 0.29$), ObsID 7710, ACIS-I detector.
    Because this cluster is disturbed, the X-ray centroid does not agree with the \redmapper{} BCG, nor should it.
    For this reason, this cluster's centroid is removed from the centering checks on \redmapper{} in \autoref{subsec:centering}.
    However the peak is kept, and the cluster's \lx{} and \tx{} values are used in the scaling relations in \autoref{sec:results}.
    (b) RM J133422.9+372836.6 (\memmatch{} 1806, $z = 0.31$), ObsID 12307, ACIS-I detector.
    Here, \matcha{} successfully determines temperatures and luminosities for both \rtfh{} and \rfh{}, but these are suspect because the cluster is right in the corner of its only observation.
    Additionally, the centroid determination is adversely affected by the proiximity to the chip edge.
    Thus, this cluster is removed from the data for both centering and scaling relations.
    (c) RM J213518.8+012527.0 (\memmatch{} 155, $z = 0.23$), ObsID 15097, ACIS-I detector.
    Here, \matcha{} successfully determines a temperature for both \rtfh{} and \rfh{}, but unfortunately it has done so for the wrong cluster---this \redmapper{} cluster at $z = 0.23$ is being masked by ABELL 2355 at $z = 0.12$.
    This cluster is removed from the data for both centering and scaling relations.
    (d) RM J222356.9$-$013459.7 (\memmatch{} 48, $z = 0.10$), ObsID 15107, ACIS-I detector.
    Here, emission from RM J222314.6$-$013936.7 (\memmatch{} 7113, $z = 0.30$, right) contaminates the background spectrum for RM J222356.9$-$013459.7 (left).
    For this reasons, RM J222356.9$-$013459.7 (left) is removed from the data for scaling relations.
    \colorinfo{}\label{fig:bad-xray}
  }
\end{figure*}

Here we present sample images of \chandra{} observations produced by \matcha{} as described in \autoref{sec:matcha}.
\autoref{fig:matcha-visual} demonstrates \matcha{} output for an asymmetric cluster, a cluster with substructure, and a low-redshift cluster.
\autoref{fig:bad-xray} demonstrates various cases in which \matcha{} gives a result which is either incorrect or not useful.
The correction for and accounting of these errors is discussed in \autoref{subsec:post-pipeline}.

\section{Flags and Data Cuts}%
\label{sec:flag-effects}
Here we present the effects on our data of each individual flag which we use in \autoref{subsec:post-pipeline}.
In the interest of reproducibility we present these effects in the $0.1 < z < 0.35$ redshift range unless otherwise noted.
This is the range which gives the most accurate \redmapper{} results (see \autoref{subsec:lambda-scaling-relations}) and is the same redshift range for which we release our data in \autoref{sec:matcha-data}.

\begin{figure*}[htbp!]
  \centering
  Flags and their Effect on Data
  \gridline{%
    \fig{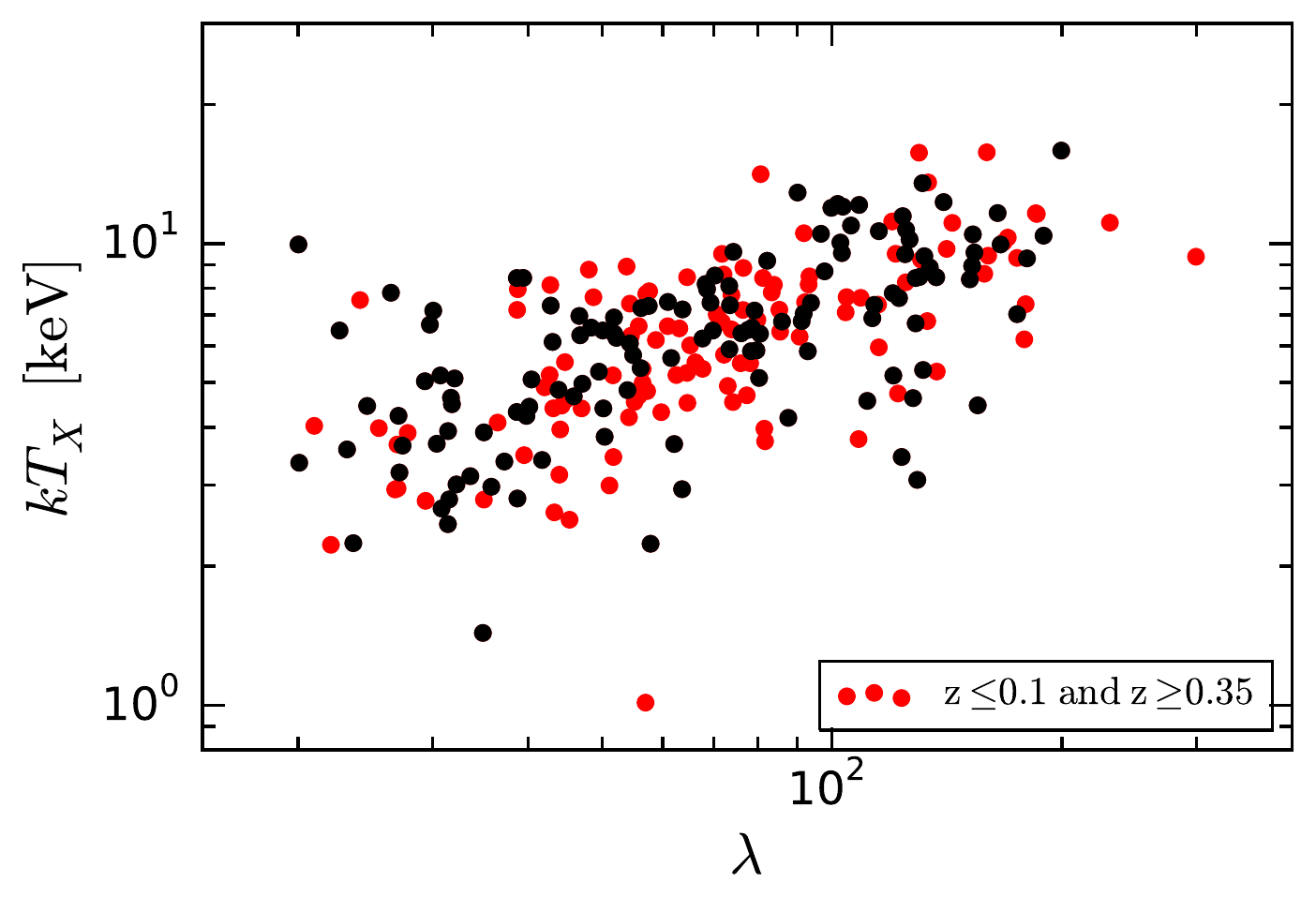}{0.5\textwidth}{(a)}
    \fig{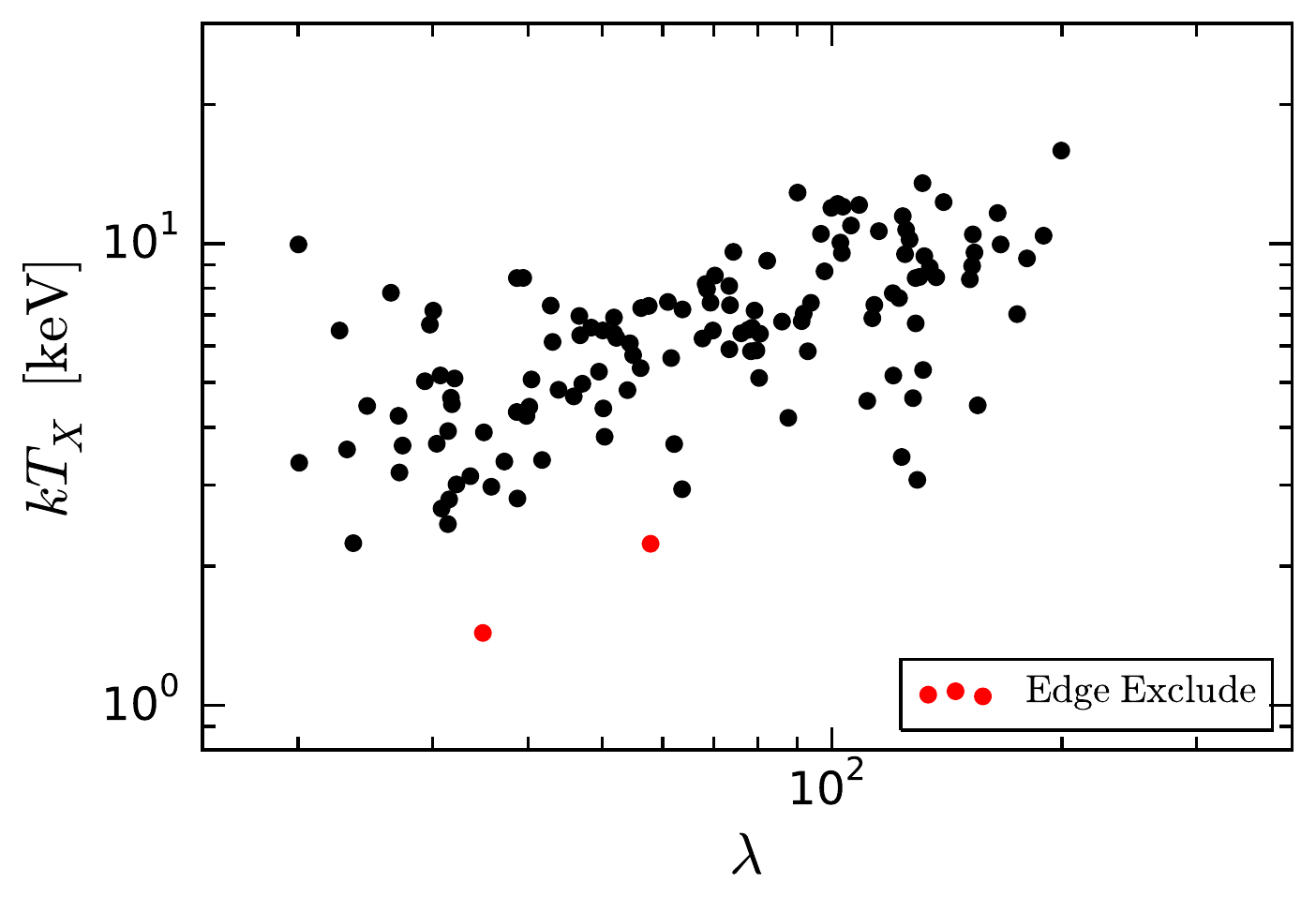}{0.5\textwidth}{(b)}
  }
  \gridline{%
    \fig{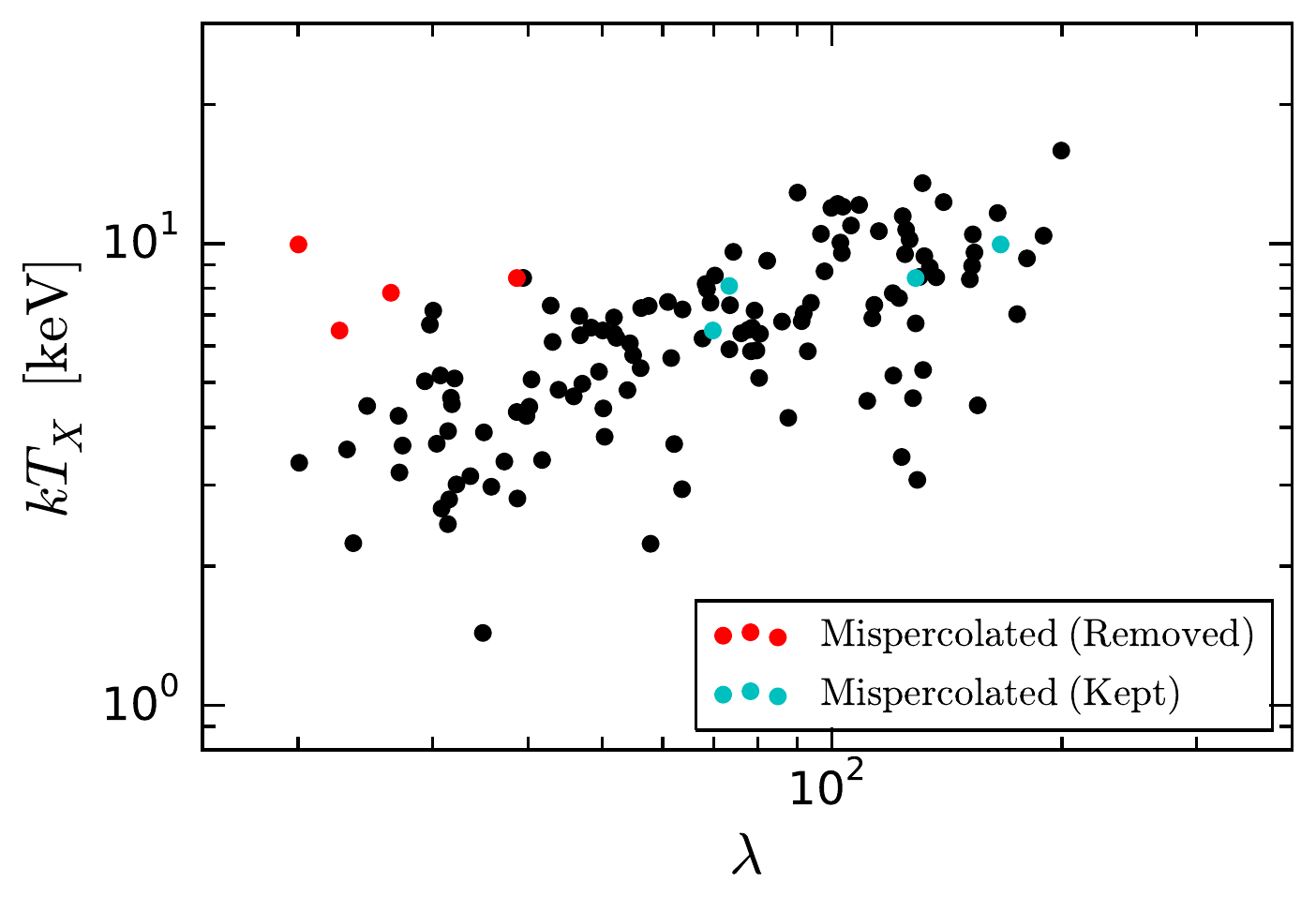}{0.5\textwidth}{(c)}
    \fig{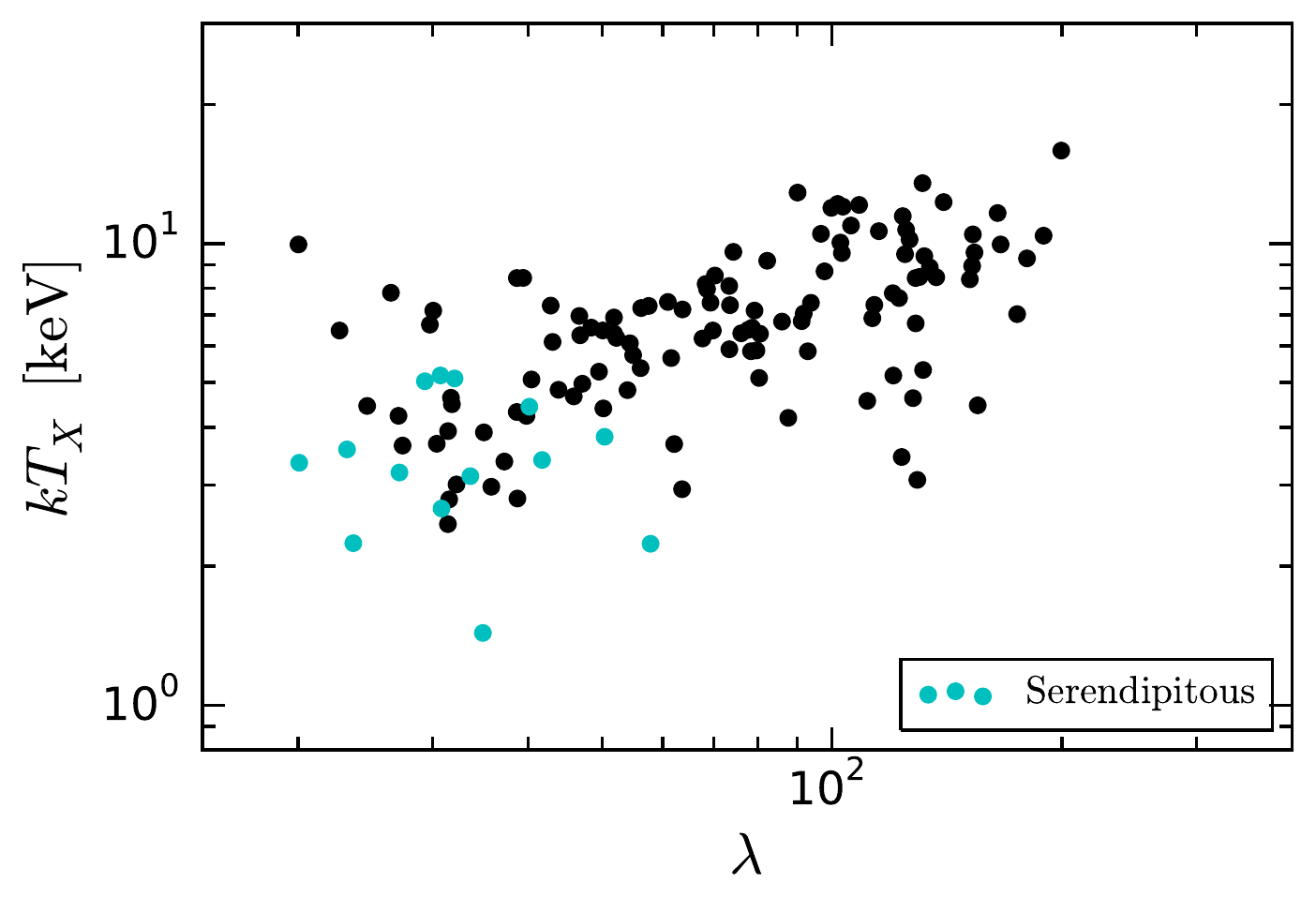}{0.5\textwidth}{(d)}
  }
  \caption{%
    In each image, the red dots mark clusters which we cut from at least one of the relations in \autoref{subsec:lambda-scaling-relations}--\autoref{subsec:x-ray-scaling-relations}, cyan dots mark interesting clusters which are not cut, and black dots mark un-flagged data points.
    (a) Here we see the effects of limiting our data to $0.1 < z < 0.35$.
    The red dots lie outside this redshift range; the black dots lie inside of it.
    The choice to limit our data to $0.1 < z < 0.35$ does not seem to preferentially select particular temperatures or richnesses, and indeed in Tables~\ref{tab:r2500-scaling-relations}--\ref{tab:r500-scaling-relations} we see no significant effect of this choice upon our fitted slope or scatter.
    This redshift cut removes a number of \tx{}-$\lambda$ outliers.
    We conjecture that these outliers are caused by issues with \redmapper{}'s richness assignment outside of $0.1 < z < 0.35$, see \citet{redmapperI} for details.
    (b) Here we see the effects of excluding clusters which are marked as being likely too close to the chip edge to reliably measure their \rtfh{} temperature.
    The two excluded clusters marked in red are clear outliers, likely with under-estimated temperatures.
    (c) Here we see the effects of excluding mispercolated halos.
    The four mispercolated halos marked in red are clear outliers with under-estimated richnesses, as expected (see \autoref{subsec:mispercolations}).
    Marked in cyan are the corresponding halos which we keep after manually adjusting their properties (see \autoref{tab:mispercolated}).
    (d) Marked in cyan here are our serendipitous clusters (see \autoref{subsec:post-pipeline}).
    These clusters primarily inhabit the low-temperature, low-richness regime; this is somewhat intuitive because more massive clusters at these redshifts are more likely to have been specifically studied in X-ray.
    Unfortunately, there are too few serendipitous clusters to use them to draw meaningful conclusions about the effects of our archival cluster selection, as briefly discussed in \autoref{subsec:selection-effects}.\label{fig:flag-effects}
  }
\end{figure*}

In the four subplots of \autoref{fig:flag-effects} we highlight, on an \rtfh{} \tx{}-$\lambda$ plot, (a) clusters outside the $0.1 < z < 0.35$ range, (b) clusters which are too close to chip edges, (c) mispercolated clusters, and (d) serendipitous clusters.
As expected, the redshift restriction does not seem to preferentially bias the data.
Additionally, the data show that proximity to an edge leads to under-estimating \tx{} and mispercolation leads to under-estimating richness.
For a discussion of serendipitous clusters, see \autoref{subsec:selection-effects}.

\section{\matcha{} Data}%
\label{sec:matcha-data}
Here we present the galaxy cluster data produced by our \matcha{} pipeline.
We include data for each cluster within $0.1 < z < 0.35$, except for those which have unusable \chandra{} data or which are masked by another cluster (see \autoref{subsec:post-pipeline}).
In \autoref{tab:data-basic}, we record each cluster's \redmapper{} \memmatch{}, name, list of \chandra{} observations, redshift, and flags.
In \autoref{tab:data-scaling}, we list the \redmapper{} \memmatch{}, richness, \tx{} data, and \lx{} data and for each cluster.
In \autoref{tab:data-centering}, we give the \redmapper{} \memmatch{} and centering information for each cluster.
These tables are available in full in machine readable format; the first five rows (arranged by \memmatch{}) of each table are shown here as a reference for their form and content.

\begin{splitdeluxetable*}{rlrrBllllrrr}
  \tablecaption{\matcha{} Cluster Name and Flags}
  \tablehead{\colhead{Mem Match ID} & \colhead{Name} & \colhead{ObsIDs} & \colhead{Redshift} & \colhead{Detected} & \colhead{On Chip Edge} & \colhead{On Off-Axis Chip} & \colhead{Serendipitous} & \colhead{500 kpc SNR} & \colhead{500 kpc SNR Error}}
  \startdata{}%
  2 & RM J164019.80+464241.50 & 896,7892,13988,14355,14356,14431,14451 & 0.23 & True & False & False & False & 470.77 & 1.09 \\
  3 & RM J131129.50$-$12028.00 & 540,1663,5004,6930,7289,7701 & 0.18 & True & False & False & False & 587.42 & 1.10 \\
  5 & RM J90912.20+105824.90 & 924,7699 & 0.17 & True & False & False & False & 114.32 & 1.07 \\
  6 & RM J133520.10+410004.10 & 3591 & 0.23 & True & True & False & False & 91.18 & 1.07 \\
  11 & RM J82529.10+470800.90 & 15159 & 0.13 & True & False & False & False & 44.48 & 1.05 \\
  \enddata{}
  \tablecomments{This table is available in full in machine readable format; the first 5 rows (arranged by \memmatch{}) are shown here as a reference for its form and content.
    The ``\memmatch{}'' column contains the cluster's \redmapper{} \memmatch{}.
    This is unique to each cluster, allowing for easy cross-referencing of clusters between tables.
    The ``Name'' column gives the \redmapper{} name of the cluster.
    The ``ObsIDs'' column gives a comma-delimited list of \chandra{} observations used in the analysis of the cluster.
    The ``Redshift'' column gives the \redmapper{}-determined redshift for the cluster.
    The ``Detected'' column is a Boolean value which is true if the cluster was detected (SNR $\geq$ 5.0).
    See \autoref{subsec:finding-tx} for details.
    The ``On Chip Edge'' column is a Boolean value which is true if the cluster is on a chip edge in all of its observations.
    Note that it is not necessarily a problem for this to be the case---one needs to refer to the relevant ``Edge Exclusion'' columns in \autoref{tab:data-scaling} and \autoref{tab:data-centering}.
    The ``On Off-Axis Chip'' column is a Boolean value which is true if the cluster is on non-aimpoint chips for all of its observations.
    The ``Serendipitous'' column is a Boolean value which is true if the cluster is never the aimpoint of an observation.
    This is somewhat subjective, so we focus on eliminating false positives.
    That is, if a cluster is marked ``serendipitous'', it is certainly not the aimpoint of any observation under consideration; if it is not marked ``serendipitous'', it may still be the case that it is not the aimpoint of any observation under consideration.
    Finally, the ``500 kpc SNR'' and ``500 kpc SNR Error'' columns contain respectively the signal-to-noise ratio within a 500 kpc aperture and its $1\sigma$-equivalent uncertainty.
    See \autoref{subsec:finding-tx}.\label{tab:data-basic}
  }
\end{splitdeluxetable*}

\begin{splitdeluxetable*}{rrrrrrrrrBrrrrrrrrrrBrrrrrrrr}
  \tabletypesize{\scriptsize}
  \tablecaption{\matcha{} Cluster Scaling Relations Information}
  \tablehead{%
    \colhead{Mem Match ID}
    & \colhead{Lambda} & \colhead{Lambda Error}
    & \colhead{\rtfh{} \lx{}} & \colhead{\rtfh{} \lx{} $-$} & \colhead{\rtfh{} \lx{} $+$}
    & \colhead{\rtfh{} Bolo \lx{}} & \colhead{\rtfh{} Bolo \lx{} $-$} & \colhead{\rtfh{} Bolo \lx{} $+$}
    & \colhead{\rtfh{} \tx{}} & \colhead{\rtfh{} \tx{} $-$} & \colhead{\rtfh{} \tx{} $+$}
    & \colhead{\rfh{} \lx{}} & \colhead{\rfh{} \lx{} $-$} & \colhead{\rfh{} \lx{} $+$}
    & \colhead{\rfh{} Bolo \lx{}} & \colhead{\rfh{} Bolo \lx{} $-$} & \colhead{\rfh{} Bolo \lx{} $+$}
    & \colhead{\rfh{} \tx{}} & \colhead{\rfh{} \tx{} $-$} & \colhead{\rfh{} \tx{} $+$}
    & \colhead{Fixed-\tx{} \lx{}} & \colhead{Fixed-\tx{} \lx{} $-$} & \colhead{Fixed-\tx{} \lx{} $+$}
    & \colhead{\lx{} Upper Limit}
    & \colhead{\rtfh{} Edge Exclusion}
    & \colhead{\rfh{} Edge Exclusion}
  }
  \startdata{}%
  2 & 199.54 & 5.30 & 8.84e+44 & 2.00e+42 & 1.90e+42 & 3.92e+45 & 1.27e+43 & 1.26e+43 & 15.91 & 0.26 & 0.26 & 1.04e+45 & 3.22e+42 & 3.92e+42 & 4.59e+45 & 2.92e+43 & 1.68e+43 & 15.36 & 0.34 & 0.38 & NULL & NULL & NULL & NULL & False & False \\
  3 & 164.71 & 4.24 & 7.25e+44 & 1.57e+42 & 1.46e+42 & 2.88e+45 & 7.76e+42 & 7.59e+42 & 11.65 & 0.11 & 0.11 & 7.80e+44 & 2.47e+42 & 1.24e+42 & 3.19e+45 & 7.97e+42 & 1.55e+43 & 12.89 & 0.20 & 0.20 & NULL & NULL & NULL & NULL & False & False \\
  5 & 174.70 & 4.95 & 1.76e+44 & 1.71e+42 & 1.65e+42 & 5.76e+44 & 7.33e+42 & 7.24e+42 & 7.03 & 0.24 & 0.24 & 2.65e+44 & 2.77e+42 & 2.45e+42 & 8.26e+44 & 1.31e+43 & 1.32e+43 & 6.28 & 0.29 & 0.28 & NULL & NULL & NULL & NULL & False & False \\
  6 & 189.18 & 5.61 & 3.81e+44 & 4.39e+42 & 4.67e+42 & 1.47e+45 & 2.37e+43 & 2.38e+43 & 10.40 & 0.63 & 0.63 & 5.08e+44 & 5.35e+42 & 7.16e+42 & 1.85e+45 & 4.08e+43 & 2.81e+43 & 8.77 & 0.47 & 0.50 & NULL & NULL & NULL & NULL & False & False \\
  11 & 131.58 & 4.81 & 8.00e+43 & 2.00e+42 & 2.02e+42 & 2.30e+44 & 6.79e+42 & 6.91e+42 & 5.32 & 0.40 & 0.42 & 1.20e+44 & 3.07e+42 & 3.51e+42 & 3.59e+44 & 1.26e+43 & 2.01e+43 & 5.85 & 0.60 & 0.86 & NULL & NULL & NULL & NULL & False & True \\
  \enddata{}
  \tabletypesize{\normalsize}
  \tablecomments{Galaxy cluster \memmatch{}s and scaling relation-related information.
    This table is available in full in machine readable format; the first 5 rows (arranged by \memmatch{}) are shown here as a reference for its form and content.
    The ``Lambda and ``Lambda Error'' columns contains the richness assigned to the cluster by \redmapper{}, and its (symmetric) $1\sigma$ uncertainty.
    The columns labeled \rtfh{}/\rfh{} \lx{}/\tx{} contain the respective values determined for the cluster (if any) and the associated uncertainties.
    \lx{} columns marked ``Bolo'' contain bolometric luminositities; the other \lx{} columns contain soft-band luminosities.
    The ``Fixed-\tx{} \lx{}'' columns contain the \lx{} value determined for the cluster if it is calculated via the method outlined in \autoref{subsec:lx-no-tx} and associated uncertainty (see \autoref{subsec:lx-no-tx}).
    The ``\lx{} Upper Limit'' column contains the upper-limit \lx{} determined for the cluster if the cluster is not considered detected (see \autoref{subsec:lx-upper-lims}).
    In this table, \lx{} values have units of ergs/s and \tx{} values have units of keV.
    The \rtfh{} / \rfh{} ``Edge Exclusion'' columns contain Boolean values which are true if proximity to the chip edge is considered to be a problem for \lx{} and \tx{} in \rtfh{} / \rfh{}.
    When these ``Edge Exclusion'' columns are true, the given cluster is removed from the relevant scaling relation (see \autoref{subsec:post-pipeline}).\label{tab:data-scaling}
  }
\end{splitdeluxetable*}

\begin{splitdeluxetable*}{rrrrrBrrrrr}
  \tablecaption{\matcha{} Cluster Centering Information}
  \tablehead{%
    \colhead{Mem Match ID}
    & \colhead{\redmapper{} RA} & \colhead{\redmapper{} Dec}
    & \colhead{\rtfh{} Centroid RA} & \colhead{\rtfh{} Centroid Dec}
    & \colhead{\rfh{} Centroid RA} & \colhead{\rfh{} Centroid Dec}
    & \colhead{X-Ray Peak RA} & \colhead{X-Ray Peak DEC} & \colhead{Edge Exclusion}
  }
  \startdata{}%
  2 & 250.082548387 & 46.7115313536 & 250.085144 & 46.708554 & 250.081223 & 46.709479 & 250.082507143 & 46.7105242857 & False \\
  3 & 197.872957171 & -1.34111627953 & 197.87343 & -1.340688 & 197.8766 & -1.337763 & 197.873123333 & -1.341575 & False \\
  5 & 137.300744635 & 10.9735949355 & 137.302175 & 10.97637 & 137.299865 & 10.984945 & 137.30308 & 10.975423 & False \\
  6 & 203.833722679 & 41.0011464409 & 203.83041 & 41.00074 & 203.82491 & 40.99719 & 203.82647 & 41.00031 & False \\
  11 & 126.371092335 & 47.1335713021 & 126.37128 & 47.13054 & 126.36986 & 47.13133 & 126.37321 & 47.13104 & False \\
  \enddata{}
  \tablecomments{Galaxy cluster \memmatch{}s and centering relation-related information.
    This table is available in full in machine readable format; the first 5 rows (arranged by \memmatch{}) are shown here as a reference for its form and content.
    The ``\redmapper{} RA'' and ``\redmapper{} Dec'' columns give the \redmapper{} BCG position for each cluster.
    The ``\rtfh{} Centroid RA'' and ``\rtfh{} Centroid Dec'' columns give the position of the \rtfh{} centroid for each cluster.
    The ``\rfh{} Centroid RA'' and ``\rfh{} Centroid Dec'' columns give the position of the \rfh{} centroid for each cluster.
    The ``X-Ray Peak RA'' and ``X-Ray Peak Dec'' columns give the position of the X-ray peak for each cluster.
    The ``Edge Exclusion'' column contains a Boolean value which is true if the given cluster's proximity to chip edges is considered a problem for centering.
    When this ``Edge Exclusion'' column is true, the given cluster is removed from the centering distribution (see \autoref{subsec:post-pipeline}).\label{tab:data-centering}
  }
\end{splitdeluxetable*}
\clearpage
\bibliography{references}{}

\end{document}